\documentclass[usenatbib,useAMS,usegraphicx]{mn2e} 

\input{journal.cls}

\newcommand{\bm}[1]{\mbox{\boldmath$ #1 $}}
\newcommand{\lskip}{\vskip \baselineskip}
\newcommand{\nskip}{\lskip \noindent}
\newcommand{\halfskip}{\vskip 0.5\baselineskip}
\newcommand{\be}{\begin{equation}}
\newcommand{\ee}{\end{equation}}
\newcommand{\equref}[1]{(\ref{#1})}

\usepackage{hyperref}
\usepackage[table]{xcolor}
\usepackage{graphicx,epsfig}
\usepackage{verbatim}
\usepackage{syntonly}
\usepackage{txfonts}

\title[Simulating relativistic AGN jets I. Steady axisymmetric jets]{Relativistic AGN jets I.
The delicate interplay between jet structure, cocoon morphology and jet-head propagation}

\author[S. Walg et al.]{
S. ~Walg, $^{1,2}$\thanks{email: \url{s.walg@astro.ru.nl}}
A. ~Achterberg, $^1$
S. ~Markoff, $^2$
R. ~Keppens, $^3$
Z. ~Meliani  $^4$\\
$^1$Astronomical Institute, Radboud University Nijmegen,
                                   Heyendaalseweg 135, 6525 AJ Nijmegen, The Netherlands\\
$^2$Astronomical Institute "Anton Pannekoek," University of Amsterdam,
                                   Science Park 904, 1098 XH Amsterdam, The Netherlands\\
$^3$Centre for mathematical Plasma Astrophysics, Department of Mathematics, KU Leuven,
                                   Celestijnenlaan 200B, 3001 Heverlee, Belgium\\
$^4$LUTH, Observatoire de Paris, France
}

\date{\today}

\pagerange{\pageref{firstpage}--\pageref{lastpage}}
\volume{0000}
\pubyear{2013}

\begin{document}

\maketitle

\label{firstpage}

\begin{abstract}
Current observations have shown that astrophysical jets reveal strong signs of radial structure. They suggest
that the inner region of the jet, the jet spine, consists of a low-density, fast-moving gas, while the outer
region of the jet consists of a more dense and slower moving gas, called the jet sheath. Moreover, if jets
carry angular momentum, the resultant centrifugal forces lead to a radial stratification. Current
observations are not able to fully resolve the radial structure, so little is known about its actual profile.
We present three AGN jet models in $2.5D$ of which two have been given a radial structure. The first
model is a homogeneous jet, the only model that doesn't carry angular momentum; the second model is a
spine-sheath jet with an isothermal equation of state; and the third jet model is a (piecewise) isochoric
spine-sheath jet, with constant but different densities for jet spine and jet sheath.
In this paper, we look at the effects of radial stratification on jet integrity, mixing between the
different jet components and global morphology of the jet-head and surrounding cocoon. We consider steady jets
that have been active for 23 Myr. All jets have developed the same number of strong internal shocks along
their jet axis at the final time of simulation. These shocks arise when vortices are being shed by the jet-head.
We find that all three jets maintain their stability all the way up to the jet-head. The isothermal jet
maintains part of its structural integrity at the jet-head where the distinction between jet spine and jet
sheath material can still be made. In this case, mixing between jet spine and jet sheath within the jet is
fairly inefficient. The isochoric jet, on the other hand, loses its structural jet integrity fairly quickly
after the jet is injected. At its jet-head, little structure is maintained and the central part of the jet
predominantly consists of jet sheath material. In this case, jet spine and jet sheath material mix efficiently
within the jet. We find that the propagation speed for all three models is less than expected from simple
theoretical predictions. We propose this is due to an enlarged cross section of the jet which impacts with the
ambient medium. We show that in these models, the effective surface area is 16 times as large in case of the
homogeneous jet, 30 times as large in case of the isochoric jet and can be up to 40 times as large in case of
the isothermal jet.
\end{abstract}


\begin{keywords}
Hydrodynamics\ -- Relativistic processes -- Intracluster medium\ -- Jets\ -- \\
Numerical methods
\end{keywords}

\section{Introduction} \label{sec:Introduction} 

Astrophysical jets are highly collimated outflows of plasma, generated near a compact object from its accretion
disk in accreting systems. Jets on parsec (pc) scales are known to arise from a stellar mass compact object in
close binaries, such as a white dwarf (WD), a neutron star (NS) or black hole (BH), while jets on kpc-Mpc scales
are associated with active galactic nuclei (AGN), where gas is accreted onto a super massive black hole (SMBH)
of $10^6-10^{10} \; M_{\odot}$.
In this paper we will only focus on jets arising from super massive black hole systems.

Observations show strong signs that astrophysical jets have a transverse (radial) structure (see for instance
\citealt*{Sol1989}; \citealt{Giroletti2004}; \citealt*{Ghisellini2005}; \citealt{Gomez2008}). It has been
suggested that most jets consist of two different regions, namely a low-density, fast-moving inner region called
the {\it jet spine}, thought to emerge from a region very close to the BH, and a denser and slower moving outer
region called the {\it jet sheath}, thought to emerge from the inner accretion disk. Numerical simulations of
accretion near black holes also show such a radial structure emerging, (e.g. \citealt*{Hardee2007};
\citealt{Porth2010}). However, the formation, properties and evolution of jet spine and jet sheath are not well
understood. In fact, whether the observed radial structure is actually the result of an onderlying spine-sheath
jet structure has not been verified by observations.

Large-scale jets are usually divided into two categories, namely FRI and FRII jets \citep{Fanaroff1974}. The
distinction is based on jet/lobe luminosity (at 178 MHz) and radio morphology. FRI jets have low luminosity
($<10^{41}$ erg s$^{-1}$) and diffusive jets/radio lobes with no prominent hot spots. FRII jets have a high
luminosity ($>10^{41}$ erg s$^{-1}$), are generally thought to be more stable and collimated and do have
prominent hot spots.

Supersonic and under-dense
\footnote{Compared to the local intergalactic medium (IGM).}
jets inflate a hot and over-pressured cocoon through which shocked jet- and ambient material flows. These jets
deposit a large amount of energy into the surrounding medium and will alter their direct environment drastically.
This phenomenon ties in closely to the study of {\it AGN feedback}, the question of how part of the energy
produced by AGNs is put back into the intergalactic medium and how this influences galaxy evolution, (e.g.
\citealt{Ciotti2007}; \citealt{Schawinski2007}; \citealt{Sijacki2007}; \citealt*{Rafferty2008};
\citealt{Fabian2012}; \citealt*{Gitti2012}).

Even though there is strong evidence of a radial structure within AGN jets, the connection between this
structure and its impact on the IGM at large scales still remains largely unknown. Since the exact form of a
transverse stratification profile might have a large influence on the evolution of the jet at large scales,
a study about this aspect is clearly called for.

  \subsection{Main focus of this research}

AGN jets generally remain collimated over huge distances, reaching lengths up to hundreds of kpc or even
several Mpc. This implies that these jets either remain very stable internally and are not easily disrupted by
instabilities such as the  Kelvin-Helmholtz instability, or are confined by external pressure forces.

In this paper, we explore three different jet models, one radially uniform jet (from this point on
referred to as the {\em homogeneous} jet) and two jets with a different type of spine-sheath jet structure. We
study the effect of radial stratification on transverse jet integrity and quantify the mixing between jet
components in detail. Also, we closely look at the flow patterns that emerge within the jet-head. Moreover,
we study how these jets (initiated as typical FRII jets) and their surrounding cocoons have evolved after
they have been active for a period of \mbox{$\sim 10^7$ yr}. It is known that a jet and its surrounding cocoon
quickly achieve approximate pressure balance as the jet penetrates into the ambient medium. As a result, the
jet adapts to pressure variations that travel down the cocoon. We will look in more detail at these pressure
waves and how they relate to the formation of strong internal shocks within the jet. Finally, we will compare
the actual propagation of the jet-head to the propagation predicted by simple theory.

  \subsection{Outline of this paper}

The outline of this paper is as follows: In Section \ref{sec:Theory} we present background theory for our models.
Then in Section \ref{sec:Method} we discuss the method, numerical schemes and the parameter regime. In Section
\ref{sec:Results} we present the results of the different simulations. Discussion and conclusions are
found in Sections \ref{sec:Discussion} and \ref{sec:Conclusions}.

\section{Theoretical background}\label{sec:Theory}

  \subsection{Motivation for this research}

A number of numerical simulations have been conducted that study the interaction of (relativistic) jets with
their ambient medium. These studies include the pure hydrodynamical case (HD), as well as the
magnetohydrodynamical case (MHD), with the jet models set up in $2D$, $2.5D$ or $3D$. See for example
\citealt{Marti1997};
\citealt{Rosen1999};
\citealt{Aloy2000};
\citealt*{Meliani2008};
\citealt{Migone2010};
\citealt*{Perucho2011};
\citealt*{Bosch-Ramon2012};
\citealt{Gilkis2012};
\citealt{Prokhorov2012};
\citealt{Refaelovich2012};
\citealt{Soker2012};
\citealt*{Wagner2012}.
The dependence of the energy feedback from a homogeneous jet to the ambient medium on
the finite opening angle of a jet has been studied in detail by \citealt*{Monceau-Baroux2012}. Moreover,
\citet{Aloy2000} have studied jets with a spine-sheath jet structure, however, these jet models do not include
angular momentum. They do, however, include magnetic fields.

A global picture of the flow patterns within a jet and its surrounding cocoon has emerged, but a more detailed
description of the flow dynamics, and the role of a spine-sheath jet structure in particular is still missing.
Having a better understanding of these flow patterns will improve our view on AGN feedback in general. Relevant
questions are: How does the jet impact the ambient medium exactly? What part of the ambient medium undergoes
strong interaction with the jet and what part is merely deflected? How much mixing is there between shocked
ambient medium and shocked jet material? What effect will a different radial stratification have on the jet
integrity and possibly the formation and development of internal shocks? And in the case of structured jets,
how does spine and sheath material mix internally within the jet, as well as in their surrounding cocoon? 

Having a better understanding of the interplay between jet, cocoon and ambient medium, as well as the effect
of radial stratification on jet integrity and mixing effects could help us to search for and compare with
observational features.

  \subsection{Jet models}\label{subsec:jetmodels}

When dealing with jets, it is convenient to express their length scales in terms of the gravitational radius
of the black hole in the 'central engine' that feeds jet activity, \mbox{$R_{\rm g} = GM_{\rm BH}/c^2$}, with
$G$ the gravitational constant and $c$ the speed of light
\footnote{To give a sense for the dimensions, the gravitational radius for a black hole with
\mbox{$M_{BH} = 10^8 M_\odot$} is \mbox{$R_{\rm g} \sim 1.48 \times 10^8$ km $\approx 1$ AU.}}.

Theoretical considerations together with some observational evidence \citep[e.g.][]{Hada2011} point at a
situation where jets in general have distinct regions, characterized by processes that take place at
different distances from the central engine. If the jet launching mechanisms for black hole binaries (BHBs) and
AGNs are intrinsically similar, then we expect the processes that take place along the jet axis to be
approximately scale invariant. In that case, these characteristic regions are located at approximately the same
distance, when measured in units of the black hole gravitational radius $R_{\rm g}$. \citet{Lobanov2011}
discusses five such distinct regions.
VLBI observations of AGNs usually probe the collimation and acceleration region, which occurs at a distance of
about \mbox{$\sim 10^3 \; R_{\rm g}$}, where magnetic fields are still thought to play a significant role. In
some cases, VLBI observations of AGN jets are able to resolve up to even much smaller distances from the central
engine (in \citealt{Hada2011}, M87 is observed only a few tens of gravitational radii from the central engine and
recently \citealt{Doeleman2012} have been able to resolve the jet base and estimate this region to lie at a
distance of \mbox{$\sim 5.5 \; R_{\rm g}$} from the supermassive black hole).
However, in our simulations we will focus on the kinetic energy flux dominated (KFD) region of the jets,
which typically occurs at \mbox{$\sim 10^6 - 10^{11}\; R_{\rm g}$}.
There, the magnetic field is weak, so it doesn't significantly affect the dynamics of the jet flow
\footnote{The magnetic field in the KFD region does of course induce the observed synchrotron emission.}.
Therefore, we will not be primarily concerned with the dynamical effect of magnetic fields.

It is often assumed that a hot and tenuous plasma is present in the innermost regions of accretion, close to
the BH horizon and the innermost stable circular orbit, with magnetic field lines threading the BH
horizon. If the BH is spinning, gas and magnetic field lines are carried along by a general relativistic
effect called 'frame dragging', extracting angular momentum from the spinning BH \citep{Blandford1977}. It
is therefore expected that if jets indeed consist of a spine-sheath jet structure, the jet spine emerges from
this region, and consists of a hot, tenuous and fast-rotating gas. 

Further out, but still within the inner accretion disk, material is thought to be less hot and more dense,
rotating at lower velocities than material in the direct vicinity of the BH. The jet sheath is likely to emerge
from this region \citep{Blandford1982}. Therefore, it is expected that the jet sheath consists of a denser and
colder flow, with lower azimuthal velocities than the jet spine. At large distances from the central engine, the
jet sheath material still has a relativistic bulk velocity, but a lower Lorentz factor than that of the jet
spine. For work relating radiative features of AGN jets to a spine-sheath jet configuration, see for instance
\citet[][]{Ghisellini2005}.

In this paper we consider both a homogeneous jet with constant density and pressure over its cross section, as
well as jets with a spine-sheath jet configuration. It should be noted that not much is known about the actual
radial structure of a spine-sheath jet, so we will assume that all jets start out in pressure equilibrium with
their ambient medium. We consider two different types of structured jets: The first model uses a polytropic index
\mbox{$\Gamma = 5/3$} and is piecewise isochoric: a constant but different density for jet spine and jet sheath,
which we will refer to as the {\em isochoric jet} from now on. The other model is set up with an isothermal
equation of state and assumes a constant temperature across the jet cross section. We will
refer to this model as the {\em isothermal jet} from now on. These two cases result in a different radial
structure, as will be discussed in section \ref{subsec:radpres}. Jet spine and jet sheath are given different
values for density, pressure and velocity, and we allow for rotation around the jet axis, so that the jet
carries angular momentum.

  \subsection{Hydrodynamics: basic equations and methods}

\begin{figure*}
$
\begin{array}{c c}
\includegraphics[width=0.5\textwidth]{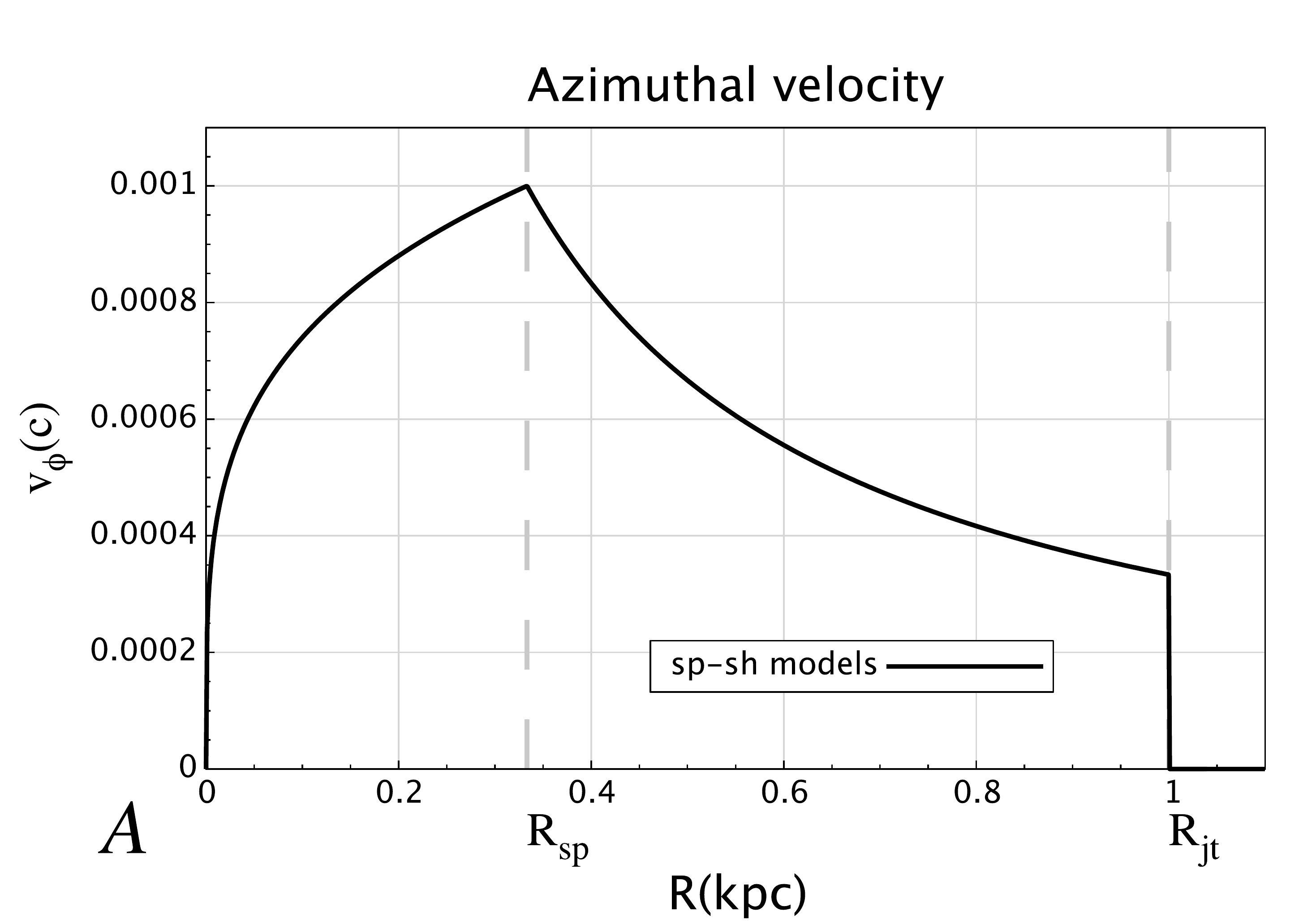} &
\includegraphics[width=0.5\textwidth]{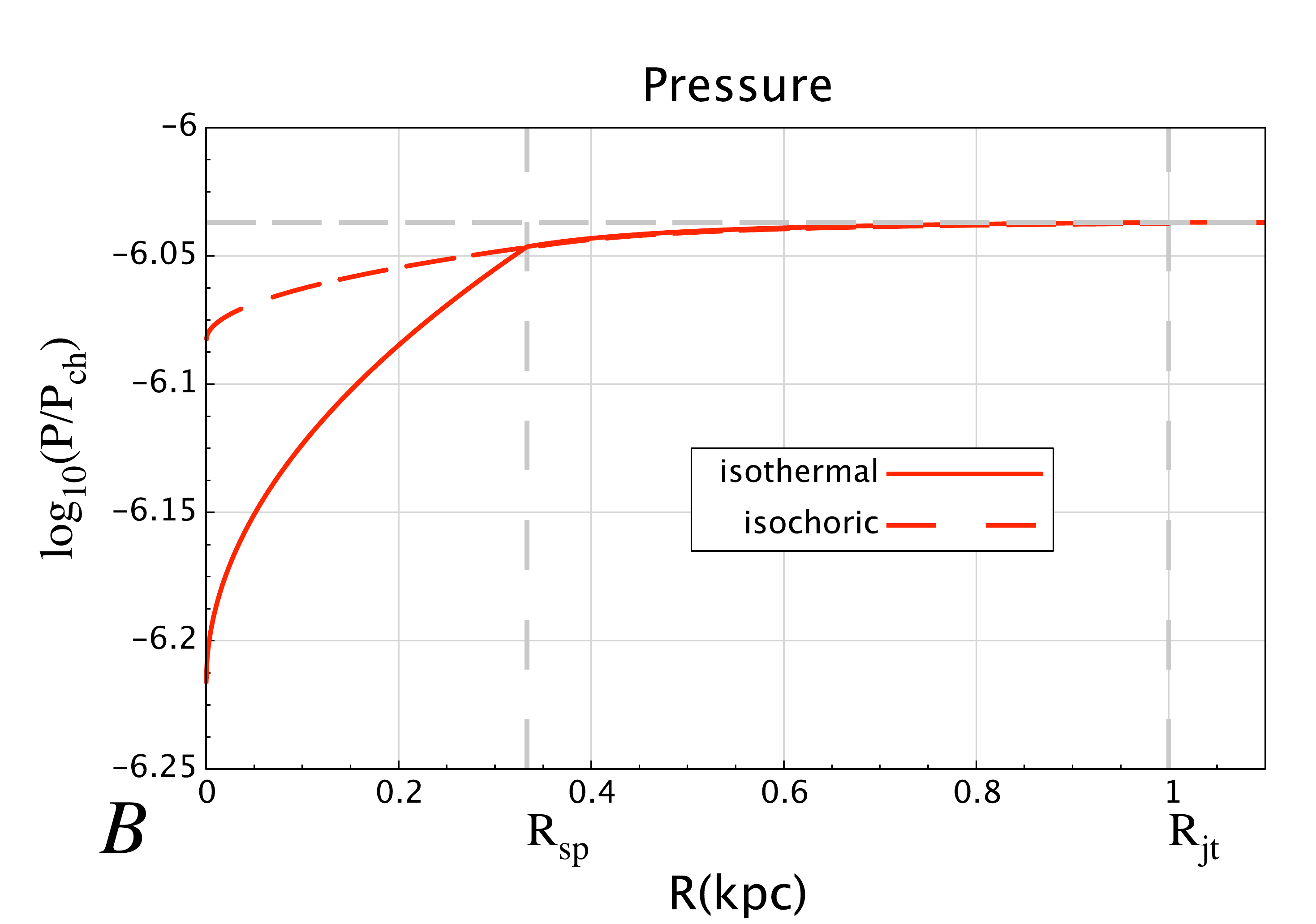} \\
\includegraphics[width=0.5\textwidth]{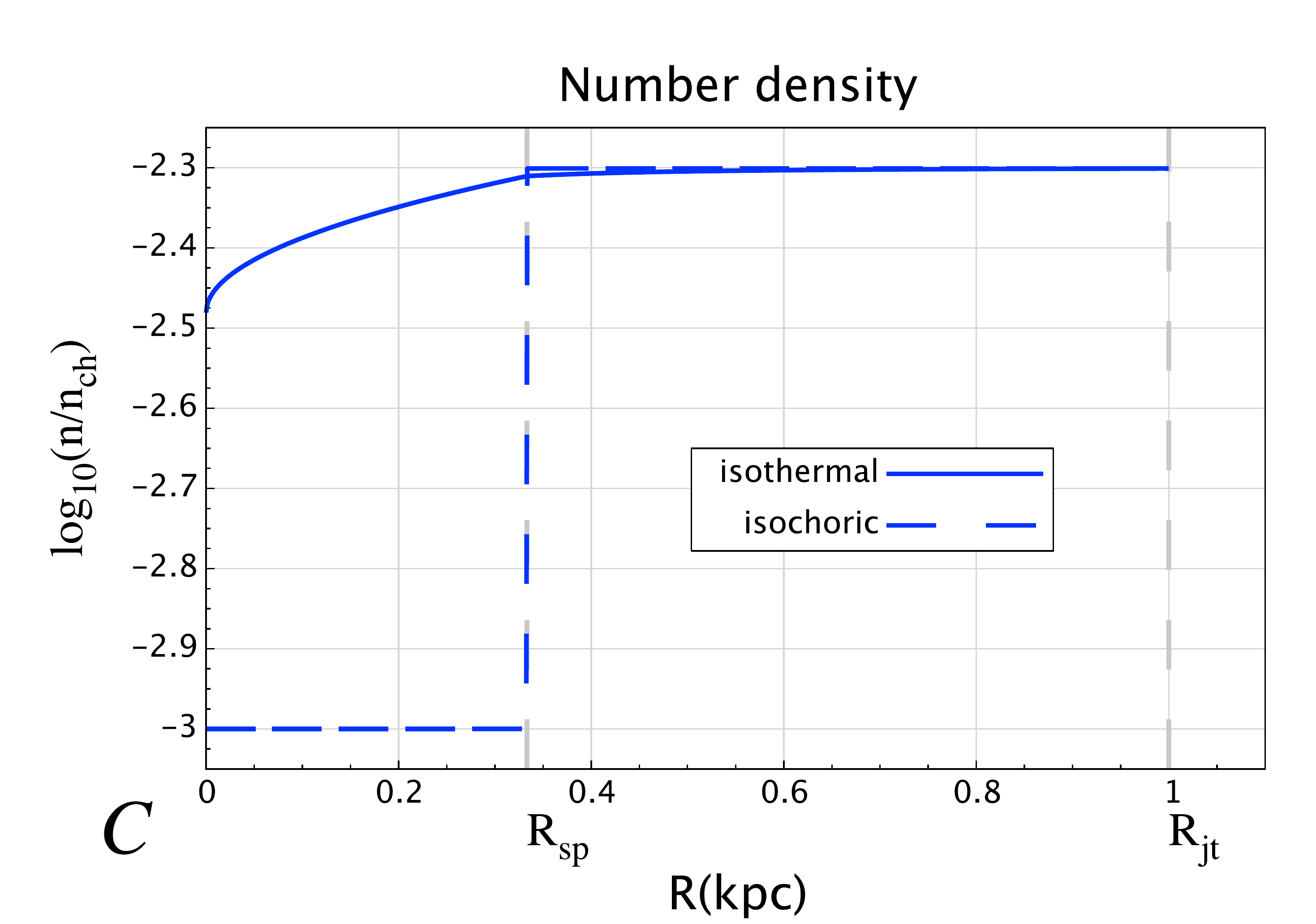} &
\includegraphics[width=0.5\textwidth]{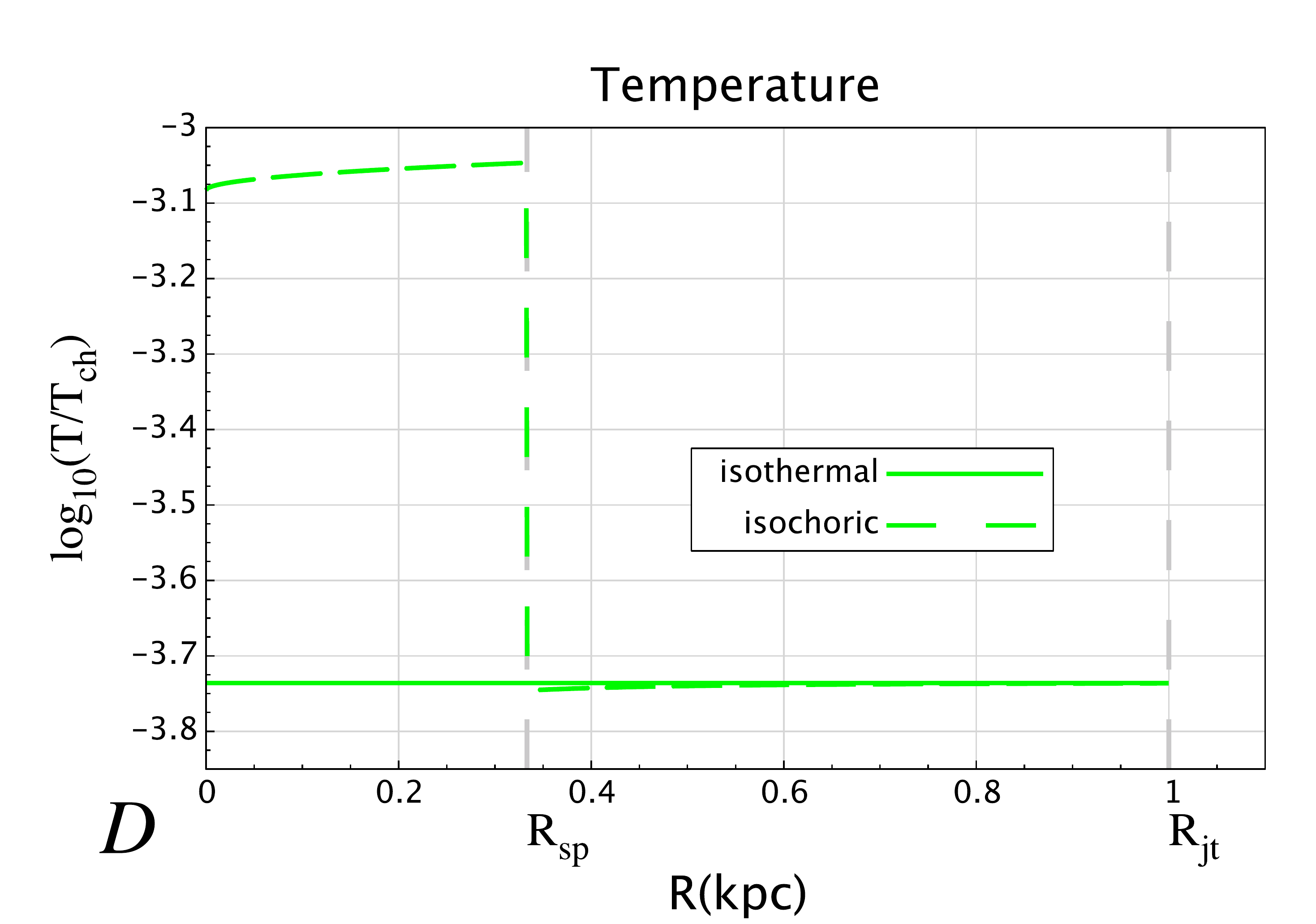}
\end{array}
$

\caption{Initial transverse jet profiles for the isothermal jet (solid lines) and the piecewise isochoric jet
(dashed lines), in case \mbox{$V_{\phi} = 1.0 \times 10^{-3} c$}. The cross cuts show in {\em black} the profile
of azimuthal rotation $v_{\phi}(R)$ \mbox{(figure {\bf A})}. This rotation profile has been used for both the
isothermal and the isochoric jet model; In {\em red}, the $\log_{10}$ of the pressure $P$ in units of the
characteristic pressure \mbox{$P_{\rm ch} = 1.50\times 10^{-6}$ erg cm$^{-3}$} \mbox{(figure {\bf B})};
In {\em blue} the $\log_{10}$ of the number density $n$ in units of the characteristic number density
\mbox{$n_{\rm ch} = 10^{-3}$ cm$^{-3}$} \mbox{(figure {\bf C})}; and in {\em green} the $\log_{10}$ of the
thermal temperature $T$ in units of the characteristic temperature \mbox{$T_{\rm ch} = 1.09\times 10^{13}$ K}
\mbox{(figure {\bf D})} of the jet. In addition, the images show the jet radius at \mbox{$R_{\rm jt} = 1$ kpc}
and the jet spine radius at \mbox{$R_{\rm sp} = R_{\rm jt}/3$} as the two vertical dashed lines. The pressure of
the ambient medium is denoted by the dashed horizontal line in \mbox{figure {\bf B}}.}

  \label{fig:Initial}
\end{figure*}

We have simulated the different jet models making use of the special relativistic, grid-adaptive
magneto-hydrodynamical code MPI-AMRVAC \citep{Keppens2012}. In classical and relativistic ideal hydrodynamics,
total mass, momentum and energy are conserved. That means that the fundamental equations can be cast in
conservative form, which in a 3+1 formulation read:

\begin{equation}
\frac{\partial U_i}{\partial t} + \nabla\cdot \bm{F}_{i} = 0 \; . \label{eq:conservationlaw}
\end{equation}
Here the $U_i$ (with $i=1-5$) are the conservative variables, and the $\bm{F_i}$ their corresponding fluxes.
These relations can be derived from the covariant formulation, and in particular from the vanishing divergence
of the energy-momentum tensor,  see for instance Weinberg (1972), Ch. 2.10. Employing units with $c=1$ from here
on, the conservative variables employed in MPI-AMRVAC are defined as:
\be
U = \left(
\begin{array}{c}
  D \\
  \bm{S} \\
  \tau
\end{array}
\right) \; \equiv
\left(
\begin{array}{c}
  \gamma\rho \\
  \gamma^2\rho h\bm{v} \\
  \gamma^2\rho h - P - \gamma\rho
\end{array}
\right) \; .
\ee
Here $\rho$ is the mass density in the jet rest frame, $\bm{v}$ is the velocity vector and
$\gamma = 1/\sqrt{1 - |\bm{v}|^2}$ the  associated Lorentz factor (proper speed: $\gamma \bm{v}$).
The vector $\bm{S}$ is the momentum density, $P$ is the pressure and $\tau$ is the kinetic energy
density that includes the kinetic energy of the bulk and thermal motion
\footnote{Or equivalently, $\tau$ is the total energy density in the lab-frame, with the lab-frame rest-mass
energy $\gamma\rho$ subtracted.}.
The (relativistic) specific enthalpy $h$ is:
\footnote{The quantity $\rho h$ is referred to as relativistic enthalpy.}

\be
	h = \frac{e + P}{\rho} \; ,
            \label{eq:enthalpy}
\ee
with $e = e_{\rm th} + \rho$, the total internal energy density, including the thermal energy density
$e_{\rm th}$ and the contribution of the rest-mass energy $\rho$. Moreover, $P$ is the gas pressure and
$\Gamma$ is the polytropic index of the gas. The corresponding fluxes are:

\be
	\bm{F} =
                 \left(
                 \begin{array}{c}
                   D \bm{v} \\
                   \bm{S \: v} + P \: \bm{\mathbfss{I}} \\
                   (\tau + P)\bm{v}
                 \end{array}
                 \right) \; ,
\ee
with $\textbfss{I}$ the $3\times 3$ identity matrix.

In order to obtain a complete description of the relativistic fluid, the system is closed with an equation of
state (EOS), relating gas pressure to mass density. Instead of simply putting
\mbox{$\Gamma \equiv {\rm d} \ln P/ {\rm d} \ln \rho$} equal to $5/3$ (for a classical ideal gas), or $4/3$ (for
a relativistically hot ideal gas), we employ the same interpolation function that was used in
\citet{Meliani2008}, describing a realistic transition between a relativistically hot gas and a 'cold'
non-relativistic gas. This interpolation function is called the Mathews approximation  \citep{Blumenthal1976}
and is based on the Synge EOS \citep{Synge1957}. The Mathews approximation uses an effective polytropic index
equal to:

\be
	\Gamma_{\rm eff} = \mbox{$\frac{5}{3}$} - \mbox{$\frac{1}{3}$} \: 
	\left[ 1 - \left(\frac{\rho}{e} \right)^2 \right].
\label{eq:SyngeEOS}
\ee
With this definition for the effective polytropic index, the relativistic specific enthalpy can be written as:
\be
	h = \mbox{$\frac{1}{3}$} \left[ 4 \frac{e}{\rho} - \frac{\rho}{e} \right] \; ,
            \label{eq:enthalpyMathews}
\ee
and the corresponding closure relation following from \equref{eq:enthalpy} becomes:

\be
	P = \mbox{$\frac{1}{3}$} \left( e - \frac{\rho^2}{e} \right) \; .
            \label{eq:closurerelation}
\ee

The total internal energy of the gas per particle is \mbox{$\epsilon = e/n$}, with $n$ the number density of the
gas. For a plasma consisting of protons, as is the case for a hadronic jet, the rest-mass energy per particle is
$m_{\rm p}$ and the thermal energy of the gas per particle is \mbox{$\epsilon_{\rm th} = e_{\rm th}/n$}. Not much
is known about the composition of an AGN jet at kpc scales. The plasma might consist of electrons and positrons
(see for example \citealt{Reynolds1996} or \citealt{Wardle1998}), but it might also be an electron-proton plasma,
or a mixture of both. We will however assume that at the length scales we are considering, a significant amount
of mixing with the ambient medium has taken place, so that the jet can effectively be described by a hadronic
plasma.

It can easily be seen that the effective polytropic index for a non-relativistically 'cold' gas
with \mbox{$\epsilon_{\rm th} << m_{\rm p}$} reduces to \mbox{$\Gamma_{\rm eff} = 5/3$}, while for a
relativistically hot gas with \mbox{$\epsilon_{\rm th} >> m_{\rm p}$}, it reduces to
\mbox{$\Gamma_{\rm eff} = 4/3$}
\footnote{For an electron-positron plasma, the energies at which the gas would become relativistic are lower
by a factor \mbox{$m_{\rm e}/m_{\rm p} \; \sim 5 \times 10^{-4}$}.}.
See \citet{Meliani2008} for a more complete description of the Mathews approximation to the Synge equation of
state.

  \subsection{Radial pressure profile for spine-sheath jets}\label{subsec:radpres}

Since AGN jets remain collimated over huge distances, they are expected to be in approximate pressure
equilibrium with their surroundings. In fact, if a jet does not start out in pressure equilibrium, unbalanced
pressure forces at its jet-ambient medium interface cause the jet to either expand, or contract until
approximate pressure equilibrium is reached. The question of {\em how} this ambient medium is defined exactly
is less clear. In the "standard model" for double radio galaxies (e.g. \citealt{Blandford1974};
\citealt{Scheuer1974}; \citealt*{Leahy1989}; \citealt{Begelman1989}; \citealt{Daly1990}), under-dense jets at
larger distances from the central engine create a strong bow-shock. This bow-shock encloses a hot and
over-pressured cocoon (compared to the undisturbed ambient medium). Since we do not know the exact conditions
for such cocoons when we start our simulations, we set the jets up in direct pressure equilibrium with the
undisturbed *ambient* intergalactic medium, which we will indicate with a subindex "am" from now on.

In case of the radially uniform, or {\em homogeneous} jet (which we call case $H$), we set the pressure
equilibrium up by equating the jet pressure to the pressure of the ambient medium. For jets with a
{\em spine-sheath jet structure} on the other hand (which we call case $A$ for the isochoric jet and case $I$ for
the isothermal jet), the pressure profile is not trivial. It can be obtained by solving the special relativistic
hydrodynamic (SRHD) radial force equation that balances the radial pressure force with the centrifugal force due
to the rotation of the fluid. We use cylindrical coordinates \mbox{$(R \: , \: \phi \: , \: z)$} with the jet
axis along the $z$-axis and neglect the lateral expansion of the jet (assumed to be slow so that
\mbox{$v_{R} \ll v_{z}$}) so that the velocity is
\mbox{$\bm{v} = \left( 0 \: , \: v_{\phi} \: , \: v_{z} \right)$}. One has:

\be
	\frac{{\rm d}P}{{\rm d}R} = \frac{\rho h \: \gamma^2 v_{\phi}^2}{R} = 
	\frac{\rho h v_{\phi}^2}{\left( 1 - v_{z}^2 - v_{\phi}^2 \right) R} \; .
\label{eq:radforcebal}
\ee

In this paper, the index "${\rm sp}$" refers to variables and constants belonging to the jet spine, whereas the
index "${\rm sh}$" refers to variables and constants belonging to the jet sheath. An analytical solution for
the SRHD radial force balance equation can be found if one assumes a self-similar rotation profile of the form: 

\be
v_{\phi}^2(R) = \left\{
\begin{array}{rl}
V_{\phi,{\rm sp}}^2 \: \left( \displaystyle \frac{R}{R_{\rm sp}} \right)^{a_{\rm sp}} &
\mbox{jet spine:} \; 0 \le R \le R_{\rm sp} \; ,\\
\phantom{} & \phantom{} \\
V_{\phi,{\rm sh}}^2 \: \left( \displaystyle \frac{R}{R_{\rm sp}} \right)^{a_{\rm sh}} &
\mbox{jet sheath:} \; R_{\rm sp} < R \le R_{\rm jt} \; .
\end{array} \right.
\label{eq:Vphi}
\ee
The same profile is used in \citet{Meliani2009}. Here $R_{\rm sp}$ is the radius of the jet spine and
$R_{\rm jt}$ is the outer radius of the jet sheath, which coincides with the jet radius. $V_{\rm \phi, sp}$
is a constant that gives the maximum rotation within the jet spine for \mbox{$R \longrightarrow R_{\rm sp}$}, and
similarly for the constant $V_{\rm \phi, sh}$. The constants $a_{\rm sp}$ and $a_{\rm sh}$ are self-similarity
constants. It can be seen immediately that the constant $a_{\rm sp}$ needs to be positive in order to avoid
singularities at \mbox{$R \longrightarrow 0$}. Furthermore, the Rayleigh criterion for stability of flows
rotating on a cylinder against axisymmetric perturbations is:

\be
\label{Rayleigh}
	\frac{{\rm d}}{{\rm d}R} \left(\gamma h R v_{\phi} \right) > 0 \; ,
\ee 
see for instance Pringle \& King (2007), Ch. 12. From this it follows that both self-similarity constants need
to satisfy the condition \mbox{$a_{\rm sp} > -2$} and \mbox{$a_{\rm sh}> -2$}. We set \mbox{$a_{\rm sh} = -2$},
corresponding to a jet sheath flow with constant specific angular momentum:
\mbox{$\lambda \equiv \gamma h R v_{\phi} = $ constant}, making it marginally stable according to Rayleigh's
criterion. As the self-similarity constant of the jet spine needs to be positive, we adopt the same value that
was used in \citet{Meliani2009} and set \mbox{$a_{\rm sp} = 1/2$}.

We solve the radial force-balance equation \equref{eq:radforcebal} for two different kinds of jets. The first jet
is given constant, but different density for jet spine and jet sheath. There, we set up the radial pressure
profile of the jet making use of a polytropic index equal to \mbox{$\Gamma = 5/3$}. The second kind of jet is the
isothermal jet where we fix the temperature of the jet by adjusting the density accordingly to the varying
pressure, and initializing the jet according to $\Gamma = 1$. Figure \ref{fig:Initial} shows the initial
transverse (radial) profiles of the azimuthal velocity, pressure, number density and temperature that were used
in these simulations. In the following sections (\ref{subsubsec:Pressure_I} and \ref{subsubsec:Pressure_A}), the
actual radial pressure profiles will be derived. The choice for the parameters of jet and ambient medium that
have been used to generate the exact jet profiles are discussed in sections \ref{subsec:JetProperties} and
\ref{sec:Method}.

    \subsubsection{Pressure profile for the isothermal jet ($I$)} \label{subsubsec:Pressure_I}

To solve the radial force balance equation for the isothermal jet, we first use the ideal gas law to write
\equref{eq:radforcebal} as:

\be
s^2 \frac{dP}{P} = \frac{v_{\phi}^2 dR}{(1-v_{z}^2 -v_{\phi}^2)R},
\ee
where the isothermal sound speed in a relativistic gas $s$ is given by:

\be
s^2 = \frac{{\cal R}T}{\mu h},
\ee
with ${\cal R}$ the gas constant and $\mu$ the particle mass in units of hydrogen mass. The temperature is taken
constant. To solve the radial force balance equation, the temperature for the jet spine and the jet sheath does
not necessarily have to be the same. However, we adopt a constant $T$ across jet spine and jet sheath here,
where we assume that any differences in temperature have been washed out at large distances from the central
engine. We will assume the vertical component of the velocity to be constant \mbox{$v_{z} = V_{z}$} (also not
necessarily the same for jet spine and jet sheath) and use the self similar azimuthal velocity profile
\equref{eq:Vphi}. In that case the pressure profile is easily integrated to:

\be
P = A\left\{1 - \alpha \left( \frac{R}{R_{\rm sp}} \right)^{a}\right\}^{-\sigma}.\label{eq:PIso}
\ee
Here, $A$, $a$, $\alpha$ and $\sigma$ are all constants with the latter three given in the jet spine by:
\be
\label{twoconstants1}
a = a_{\rm sp} \; \; , \; \; \alpha_{\rm sp} = \frac{V_{\phi,{\rm sp}}^2}
    {1 - V_{z,{\rm sp}}^2} \; \; , \; \; \sigma_{\rm sp} = \frac{1}{a_{\rm sp} s_{\rm sp}^2} \; .
\ee
Expressions in the jet sheath are analogous and can be found by changing the subscript
${\rm sp} \: \longrightarrow \: {\rm sh}$. The constant $A_{\rm sp}$ in the jet spine and the corresponding
constant $A_{\rm sh}$ in the jet sheath follow from requiring [1] pressure balance at the jet spine-sheath
interface at $R = R_{\rm sp}$ and [2] requiring pressure balance with the pressure $P_{\rm am}$ of the
surrounding medium at the jet outer radius $R = R_{\rm jt}$.
\nskip
This leads to two conditions:
\be
\begin{array}{l c l}
\label{balancecond1}
A_{\rm sp} \: \left\{ 1 - \alpha_{\rm sp} \right\}^{-\sigma_{\rm sp}} &
= &
A_{\rm sh} \: \left\{ 1 - \alpha_{\rm sh} \right\}^{-\sigma_{\rm sh}} \; ,
\nonumber \\
\phantom{} & \phantom{} & \phantom{} \\
A_{\rm sh} \: \left\{ 1 - \alpha_{\rm sh} \left( \displaystyle{\frac{R_{\rm jt}}{R_{\rm sp}}}
\right)^{a_{\rm sh}} \right\}^{-\sigma_{\rm sh}} &
= &
P_{\rm am} \; . \nonumber
\end{array}
\ee
These two relations determine $A_{\rm sp}$ and $A_{\rm sh}$. Moreover, the pressure at the centre of the jet
$P(R = 0) = P_{0}$ also determines the constant $A_{\rm sp}$ by:
\be
A_{\rm sp} = P_{0}\; .
\ee
The requirement for the pressure to remain positive throughout the jet's cross section is satisfied when
$\alpha_{\rm sp} < 1$ and $\alpha_{\rm sh} < 1$, which leads to the physically obvious condition
\mbox{$V_{z}^2 + V_{\phi}^2 < 1$}, the total speed at the interfaces must be less than the speed of light.

    \subsubsection{Pressure profile for the piecewise isochoric jet ($A$)}\label{subsubsec:Pressure_A}

Instead of assuming a constant temperature $T \propto P/\rho$, we now assume a piecewise isochoric (or constant
density) jet with in the jet spine a density $\rho_{\rm sp}$, polytropic index $\Gamma_{\rm sp}$ and speed
\mbox{$v_{z} = V_{z,{\rm sp}}$}, and similarly $\rho_{\rm sh}$, $\Gamma_{\rm sh}$ and $V_{z,{\rm sh}}$ in the
jet sheath. Then the radial force balance equation \equref{eq:radforcebal} can be rewritten as:

\be
R \: \frac{{\rm d} \tilde{P}}{{\rm d}R} - \frac{\Gamma}{\Gamma - 1} \:
\frac{v_{\phi}^2}{\left( 1 - v_{z}^2 - v_{\phi}^2 \right)} \: \tilde{P} = 0 \; .
\ee
Here
\be
\tilde{P}(R) \equiv \frac{\Gamma -1}{\Gamma} \:\rho h = P(R) + \frac{\Gamma -1}{\Gamma} \: \rho \; .
\ee
\halfskip
\noindent
Using rotation profile \equref{eq:Vphi} one can solve this equation:
\be
\tilde{P}(R) = \tilde{A} \: \left\{ 1 - \alpha \: \left( \frac{R}{R_{\rm sp}} \right)^{a} \right\}^{-\tau} \; .
\label{eq:PAdiab}
\ee
Here $a$ and $\alpha$ have the same meaning as in the isothermal case. The constant $\tilde{A}$ is determined
from requiring pressure equilibrium at the interfaces $R_{\rm sp}$ and $R_{\rm jt}$, as was required in the
isothermal case. Moreover, $\tau$ is a constant, which for the jet spine is given by:
\be
\label{twoconstants2}
\tau_{\rm sp} = \frac{\Gamma_{\rm sp}}{a_{\rm sp}(\Gamma_{\rm sp}-1)} \; .
\ee 
As before, expressions in the jet sheath are analogous to the expressions in the jet spine and can be found
by changing the subscript ${\rm sp} \: \longrightarrow \: {\rm sh}$. The constant $\tilde{A}_{\rm sp}$ in jet
spine (and the corresponding constant $\tilde{A}_{\rm sh}$ in the jet sheath) in this case are determined by
solving:
\be
\begin{array}{l}

\tilde{A}_{\rm sp} \: \left\{ 1 - \alpha_{\rm sp} \right\}^{-\tau_{\rm sp}} -
\displaystyle{\frac{\Gamma_{\rm sp} - 1}{\Gamma_{\rm sp}}} \: \rho_{\rm sp} \; =\\
\phantom{} \\
\tilde{A}_{\rm sh} \: \left\{ 1 - \alpha_{\rm sh} \right\}^{-\tau_{\rm sh}} -
\displaystyle{\frac{\Gamma_{\rm sh} -1}{\Gamma_{\rm sh}}} \: \rho_{\rm sh} \; ,\\
\nonumber \\
\phantom{} \nonumber \\
\tilde{A}_{\rm sh} \: \left\{ 1 - \alpha_{\rm sh} \left(\displaystyle{\frac{R_{\rm jt}}{R_{\rm sp}}}\right)^
{a_{\rm sh}} \right\}^{-\tau_{\rm sh}} \;
= \; P_{\rm am} + \displaystyle{\frac{\Gamma_{\rm sh} -1}{\Gamma_{\rm sh}}} \: \rho_{\rm sh} \; . \nonumber
\label{balancecond2}
\end{array}
\ee
These two relations determine $\tilde{A}_{\rm sp}$ and $\tilde{A}_{\rm sh}$, with the constraint that physically
allowed solutions must have $P_0 \ge 0$, or equivalently:
\be
\tilde{P}_0 \: = \: \tilde{A}_{\rm sp} \ge \frac{\Gamma_{\rm sp} -1}{\Gamma_{\rm sp}} \: \rho_{\rm sp} \; .
\ee

  \subsection{Jet properties: density ratio and kinetic luminosity}\label{subsec:JetProperties}

Observations of AGN jets yield a few basic parameters, such as the jet length and diameter, the luminosity of
jets, lobes and (in FRII sources) hot spots and possibly the synchrotron age based on the observed spectrum of
the non-thermal radiation. In addition, it is possible to derive cocoon parameters from the X-ray cavities observed
around some of the stronger sources. In principle one can estimate the advance speed of the jet from these data
and, using a model, get clues on jet composition, e.g. the question of an electron-positron jet plasma vs. a
hydrogen plasma.

In this section we will explain how observed data can be used to calculate the mass density ratio between jet
material and material of the ambient medium. Then in section \ref{subsec:JHAS}, we will use this to estimate
the jet-head advance speed for a radially uniform jet.

In order to do so, we first define the kinetic luminosity of a jet $L_{\rm jt}$ as the total power
$L_{\rm tot}$ that is produced by the jet, with its rest mass energy discharge through the jet subtracted.
Still working in units where \mbox{$c = 1$}:

\be
L_{\rm jt} = L_{\rm tot} - \dot{M}. \label{eq:Ltot}
\ee
The total power $L_{\rm tot}$ for a radially uniform jet is given by:
\be
L_{\rm tot} = A_{\rm jt}^2 n_{\rm jt} m_{\rm jt} h_{\rm jt} \gamma_{\rm jt}^2 v_{\rm jt},
\ee
and the rest mass energy discharge through the jet $\dot{M}$ by:

\be
\dot{M} = A_{\rm jt}^2 n_{\rm jt} m_{\rm jt} \gamma_{\rm jt} v_{\rm jt}.
\ee
Here $A_{\rm jt} = \pi R_{\rm jt}^2$ is the cylindrical radial cross section of the jet, $n_{\rm jt}$ is the
number density of the jet material, $m_{\rm jt}$ is the averaged mass of the particles in the jet and
$h_{\rm jt}$ is the specific relativistic enthalpy of the jet material, see eq. \equref{eq:enthalpy}. Therefore,
the kinetic luminosity of a radially uniform jet can be written as:

\be
L_{\rm jt} = A_{\rm jt} n_{\rm jt} m_{\rm jt} \gamma_{\rm jt} v_{\rm jt} (h_{\rm jt} \gamma_{\rm jt} -1).
\label{eq:Lj} 
\ee
In case of a structured spine-sheath jet, we approximate its kinetic luminosity by adding the contributions
from the jet spine and the jet sheath to the kinetic luminosity separately:

\be
L_{\rm jt} =  L_{\rm sp} + L_{\rm sh},
\ee
where $L_{\rm sp}$ and $L_{\rm sh}$ are defined in the same way as \equref{eq:Lj}, but with their indices
referring to the corresponding components. In the rest of this derivation we will just focus on the case of the
radially uniform, homogeneous jet.

Now suppose that we know the following jet parameters from observations for a particular AGN jet: kinetic
luminosity, jet radius and jet velocity (or equivalently a Lorentz factor $\gamma_{\rm jt}$). Suppose that
we can also determine a number density $n_{\rm am}$ and temperature $T_{\rm am}$ of the ambient medium (from
which we can derive the ambient medium pressure $P_{\rm am}$ with the ideal gas law). With these parameters it
is possible to calculate the density ratio between jet material and ambient medium. 

First we assume pressure equilibrium at the interface between jet and ambient medium. Then, using equations
\equref{eq:enthalpy} and \equref{eq:Lj}, one can show that the ratio of number density can be written as:

\be
\frac{n_{\rm jt}}{n_{\rm am}} = \frac{(\Gamma -1)\left({\displaystyle
\frac{L_{\rm jt}}{\pi R_{\rm jt}^2 n_{\rm am} m_{\rm jt}}}
\right) - \Gamma\gamma_{\rm jt}\sqrt{\gamma_{\rm jt}^2 -1}\left({\displaystyle
\frac{k_{\rm b} T_{\rm am}}{m_{\rm jt}}}\right)}
{(\Gamma -1)(\gamma_{\rm jt} -1)\sqrt{\gamma_{\rm jt}^2 -1}},\label{eq:densityratio}
\ee
with $k_{\rm b}$ the Boltzmann constant.

As we mentioned before, we assume the jet to be hadronic, so that the number density ratio can be written as a
(proper) mass density ratio, given by:
\footnote{Note that the  inertia of the material in the lab frame scales as $\gamma^2 nm$ for a given particle
mass $m$.}

\be
\eta_{\rm R} = \frac{\rho_{\rm jt}}{\rho_{\rm am}} = \frac{n_{\rm jt}}{n_{\rm am}}.
\ee

Jets with $\eta_{\rm R} < 1$ are called {\em under-dense} and jets with $\eta_{\rm R} > 1$ are called {\em
over-dense}. Under-dense jets are less stable than over-dense jets and develop internal (diamond-shaped) shocks
more easily. For under-dense jets the propagation speed of the jet's head is much lower than the velocity of
the bulk material of the jet (see section \ref{subsec:JHAS}). For these jets, at the jet's head the jet flow
is terminated by a strong shock called the Mach disk.

The intergalactic medium (IGM) in the vicinity of galaxies and inside clusters of galaxies (the so-called
intra-cluster medium, or ICM) is usually denoted as a warm-hot intergalactic medium (WHIM). In these regions
number densities range from \mbox{$\sim 5 \times 10^{-6}\: {\rm cm^{-3}}$} to
\mbox{$\sim 10^{-3}\: {\rm cm^{-3}}$} and temperatures are on the order of \mbox{$10^5 - 10^7$ K}
(see e.g. \citealt{Dave2001}; \citealt{Dave2010} and \citealt{Kunz2011}).

Since many powerful AGN jets are formed inside clusters of galaxies (e.g. \citealt*{Begelman1984};
\citealt{Smith2002}), we choose to focus on the intra-cluster medium as the ambient medium for our jets and
take for the number density \mbox{$n_{\rm am} = 1 \times 10^{-3}\: {\rm cm}^{-3}$} and fix the temperature
of the ambient medium to \mbox{$T_{\rm am} = 10^7\: {\rm K}$}.

For the jet we will take a powerful radio source, with a luminosity of
\mbox{$L_{\rm jt} = \rmn{a\; few} \times 10^{46}\: {\rm erg \; s^{-1}}$} (a typical luminosity for FRII and BL
Lac sources, see for instance \citealt[][]{Ito2008} or \citealt[][]{Ma2008}). Also, the bulk material of
the jets in our simulations is cold (by which we mean that the gas satisfies a classical equation of state,
\mbox{$\Gamma = 5/3$}). We will take for the radius of the jet \mbox{$R_{\rm jt} = 1$ kpc}, corresponding to a
jet with a typical half-opening angle of $1^\circ$ \citep{Pushkarev2009} at a distance of $57$ kpc from the
central engine
\footnote{At this distance, the jet is dominated by kinetic energy flux.}. And finally we will take this jet to
be trans-relativistic with a moderate Lorentz factor of $\gamma_{\rm jt} = 3$.

Substituting these values into \equref{eq:densityratio} we find a mass density ratio on the order of
\mbox{$\eta_{\rm R} \sim 10^{-3}$}, corresponding to very under-dense jets. Table \ref{tab:parameters} shows
the exact jet parameters that are used for the jet models in this paper
\footnote{It is worthwhile to note that the choice in parameter space is fairly large and that different choices
for $L_{\rm jt}$, $T_{\rm am}$ or $\gamma_{\rm jt}$ could in principle result easily in different density ratios.
However, it turns out that for most sets of realistic parameters, the density ratio will in general lie in the
range of \mbox{$\eta_{\rm R}\sim 10^{-3}-1$}, most of which correspond to under-dense jets. Our choice is
therefore reasonable and corresponds to an under-dense jet at the lower end of the spectrum.}.
Some properties of under-dense jets will be treated in the section \ref{subsec:JHAS}.

  \subsection{Jet-head advance speed}\label{subsec:JHAS}

The velocity with which a jet penetrates into the ambient medium is less than the bulk velocity of jet material.
This is especially true for under-dense jets. Near the point where the jet impacts with the ambient
intergalactic/interstellar medium, a structure forms including a forward bow shock that precedes the jet, a
contact discontinuity separating shocked ambient gas from shocked jet material and a reverse shock (Mach disk)
that decelerates the jet flow. This whole system comprises the \emph{jet-head}.

The jet-head advance speed can actually be estimated from ram pressure arguments in the rest frame of the head,
where the flow is more-or-less steady (\citealt{Marti1997}; \citealt{Rosen1999}).
The jet-head advance speed found in this way equals:
\be
\beta_{\rm hd} = \frac{\gamma_{\rm jt}\sqrt{\eta_{\rm R}}\; \beta_{\rm jt}}{\displaystyle 1 +
\gamma_{\rm jt}\sqrt{\eta_{\rm R}}} \; ,
\label{eq:JHAS}
\ee
where again \mbox{$\eta_{\rm R} = \rho_{\rm jt}/\rho_{\rm am}$} is the ratio of mass density of jet material
and mass density of ambient medium material. In the case where the gas is relativistically hot, so that
\mbox{$h > 1$}, the same expression holds, but then the ratio of mass densities is substituted by the ratio of
relativistic enthalpies
\mbox{$\eta_{\rm R} \longrightarrow \rho_{\rm jt} h_{\rm jt} / \rho_{\rm am} h_{\rm am}$}.

From equation \equref{eq:JHAS} it is immediately clear that under-dense jets with \mbox{$\eta_{\rm R} << 1$} have
propagation speeds much less than their bulk velocities, unless they are very relativistic with
\mbox{$\gamma_{\rm jt} >> 1$}. Using the same parameters as we did in section \ref{subsec:JetProperties}
(resulting in a density ratio of \mbox{$\eta_{\rm R} \sim 10^{-3}$} and a Lorentz factor of
\mbox{$\gamma_{\rm jt} \sim 3$} with corresponding \mbox{$\beta_{\rm jt} = 0.943$}), we find that the jet-head
propagation speed is approximately $\beta_{\rm hd} \sim 8\times 10^{-2}$. The jet-head advance speed, together
with the length of the jet, yields an estimate for the time that the central engine has been active. We will
use this method for analytically predicting the jet-head advance speed to compare with our simulations in
\mbox{section \ref{sec:Results}}.

  \subsection{Jet properties: Rotation}\label{subsec:Rotation}

    \subsubsection{Jet angular momentum}
    \label{subsubsec:JetAngularMomentum}

In steady, axisymmetric hydrodynamic flows the specific angular momentum
\mbox{$\lambda \equiv \gamma h RV_{\phi}$} (neglecting general-relativistic corrections) is conserved. Its value
is set by the rotation of the wind source. Then, the azimuthal four-velocity decays as

\be
\label{hydrocase}
	\gamma h V_{\phi} = \frac{\lambda}{R} \; .
\ee
In axisymmetric and ideal MHD flows with a magnetic field
\mbox{$\bm{B} = \bm{B}_{\rm p} + B_{\phi} \: \bm{\hat{e}}_\phi$}, the situation is different. There, the angular
velocity $\Omega$ of poloidal field/flow lines, formally defined by

\be
	\Omega = \frac{V_{\phi}}{R} - \kappa \frac{B_{\phi}}{R} \; ,
\ee
is constant along flow lines. Its value is set by conditions at the source of the wind. Here,
\mbox{$\kappa \equiv V_{\rm p}/B_{\rm p}$} is the ratio of the poloidal velocity and magnetic field, again a
constant along flow lines. 

Such axisymmetric MHD winds behave roughly as follows: close to the source, where the wind is sub-Alv\'enic in
the sense that
\mbox{$\gamma V_{\rm p} \ll \left(B_{\rm p}/\sqrt{4 \pi \rho_{0} h} \right)\left(1 - \Omega^2 R^2/c^2 \right)$}
with \mbox{$\rho_{0} = \rho/\gamma$} the proper density, the wind rotates almost rigidly with

\be
	V_{\phi} \sim \Omega R \; .
\ee
This solid rotation is enforced by strong magnetic torques on the wind material. Although one can define a
conserved specific angular momentum $\lambda$ that has a mechanical, as well as a magnetic contribution, the
{\em mechanical} angular momentum is obviously not conserved! 

Well beyond the so-called Alfv\'en point, the point on a flow line where
\mbox{$\gamma V_{\rm p} = \left(B_{\rm p}/\sqrt{4 \pi \rho_{0} h} \right)\left(1 - \Omega^2 R^2/c^2 \right)$},
the flow speed is super-Alfv\'enic and magnetic torques become dynamically unimportant. There, the wind satisfies
(\ref{hydrocase}), but with the value of $\lambda$ now set by $\Omega$ and the radius $R_{\rm A}$ of the Alfv\'en
point:

\be
	\lambda = \mu \Omega \: R_{\rm A}^2 \; ,
\ee 
where \mbox{$\mu \equiv {\cal E}/c^2 \ge 1$} with ${\cal E}$ the conserved total energy per unit mass in the
wind. This means that the value of $\lambda$ can be much higher than in the hydrodynamic case, leading to a
larger rotation speed far from the source.

    \subsubsection{Continuous rotation profile}
    \label{subsubsec:ContinuousRotation}

Valid solutions of the radial force balance equation \equref{eq:radforcebal} allow for different values of the
constants $V_{\rm \phi, sp}$ and $V_{\rm \phi, sh}$. Giving these constants a different value will result
in a discontinuous rotation profile, where the most realistic scenario is the one where
\mbox{$V_{\rm \phi, sp} > V_{\rm \phi, sh}$}. Close to the central engine, where the different jet regions
(spine and sheath) are thought to be driven by different mechanisms, such a rotation profile seems a
reasonable one. However, as the jet propagates through the ambient medium, mixing effects between jet spine and
jet sheath are likely to wash out the discontinuity occurring at the jet spine-sheath interface. Therefore,
the rotation profile at larger distances from the central engine is likely to be continuous. This leads us to
choose the rotation constants equal to one maximum value:
\mbox{$V_{\rm \phi,sp} = V_{\rm \phi,sh} \equiv V_{\phi}$}.

\section{Method} \label{sec:Method} 

{\renewcommand{\arraystretch}{1.3}
\begin{table*}
\caption{Free parameters that were used for the jet inflow properties and the initialization of the ambient
medium for the three jet models $H$, $I$ and $A$.}

\begin{tabular}{l c c c c c c} \hline
{\bf Models} &
$L_{\rm jt} \; [10^{46}  \; \rmn{erg \; s^{-1}} ]$ &
$n         \; [10^{-6}  \; \rmn{cm^{-3}}       ]$ &
$\gamma$                                             &
$V_{\phi}  \; [10^{-3}  \; c]$                       &
$\Gamma$                                             &
$P         \; [10^{-12} \; \rmn{erg \; cm^{-3}}]$
\\

\phantom{}                                           &
\begin{tabular}{c c c} \hline
sp & $|$ & sh
\end{tabular}                                        &
\begin{tabular}{c c c} \hline
sp & $|$ & sh
\end{tabular}                                        &
\begin{tabular}{c c c} \hline
sp & $|$ & sh
\end{tabular}                                        &
\begin{tabular}{c c c} \hline
sp & $|$ & sh
\end{tabular}                                        &
\begin{tabular}{c c c} \hline
sp & $|$ & sh
\end{tabular}                                        
&
\phantom{}
\\ \hline

\rowcolor{lightgray}
      $\mathbf{H}$ {\bf (homogeneous)} & 3.82 & 4.55    & 3.11  & 0.0 & 1 & 1.38
      \\
      $\mathbf{I}$ {\bf (isothermal)} & 
                                         \begin{tabular}{c c c}
                                             1.82 & \phantom{} & 3.35
                                         \end{tabular}
                                         &
                                                $P/\rho = $ constant
                                            & 
                                            \begin{tabular}{c c c}
                                                6.0 & \phantom{} & 3.0
                                            \end{tabular}
                                              &
                                              \begin{tabular}{c c c}
                                                  1.0 & \phantom{} & 1.0
                                              \end{tabular}
                                                  &
                                                  \begin{tabular}{c c c}
                                                      5/3 & \phantom{} & 5/3
                                                  \end{tabular}
                                                      & according to eq. \equref{eq:PIso}
                                                      \\
  \rowcolor{lightgray}
      $\mathbf{A}$ {\bf (isochoric)} & 
                                         \begin{tabular}{c c c}
                                             0.44 & \phantom{} & 3.39
                                         \end{tabular}
                                         &
                                      \begin{tabular}{c c c}
                                          1.0 & \phantom{} & 5.0 
                                      \end{tabular}
                                          & 
                                          \begin{tabular}{c c c}
                                              6.0 & \phantom{} & 3.0
                                          \end{tabular}
                                              &
                                              \begin{tabular}{c c c}
                                                  1.0 & \phantom{} & 1.0
                                              \end{tabular}
                                                  &
                                                  \begin{tabular}{c c c}
                                                      5/3 & \phantom{} & 5/3
                                                  \end{tabular}
                                                      & according to eq. \equref{eq:PAdiab}
                                                      \\
      {\bf External medium} & - & $1.0\times 10^{3}$ & - & - & 5/3 & 1.38\\
      \hline
    \end{tabular}

\medskip
Kinetic luminosity ($L_{\rm jt}$), number density ($n$), Lorentz factor ($\gamma$), azimuthal
velocity ($V_{\phi}$), polytropic index ($\Gamma$), gas pressure ($P$). In case of model $H$,
the jet is homogeneous in the radial direction and is described by single-valued quantities. 
The pressure in the ambient medium follows from the number density $n_{\rm am}$ and
assuming a temperature of the ambient medium of $T_{\rm am} = 10^7$ K.
The parameters for models $I$ and $A$ are initialized separately for jet spine (denoted as "sp" in
the table) and jet sheath (denoted as "sh" in the table). In case of the models $I$ and $A$, the pressure
varies radially, as indicated. In case of the $I$ model, the density varies radially in order to
keep the temperature constant.

    \label{tab:parameters}

\end{table*}
}


  \subsection{The models, setup and initial conditions}

In this paper, we simulate AGN jets with moderate Lorentz factors of \mbox{$\gamma \sim$ a few}, putting
them into the trans-relativistic regime. We simulate a continuously driven homogeneous ($H$) jet and
two structured spine-sheath jets (an isothermal ($I$) jet and a piecewise isochoric ($A$) jet) of which the
radial profiles are treated in \mbox{section \ref{subsec:radpres}}.

All jets have constant and similar luminosity during their entire evolution. In a follow-up
paper, we will be concerned with the case of two distinct episodes of jet activity for the same jet models $H$,
$I$ and $A$. In order to make a clear distinction between the two cases, we introduce an index '1' for the
steady case and introduce an index '2' for the case of episodic activity. Therefore, this paper will treat the
simulations \mbox{$H1$, $I1$ and $A1$}.

The simulations have been performed on the same spatial domain for a duration of
\mbox{$\sim$ 23 Myr (22.8 Myr)} with a kinetic luminosity of
\mbox{$L_{\rm jt} \sim 4 - 5\times 10^{46}\; {\rm erg\:s^{-1}}$}. The jets are injected into a warm-hot
intergalactic medium with constant density (\mbox{$n_{\rm am} = 10^{-3} \; \rmn{cm}^{-3}$}) and constant
temperature (\mbox{$T_{\rm am} = 10^{7} \; \rmn{K}$}), which is a reasonable approximation for the conditions
inside a cluster of galaxies, at large distances from the central engine. The time steps are dynamically
determined by the code, but are on the order of \mbox{270 yr}.

Our jets are cylindrically symmetric with their jet axis along the $Z-$axis. At the start of the simulation
the jet protrudes along its axis into the computational domain over a distance equal to its initial radius,
which we choose $R_{\rm jt}$ $=1$ kpc for all three models.
In the case of structured jets, this is equivalent to the outer jet sheath radius. For these jets we choose (in
absence of observational constraints, and in accordance with \citealt[][]{Meliani2009}) the radius of the jet
spine equal to $R_{\rm sp} = R_{\rm jt}/3$.

We choose the maximum rotation of the structured jets to be $V_{\phi} \sim 1 \times 10^{-3}$
\footnote{This is a fairly conservative choice compared to the value of the critical rotation for the isochoric
jet, see section \ref{subsubsec:CriticalAzimuthalV}. Moreover, note that even though we simulate purely
hydrodynamical jets at kpc scales, we assume they have all started out as fully magnetohydrodynamical jets.}.

The jets start out in pressure equilibrium with their surrounding, as described in section \ref{subsec:radpres}.
After initialization, the jet flow is created by letting material flow into the computational domain through the
boundary cells at the \mbox{$Z = 0$} axis, between \mbox{$R = 0$} and \mbox{$R = R_{\rm jt}$}. Except for the
cells involved in injecting the jet material, all other cells in the lower boundary are free {\em outflow}
boundaries. In addition, the {\em inflow} velocity of these cells is reduced to $20\%$ of their original value,
in order to avoid spurious numerical effects next to the jet inlet.

The size of our computational domain is $(250 \times 500)$ kpc$^2$. We choose a basic resolution
of $(120 \times 240)$ grid cells and allow for four additional refinement levels. This results in an effective
resolution of $(1920 \times 3840)$ grid cells. Therefore, we can resolve details up to
\mbox{$(65 \times 65)$ pc$^2$}.

Table \ref{tab:parameters} gives an overview of the free parameters that were used for these simulations.
Moreover, table \ref{tab:charvars} shows a list of characteristic variables that are used throughout the paper,
and which apply to the plots. 

  \subsection{MPI-AMRVAC and numerical schemes}

{\renewcommand{\arraystretch}{1.3}
\begin{table}
\caption{List of characteristic quantities shown in cgs units. These characteristic quantities
         apply throughout the paper.}
\rowcolors{1}{white}{lightgray}
    \begin{tabular}{ l  c  c } \hline
       {\bf Char. quantities} & {\bf symbol} & {\bf cgs units}                              \\
       \hline
                      Number density   & $n_{\rm ch}$ & $10^{-3}\; {\rm cm}^{-3}$                    \\
       
                            Pressure   & $P_{\rm ch}$ & $1.50 \times 10^{-6}\; {\rm erg \; cm}^{-3}$ \\
      
                         Temperature   & $T_{\rm ch}$ & $1.09 \times 10^{13}\; {\rm K}$              \\
       \hline
    \end{tabular}
    \label{tab:charvars}
\end{table}
}

Our simulations employ the code MPI-AMRVAC \citep{Keppens2012}. It is a versatile code that
allows for various discretization schemes, involving the use of different limiters in the reconstructions from
cell centre to cell edge. It allows for adaptive mesh refinement and can be run parallel on multiple processors.

The simulations are performed with a special relativistic hydrodynamical module. We choose a 4-step 'Runge-Kutta'
time-discretization scheme, in combination with a second order spatial Total Variation Diminishing
Lax-Friedrichs scheme with a Koren limiter. This combination captures shocks well without exhausting
computational resources.

MPI-AMRVAC can be initialized using conservative variables, which are advected as according to their fluxes
calculated through \equref{eq:conservationlaw}. However, the variables can also be initialized as {\em primitive}
variables, which MPI-AMRVAC then converts back to conservative variables. We choose to do the latter.
In that case, the free parameters of the models are the mass-density $\rho$, the velocity $\bmath{v}$
and the pressure $P$. Finally, MPI-AMRVAC needs to be initiated with a maximum value for the polytropic index
$\Gamma$. We initialized the polytropic index as $\Gamma = 5/3$. This choice for $\Gamma$ is consistent with the
Mathews approximation (equations \equref{eq:SyngeEOS}, \equref{eq:enthalpyMathews} and
\equref{eq:closurerelation}) for these parameters.

  \subsection{Tracers of jet material}

In jets with radial structure, or in cases where jet activity is episodic, it is important to keep track of
the various constituents (for example, jet, jet spine, jet sheath, or ambient medium). To that end we employ
{\em tracers}, \mbox{$\theta_{\rm A}(t,\bmath{r})$},
\footnote{The index $A$ refers to a certain constituent $A$ in the simulation.} 
that are passively advected by the flow from cell to cell. Appendix \ref{sec:Tracers} treats the definition of
the tracers employed here. The number of tracers that were used for each simulation varies from case to case.
Basically, every constituent we would like to trace is initialized to \mbox{$\theta = \theta_{\rm max} = +1$} in
the region where this constituent is injected into the system. We put its value equal to
\mbox{$\theta = \theta_{\rm min} = -1$} elsewhere. For completeness sake, we will list the exact values for each
simulation below.

\begin{description}
\item {\em H1}: For the homogeneous steady jet we use one tracer, $\theta$. We initialize this tracer to
\mbox{$\theta = +1$} for jet material, and \mbox{$\theta = -1$} for the
ambient medium.

\item {\em A1} and {\em I1}: For steady jets with structure, we employ two tracers; $\theta^{\rm sp}$ for
material from the jet spine and $\theta^{\rm sh}$ for material from the jet sheath. The tracer $\theta^{\rm sp}$
is initialized as \mbox{$\theta^{\rm sp} = +1$} for material in the jet spine and
\mbox{$\theta^{\rm sp} = -1$} elsewhere. Equivalently, tracer $\theta^{\rm sh}$ was
initialized as \mbox{$\theta^{\rm sh} = +1$} for material in the jet sheath and
\mbox{$\theta^{\rm sh} = -1$} elsewhere.

\end{description}

Despite the fact that the tracers are initiated with values \mbox{$\theta_{\rm A}(t,\bmath{r}) = \pm 1$}, as soon
as they are advected, actual mixing as well as effects from numerical discretization will yield tracer values
within a volume element $\delta V(t,\bmath{r})$ between \mbox{$-1 \le \theta_{\rm A}(t,\bmath{r}) \le +1$}. We
will interpret the tracer value $\theta_{\rm A}(t,\bmath{r})$ to directly correspond to the amount of constituent
$A$ in that volume element.

  \subsection{Mixing effects for various constituents} \label{subsubsec:Mixing}

Based on the amount of various constituents in a given volume element $\delta V(t,\bmath{r})$, we are able to
study the amount of mixing between different constituents. The following sections give a detailed description of
how mixing can be quantified.

The first type of mixing is called {\em absolute mixing} ($\Delta$) and deals with the exact mass fractions of
those constituents in a volume element. In that case \mbox{$\Delta = 0$} means that the constituents have not
mixed at all, while \mbox{$\Delta = 1$} means that the mass fractions of the constituents in a volume element
are equal, regardless of what those mass fractions are.

The second type of mixing is called {\em mass-weighted mixing} ($\Lambda$). For this type of mixing, the
mass fraction of a constituent in a volume element is divided by the {\em total mass} of that constituent in
the computational domain. It is therefore a measure of homogeneity: \mbox{$\Lambda = 0$} means no mixing, while
\mbox{$\Lambda = 1$} means a completely homogeneous mixture.

    \subsubsection{Mass fractions of multiple constituents in a volume} \label{subsubsec:MassFractions}

When considering fluid volume elements, all material within one element $\delta V(t,\bmath{r})$ (one grid cell) is
the sum of all its constituents. While some models treat the contents of a volume element with a multiple fluid
approach (i.e. different constituents having a different temperature, density, velocity, etc.), our numerical
method averages these quantities out, so that each grid cell can be characterized by one mass density, one
pressure, one velocity vector, etc., known as the {\em one-fluid approximation}. In that case, the total mass
density $\rho(t,\bmath{r})$ within $\delta V(t,\bmath{r})$ is the sum of mass densities of the different
constituents $\rho_{k}(t,\bmath{r})$. For a system with $N$ constituents, this can be written as:

\be
\rho(t,\bm{r}) = \sum_{k=1}^N \rho_{k}(t,\bm{r}) = \rho(t,\bm{r})\sum_{k=1}^N \delta_{k}(t,\bm{r})\; ,
\ee
where $\delta_{k}(t,\bmath{r})$ is the mass fraction of constituent $k$ within $\delta V(t,\bmath{r})$, so that:

\be
\sum_{k=1}^N \delta_{k}(t, \bm{r}) = 1 \; .\label{eq:sumdelta}
\ee

We are interested in the effect of mixing of two certain constituents $A$ and $B$ (e.g. jet spine material and
jet sheath material, or jet material and ambient medium material). In that case we can write the sum of the
mass fractions of the constituents as:

\be
\sum_{k=1}^N \delta_{k}(t, \bm{r}) = \delta_{\rm A}(t,\bm{r}) + \delta_{\rm B}(t,\bm{r}) +
\delta_{\rm \Sigma}(t,\bm{r}) = 1 \; ,
\ee
where $\delta_{\rm \Sigma}(t,\bmath{r})$ is the sum of all other components within $\delta V(t,\bmath{r})$
\footnote{Note that this can either simply be {\em the ambient medium}, but it could in theory also be a
whole collection of other constituents.}.
In the simple case where the system only consists of two constituents (e.g. jet and ambient medium), one has
\mbox{$\delta_{\Sigma}(t,\bmath{r}) = 0$}. Since this derivation applies to all individual grid cells, we will
drop the index $(t,\bmath{r})$ from now on.

    \subsubsection{Quantifying the amount of {\bf absolute} mixing $\Delta$} \label{subsubsec:AM}

To study the amount of mixing between the two constituents $A$ and $B$, it is useful to define an
{\em absolute mixing factor} $\Delta_{\rm AB}$, which considers the absolute amount of the mass fractions
within that cell. We choose $\Delta_{\rm AB} = 0$ in the case of no mixing by which we mean that only one of the
two components $A$ or $B$ is present within $\delta V$ and therefore \mbox{$\delta_{\rm A} = 0$} {\em or}
\mbox{$\delta_{\rm B} = 0$}. We choose $\Delta_{\rm AB} = 1$ in the case of maximum absolute mixing by which we
mean the same amount of constituents $A$ and $B$ are present within $\delta V$, so
\mbox{$\delta_{\rm A} = \delta_{\rm B}$}. Furthermore, we impose a linear scaling between the mass fractions
within the cell and the amount of absolute mixing. These assumptions completely determine the definition of
the absolute mixing factor for two different constituents in a cell:
\footnote{One subtle point must be made regarding this definition of the absolute
mixing factor. If a volume element $\delta V$ would contain neither of constituent $A$ or $B$ (and therefore
\mbox{$\delta_{\rm A} = \delta_{\rm B}= 0$}), the absolute mixing factor is not clearly defined since the
second term would yield a value \mbox{$0/0$}.

In this particular case, we define the absolute mixing factor $\Delta_{\rm AB}$ as first taking
one of the two mass fractions equal to 0 (so say \mbox{$\delta_{\rm A} = 0$}) and then formally taking the other
mass fraction equal to 0 (so say \mbox{$\delta_{\rm B} = 0$}). After having taken the first mass fraction equal
to 0, one is left with \mbox{$1 - \left | \pm \frac{\delta_{\rm B}}{\delta_{\rm B}}\right |$}, which always
yields a value of $0$, regardless of the value of the mass fraction $\delta_{\rm B}$. Therefore, the absolute
mixing in the case of absence of both constituents is per definition equal to \mbox{$\Delta_{\rm AB} = 0$.}}

\be
\Delta_{\rm AB} \equiv 1 - \left | \frac{\delta_{\rm A} - \delta_{\rm B}}{\delta_{\rm A} + \delta_{\rm B}}
\right | \; .\label{eq:AMfactor}
\ee

In theory, one can also consider the more general case of mixing between two {\em sets} of constituents; one
with a total mass fraction $\delta_{\Sigma_{\rm 1}}$ and a second with a total mass fraction
$\delta_{\Sigma_{\rm 2}}$. In that case formula \equref{eq:AMfactor} still applies, however, the indices $A$ and
$B$ will then replaced by the indices $\Sigma_{\rm 1}$ and $\Sigma_{\rm 2}$. In this paper, we will only be
concerned with the mixing of individual constituents.

    \subsubsection{Quantifying the amount of {\bf mass-weighted} mixing $\Lambda$} \label{subsubsec:MM}

Absolute mixing is a useful concept for situations where one is interested in the exact amounts of the
constituents within that volume. It will, however, not always give an intuitive sense for the amount of
homogeneity of the mixture.

To illustrate this, consider a fixed volume $V$ which is made up of two constituents $A$ and $B$, with total
masses $M_{\rm A}$ and $M_{\rm B}$, and their sum $M = M_{\rm A} + M_{\rm B}$. At first, these two constituents
are unmixed and separated by a wall, dividing $V$ into equal two parts $\frac{1}{2} V$. We then remove the wall
and stir up the constituents. When the constituents have had the time to settle down and maximally mix with each
other, the resulting mixture is homogeneous with mass density $\rho = \frac{M}{V}$. The two mass fractions in
every single cell in this case are \mbox{$\delta_{\rm A} = \frac{M_{\rm A}}{M}$} and
\mbox{$\delta_{\rm B} = \frac{M_{\rm B}}{M}$}. If $M_{\rm A}$ and $M_{\rm B}$ were not equal to begin with, even
though the mixture is completely homogeneous, the absolute mixing will be unequal to one:

\be
\mbox{$\Delta_{\rm AB} = 1 - \left | \frac{M_{\rm A} - M_{\rm B}}{M_{\rm A} + M_{\rm B}} \right |\; \ne 1$}.
\ee

It is possible to introduce another quantity which will yield a value of one for homogeneous mixtures. To this
end, we define the {\em mass-weighted mixing factor} $\Lambda_{\rm AB}$ in the same way as the absolute mixing
factor, but now the mass fractions $\delta_{\rm A}$ and $\delta_{\rm B}$ are weighted by their {\em total mass}
$M_{\rm A}$ and $M_{\rm B}$ contained in the total volume $V$:

\be
\Lambda_{\rm AB} \equiv 1 - \left | \frac{\delta_{\rm A} - \mu_{\rm AB} \delta_{\rm B}}
                                         {\delta_{\rm A} + \mu_{\rm AB} \delta_{\rm B}} \right |
\; ,\label{eq:MMfactor}
\ee
with $\mu_{\rm AB} = \frac{M_{\rm A}}{M_{\rm B}}$ the mass ratio of constituent $A$ and $B$. If we start out with
unequal amounts of mass $M_{\rm A}$ and $M_{\rm B}$, the mass-weighted mixing factor will yield a value of
$\Lambda_{\rm AB} = 1$ when the mixture has reached a homogeneous state. In that case the intuitive meaning of
{\em mixed well} simply means \mbox{$\Lambda \longrightarrow 1$}.

  \subsection{Absolute mixing and mass-weighted mixing from tracer values}
    \label{subsec:MixingfromTracer}

In the previous sections we calculated the amount of absolute mixing and mass-weighted mixing, based on the
mass fractions $\delta_{\rm A}$ and $\delta_{\rm B}$ within a volume element $\delta V$. In this section, we
express the amount of absolute and mass-weighted mixing in terms of the tracer values $\theta_{\rm A}$ and
$\theta_{\rm B}$ in each grid cell. In this way, we are able to find the amount of absolute and mass-weighted
mixing between different constituents in the jet simulations.

    \subsubsection{Absolute mixing from tracer values}

As a tracer $\theta_{\rm A}(t,\bmath{r})$ is advected by the flow, it obtains values within the range
\mbox{$\theta_{\rm min} \le \theta_{\rm A}(t,\bmath{r}) \le \theta_{\rm max}$}. Here $\theta_{\rm min}$
corresponds to the absence of constituent $A$ within this cell, whereas $\theta_{\rm max}$ corresponds to a cell
purely containing the constituent $A$. Since we interpret a tracer value to correspond directly to the amount
of that constituent in a linear way, the mass fraction of $A$ within grid cell $\delta V(t,\bmath{r})$ is
expressed by:

\be
\delta_{\rm A}(t,\bm{r}) = \left | \frac{\theta_{\rm A}(t,\bm{r}) - \theta_{\rm min}}
                                         {\theta_{\rm max} - \theta_{\rm min}} \right |.
\ee
With our choice of \mbox{$\theta_{\rm min} = -1$} and \mbox{$\theta_{\rm max} = +1$}, the mass fraction of
constituent $A$ equals (dropping the index $(t,\bmath{r})$ again):

\be
\delta_{\rm A} = \frac{1}{2} \left | \theta_{\rm A} + 1 \right | \; .\label{eq:deltaA}
\ee
The mass fraction $\delta_{\rm B}$ for constituent $B$ is found by changing the label
\mbox{$A \longrightarrow B$}. Therefore, the absolute mixing factor between constituents $A$ and $B$ in terms
of their tracer values in grid cell $\delta V$ can be written as:

\be
\Delta_{\rm AB} = 1 - \left | \frac{| \theta_{\rm A} + 1| - |\theta_{\rm B} + 1|}{| \theta_{\rm A} + 1| +
                  |\theta_{\rm B} + 1|} \right | \; . \label{eq:DeltaAB}
\ee

    \subsubsection{Absolute mixing between jet and shocked ambient medium}

In case of the homogeneous jet $H1$, there is just one jet constituent present and so we have one tracer $\theta$.
From equations \equref{eq:sumdelta} and \equref{eq:deltaA}, we find that the mass fraction of the (shocked)
ambient medium in terms of the jet tracer value $\theta$ in this case equals:

\be
\delta_{\rm am} = 1 - \delta_{\theta} = 1 -\frac{1}{2} \left | \theta + 1 \right |.
\ee

Substituting these values into the absolute mixing factor \equref{eq:AMfactor} leads to the absolute mixing
$\Delta$ between jet material and shocked ambient medium for the homogeneous jet:

\be
\Delta = 1 - \big| |\theta + 1| - 1 \big|.\label{eq:Delta}
\ee

    \subsubsection{Mass-weighted mixing from tracer values}

When dealing with two different jet constituents, the mass-weighted mixing factor in terms of tracer values
translates to:

\be
\Lambda_{\rm AB} = 1 - \left | \frac{| \theta_{\rm A} + 1| - \mu_{\rm AB}|\theta_{\rm B} + 1|}
                                    {| \theta_{\rm A} + 1| + \mu_{\rm AB}|\theta_{\rm B} + 1|} \right |\; .
\label{eq:LambdaAB}
\ee

Using the same reasoning as before, the mass-weighted mixing factor between a homogeneous jet and the
shocked ambient medium can be written as:

\be
\Lambda = 1 - \left | \frac{(1+\mu_{\rm jt-am})| \theta + 1| - 2\mu_{\rm jt-am}}
                           {(1-\mu_{\rm jt-am})| \theta + 1| + 2\mu_{\rm jt-am}} \right |\; ,\label{eq:Lambda}
\ee
with $\mu_{\rm jt-am} = \frac{M_{\rm jt}}{M_{\rm am}}$. Here, $M_{\rm jt}$ is the total mass that is injected
by the jet into the cocoon and $M_{\rm am}$ is the mass of the shocked ambient medium contained in the
cocoon at time $t$. The approximations of $M_{\rm jt}$ and $M_{\rm am}$ will be calculated in the next two
sections.

    \subsubsection{Total mass-energy discharge of the jet}

The total mass $M_{\rm jt}$ is equal to the energy-mass discharge through the Mach disk, integrated over time t
\footnote{We assume the energy-mass discharge through the Mach disk to remain constant during the entire
simulation. Therefore, we will use the initial conditions at the jet inlet in order to calculate the
energy-mass discharge of the jet.}.
Using the rules for velocity addition in special relativity, it can be shown that the energy-mass discharge
through the Mach disk for a steady homogeneous jet is:

\be
M_{\rm jt} = A_{\rm jt} \rho_{\rm jt} \gamma_{\rm jt}\gamma_{\rm hd} \left( v_{\rm jt} - v_{\rm hd} \right) t \; ,
\ee
with $A_{\rm jt}$ the surface of discharge which we take equal to $A_{\rm jt} = \pi R_{\rm jt}^2$ and as before,
$\rho_{\rm jt}$ the proper mass density, $\gamma_{\rm jt}$ and $\gamma_{\rm hd}$ the Lorentz factors of bulk
jet material and jet-head, respectively (measured in the observer's frame) and equivalently for the velocities
$v_{\rm jt}$ and $v_{\rm hd}$.

We consider very under-dense jets where the jet-head propagation speed is small compared to the bulk
velocity (\mbox{$v_{\rm hd} \ll v_{\rm jt}$}). In that case, the total mass injected by the jet into the cocoon
is approximated by:

\be
M_{\rm jt} \approx A_{\rm jt} \rho_{\rm jt} \gamma_{\rm jt} v_{\rm jt} t \; .
\ee

The total energy-mass discharge for jet spine and jet sheath are calculated in similar fashion. In that case the
indices "jt" are replaced by either "sp" for the jet spine, or "sh" for the jet sheath, and the correct
corresponding surface areas need to be considered. With these expressions for the total masses $M_{\rm sp}$ and
$M_{\rm sh}$, we find a mass ratio between jet spine and jet sheath material equal to:

\be
\mu_{\rm sp-sh} = \frac{M_{\rm sp}}{M_{\rm sh}} =
\frac{\rho_{\rm sp} \gamma_{\rm sp} v_{\rm sp}}
{(A_{\rm jt}/A_{\rm sp} - 1) \rho_{\rm sh} \gamma_{\rm sh} v_{\rm sh}} =
\frac{\rho_{\rm sp} v_{\rm sp}}{4 \rho_{\rm sh} v_{\rm sh}}\; .
\ee
Substituting the exact values for the isochoric jet results in \mbox{$\mu_{\rm sp-sh} = 0.052$}. In case of
the isothermal jet, we interpolate for the average mass density in the jet spine and the jet sheath to find
\mbox{$\mu_{\rm sp-sh} = 0.217$}.

    \subsubsection{Total mass of the shocked ambient medium}

Finally, we approximate the total mass of the shocked ambient medium $M_{\rm am}$ by the volume containing this
shocked material $V_{\rm co}^{\rm am}$, multiplied by the average local mass density in this volume
$\rho_{\rm co}^{\rm am}$. For $V_{\rm co}^{\rm am}$ we take the volume of the cocoon $V_{\rm co}$ minus the
volume bounded by the contact discontinuity $V_{\rm cd}$:

\be
V_{\rm co}^{\rm am}=V_{\rm co}-V_{\rm cd} \approx \frac{3}{4} V_{\rm co} \; ,
\ee
where we have approximated $V_{\rm cd} \approx \frac{1}{4}V_{\rm co}$
\footnote{In the "ideal case" where there would be no instabilities causing turbulent mixing, the cocoon would
consist of two regions separated by the contact discontinuity; the inner region $V_{\rm cd}$ containing purely
shocked jet material and the outer region $V_{\rm co}^{\rm am}$ containing purely shocked ambient medium. In the
realistic case where instabilities and turbulent mixing do occur, this outer region could in principle
mix with shocked jet material. That is why we consider the volume $V_{\rm co}^{\rm am}$ for calculating the mass
of the interacting ambient medium.}.
This material is actually {\em shocked} ambient medium, where the density is approximately compressed by a factor
of \mbox{$r \approx 4$}. This yields:

\be
M_{\rm am} = \rho_{\rm co}^{\rm am} V_{\rm co}^{\rm am} \approx 3 \rho_{\rm am} V_{\rm co}\; .
\ee

With these expressions for the total masses $M_{\rm jt}$ and $M_{\rm am}$, we find a mass ratio between shocked
jet and shocked ambient medium material equal to:

\be
\mu_{\rm jt-am} = \frac{M_{\rm jt}}{M_{\rm am}} \approx
\frac{A_{\rm jt} \rho_{\rm jt} \gamma_{\rm jt} v_{\rm jt} t}{3 \rho_{\rm am} V_{\rm co}} \; .
\ee
Taking the values from the simulation of the homogeneous jet, as they occur after 23 Myr, we find a mass fraction
ratio of \mbox{$\mu_{\rm jt-am} \approx 3.06 \times 10^{-6}$}. The mass ratios $\mu_{\rm jt-am}$ for the
homogeneous jet and $\mu_{\rm sp-sh}$ for the isothermal jet and for the isochoric jet will be used extensively
in the \mbox{section \ref{sec:Results}} to determine the level of homogeneity of mixing between different
constituents.

  \subsection{The relation between the effective polytropic index $\Gamma_{\rm eff}$ and the temperature T}
\label{subsec:GeffTemp}
\begin{figure}
\includegraphics[clip=true,trim=2cm 2cm 1cm 1cm,width=0.5\textwidth]{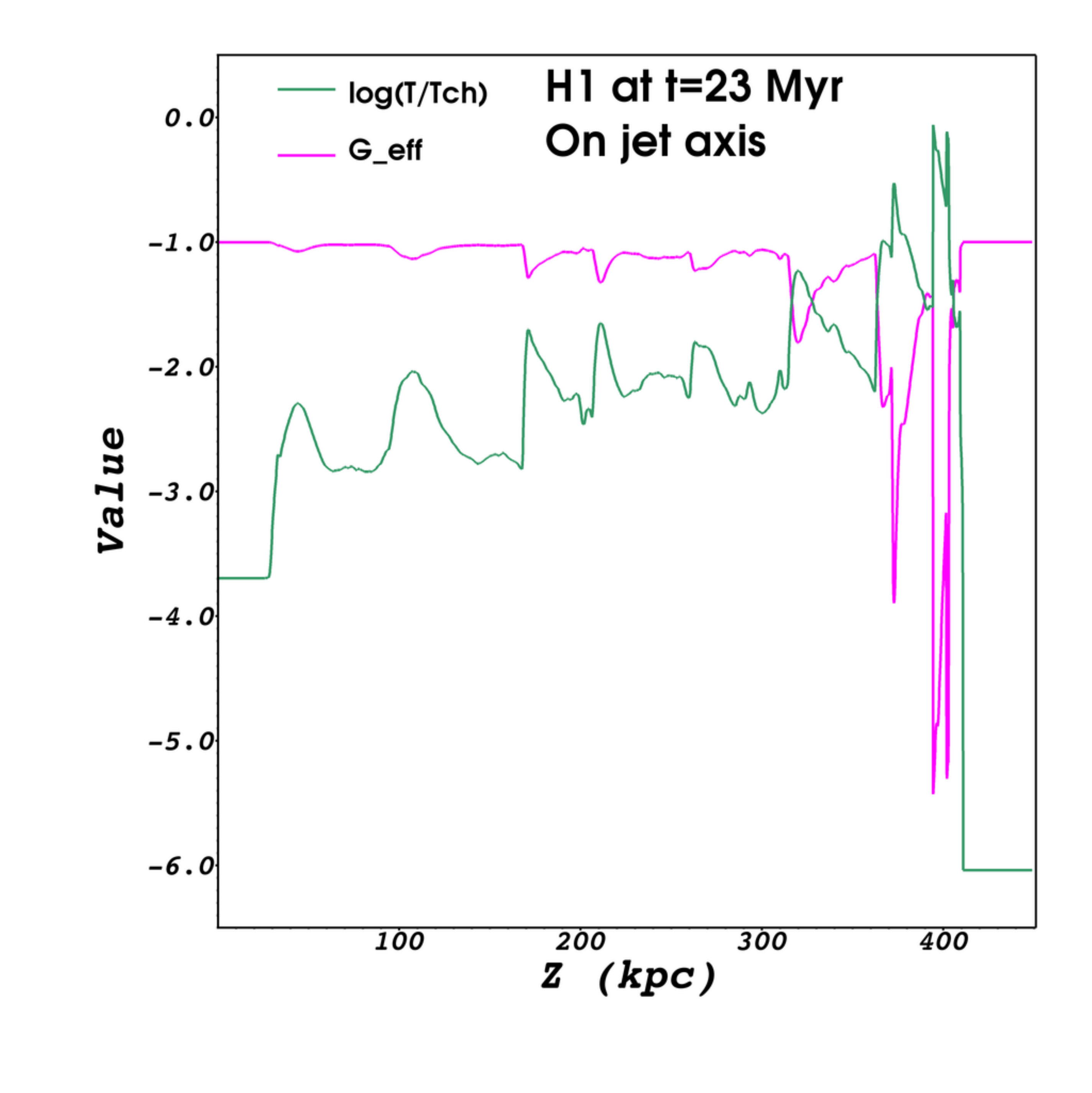}
\caption{A cut along the jet axis of a homogeneous jet after 22.8 Myr. The plot shows the $\log_{10}$ of
the thermal temperature $T$ in units of the characteristic temperature $T_{\rm ch} = 1.09 \times 10^{13}$ K
(green) and the rescaled effective polytropic index $G_{\rm eff}$ (pink) which represents a non-relativistic
equation of state for $G_{\rm eff} = -1$ and a relativistic equation of state for $G_{\rm eff} = -6$ (occurring
at $T \sim T_{\rm ch}$).}
  \label{fig:geffTemp}
\end{figure}
In equation \equref{eq:SyngeEOS} we've already seen the effective polytropic index $\Gamma_{\rm eff}$ that
describes a realistic transition between a non-relativistic gas and a relativistic gas, based on the
particle rest-mass energy $m_{\rm p} c^2$ (re-introducing $c$ for the moment) and the average thermal energy
per particle $\epsilon_{\rm th}$. When this energy $\epsilon_{\rm th}$ becomes comparable to the rest-mass energy
of the particle, the gas becomes relativistic. We therefore define the transition from non-relativistic to
relativistic at the point where \mbox{$k_{\rm b} T = m_{\rm p} c^2$}. This implies that the gas becomes
relativistic when the effective polytropic index drops below \mbox{$\Gamma_{\rm eff} = 1.417$} (which comes from
putting \mbox{$\epsilon_{\rm th} = m_{\rm p}c^2$} in equation \equref{eq:SyngeEOS}). Since there is a one-to-one
correspondence between $\Gamma_{\rm eff}$ and the temperature $T$, this also introduces a thermal temperature
at which the proton gas becomes relativistic, namely
\mbox{$T_{\rm ch} \equiv \frac{m_{\rm p} c^2}{k_{\rm b}} = 1.09 \times 10^{13}$ K}.

Figure \ref{fig:geffTemp} shows a cut along the jet axis of the steady homogeneous jet after 22.8 Myr. At
this time, several internal shocks have developed, heating the jet material. Up to the point where the
temperature is approximately $10^{12}$ K, the (proton) gas can still be described as non-relativistic. Only
near the Mach disk at $Z \sim 400$ kpc, the temperature approaches $10^{13}$ K and there indeed the effective
polytropic index drops to relativistic values.

Note that we have actually plotted a quantity $G_{\rm eff}$ here, which is a rescaled version of the effective
polytropic index:

\be
G_{\rm eff} = 5\left[3\left(\Gamma_{\rm eff} - \frac{4}{3}\right)\right] -6 \; ,
\ee
in order to make the variations of the effective polytropic index more clear. The rescaled polytropic index has
values that lie between \mbox{$-6 \le G_{\rm eff} \le -1$}, such that \mbox{$G_{\rm eff}=-1$} corresponds to the
classical equation of state \mbox{$\Gamma = 5/3$} and \mbox{$G_{\rm eff} = -6$} to a relativistic equation of
state \mbox{$\Gamma = 4/3$}.

\section{Results}\label{sec:Results}

  \subsection{Jet-head advance speed from the simulations}
\label{subsec:JHASsim}

\begin{figure}
\includegraphics[clip=false,width=0.5\textwidth,angle=0]{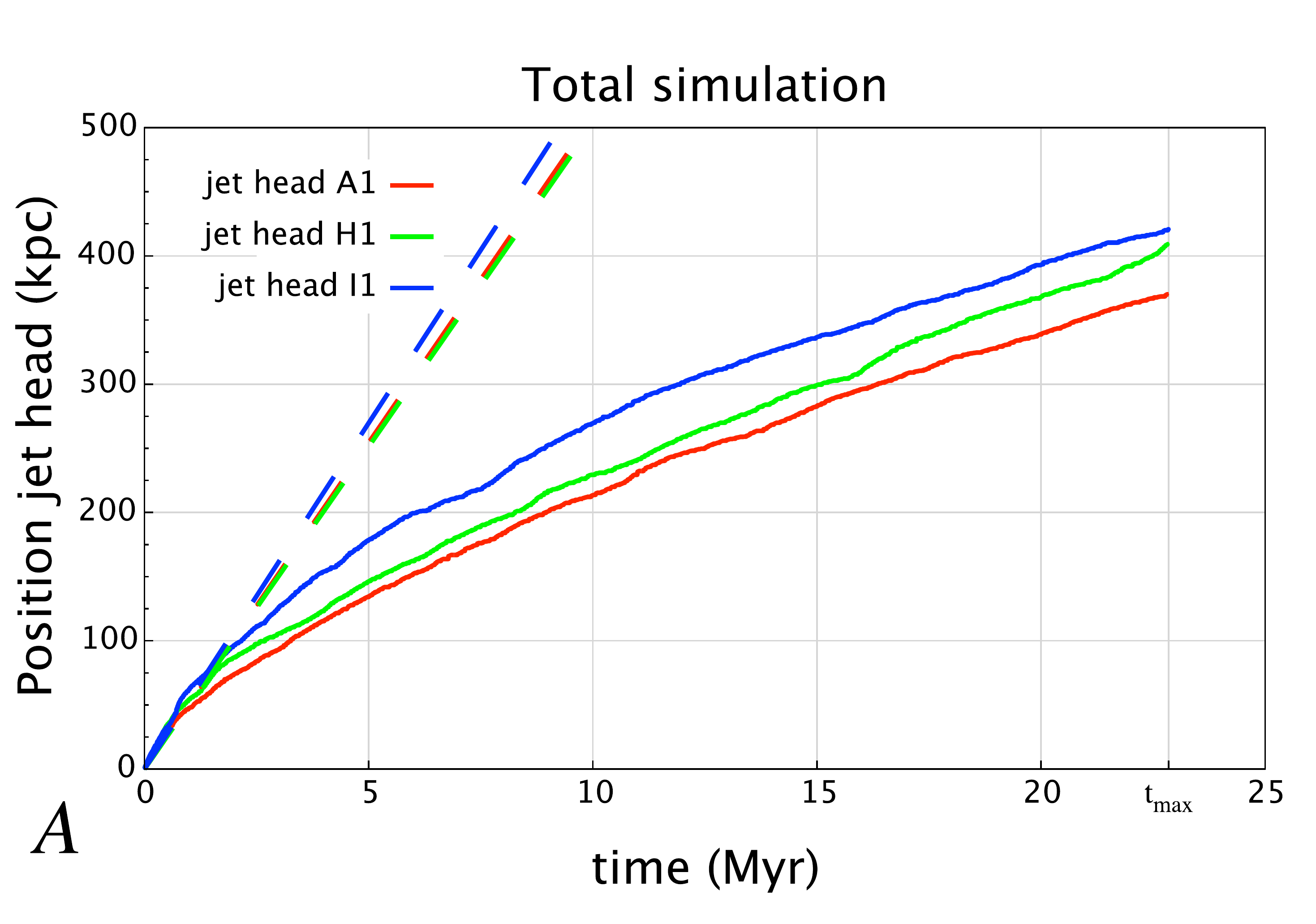} \\
\includegraphics[clip=false,width=0.5\textwidth,angle=0]{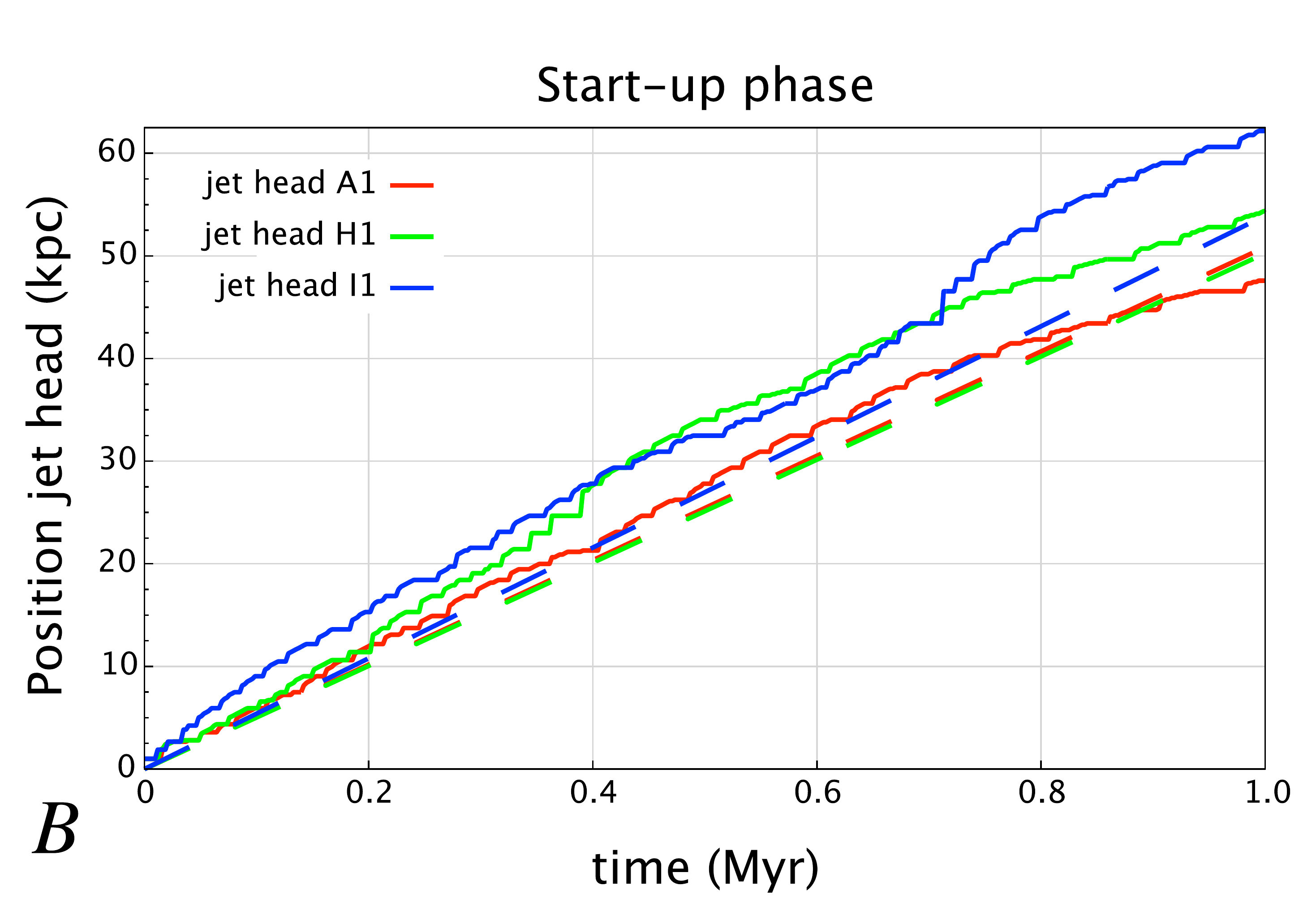}
\caption{Jet-head propagation for all the steady jets $H1$, $I1$ and $A1$. The solid lines show the results
from the simulations, the dashed lines show the analytically predicted values. Figure {\bf A} shows the
full duration of the simulation (up to \mbox{$t_{\rm max} = 22.8$ Myr)}. The jet-head propagation is not
constant, but decelerates after a start-up phase of \mbox{$\sim$ 1 Myr}. The start-up phase of the
simulations can be seen in figure {\bf B}.}
  \label{fig:JHAS-1}
\end{figure}

In section \ref{subsec:JHAS}, we showed that the jet-head advance speed can be estimated from ram-pressure
arguments. There, we assumed both the Mach disk and the bow shock of the jet-head to be strong shocks.
In addition, we assumed the pressure of the jet material behind the Mach disk and the ambient medium behind
the bow shock to be equal. These are reasonable approximations for a homogeneous jet. However, in case of a jet
with a fast moving jet spine and slower moving jet sheath, or when instabilities are taken into account, a simple
(quasi-1$D$) analytical derivation will no longer be possible because of a complex flow structure near the head
of the jet, see figures \mbox{\ref{fig:jetheads} ({\bf A, B} and {\bf C})}.

Despite the numerous processes that can influence the velocity of the jet-head, we would like to make a
comparison between the simulations and the simple theory in section \ref{subsec:JHAS}. In figure \ref{fig:JHAS-1},
we plotted the position of the jet-head as a function of time for all three different models. Plot
\mbox{\ref{fig:JHAS-1} {\bf A}} shows the position of the jet-head for the full evolution of the simulations
and plot \mbox{\ref{fig:JHAS-1}} {\bf B} shows the start-up phase \mbox{($\le$ 1.0 Myr)}. It can clearly be
seen that for all three models the jet-head velocities during the start-up phase are higher than during the rest
of the simulation. During this phase, relatively little turbulence and instabilities have occurred in the cocoons.
As can be seen in \ref{fig:JHAS-1} {\bf B}, the jet-head propagation for all three models during the start-up
phase compares quite well with the analytically predicted values.

The analytically predicted value for the jet-head advance speed for the homogeneous jet $H1$ can directly be
calculated from equation \equref{eq:JHAS} and the values listed in table \ref{tab:parameters}. In case of the
isochoric jet $A1$, we calculate the jet-head advance speed, based on the parameters for jet spine and jet sheath
separately and then take the area-weighted average of these velocities over the cross section of the jet. In
case of the isothermal jet $I1$, we first calculate the average density of the jet spine and the average density
of the jet sheath separately
\footnote{The isothermal jet has a density profile that depends on radius.}.
Then we calculate the advance speed for jet spine and jet sheath separately and finally take the area-weighted
average. The jet-head advance speed from these theoretical calculations, the actual advance speed of the start-up
phase of the simulations; and the actual advance speed at the final 2.8 Myr of the simulations are shown in table
\ref{tab:JHAS}. We will discuss the slow-down of the jet-head advance speed from start-up phase to final phase
in section \ref{subsec:EffImpactArea}.

{\renewcommand{\arraystretch}{1.3}
\begin{table}
\caption{Jet-head advance speed (JHAS), jet radius before jet-head ($R_{\rm hd}$) and effective impact radius
($R_{\rm eff}$) for the models $H1$, $I1$ and $A1$.}
  \rowcolors{1}{white}{lightgray}
    \begin{tabular}{l c c c c} \hline
      {\bf JHAS, $\mathbf{R_{\rm \mathbf{hd}}}$ and $\mathbf{R_{\rm \mathbf{ext}}}$}
                                                           & {\bf units} & {\bf H1} & {\bf I1} & {\bf A1}
       \\ \hline
      {\em Analytical prediction}                          & [c]         & 0.164    & 0.176    & 0.164
      \\
      {\em Start-up phase}             (1.0 Myr)           & [c]         & 0.174    & 0.199    & 0.152
      \\
      {\em Final phase}                (2.8 Myr)           & [c]         & 0.047    & 0.032    & 0.035
      \\
      {\em Jet radius, just before head} $R_{\rm hd}$      & [kpc]       & $\sim$ 5 & $\sim$ 6 & $\sim$ 7
      \\
      {\em Eff. impact radius}          $R_{\rm eff}$      & [kpc]       & 4.02     & 6.27     & 5.46
      \\ \hline
    \end{tabular}

\medskip
The first row shows the analytically predicted value for the jet-head advance speed. The second row shows the
average advance speed over the first \mbox{1 Myr}, where relatively little turbulence and instabilities have
formed. The third row shows the average advance speed over the final 2.8 Myr. The fourth row shows the jet
radius, just before the jet-head. The bottom row shows the effective impact radius of the area of the ambient
medium that effectively impacts the jet.
    \label{tab:JHAS}
\end{table}
}

  \subsection{Internal shocks along the jet axis}\label{subsec:ShocksAxis}

\begin{figure*}
$
\begin{array}{c c c}
\includegraphics[clip=true,trim=2.5cm 1.5cm 1cm 1cm,width=0.333\textwidth]{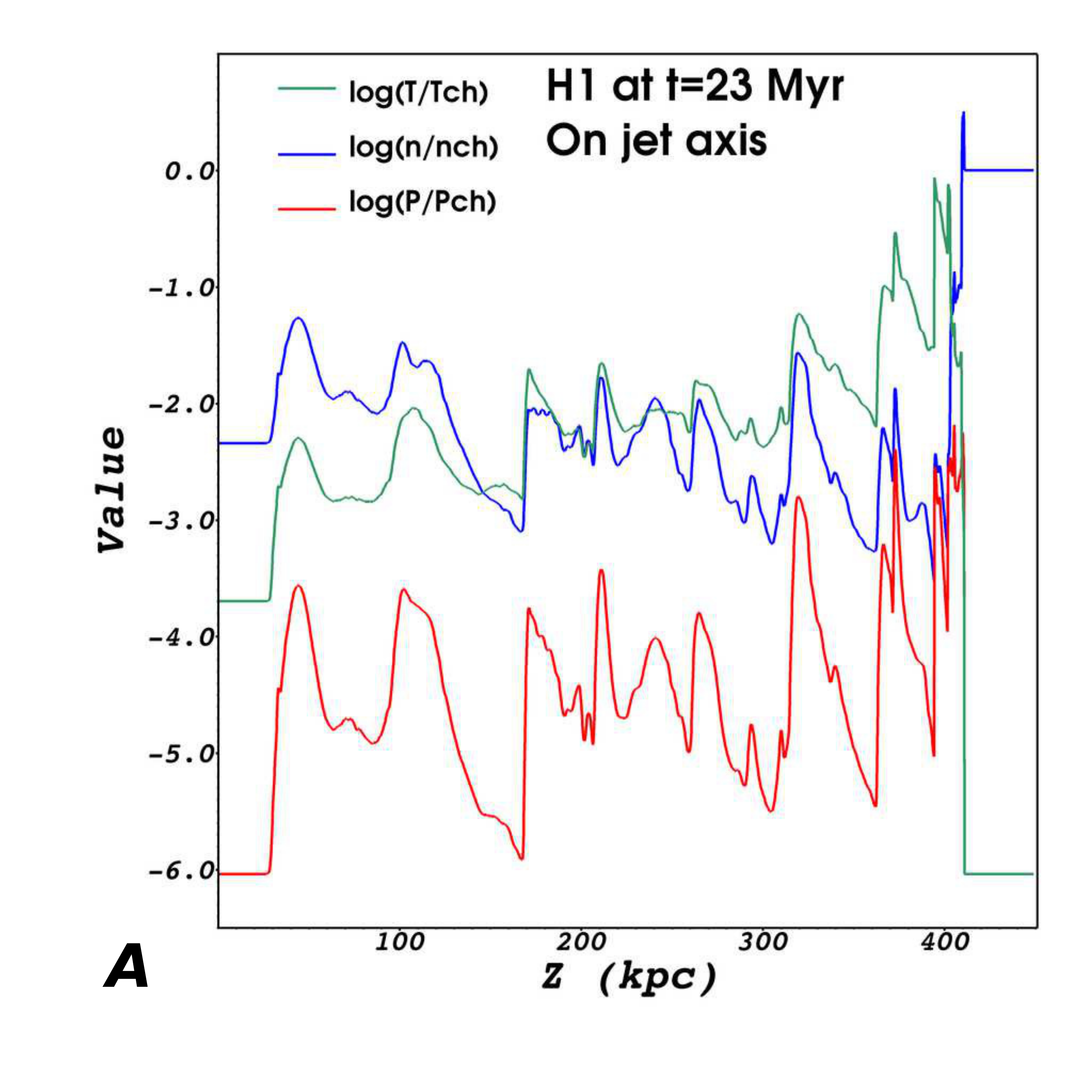} &
\includegraphics[clip=true,trim=2.5cm 1.5cm 1cm 1cm,width=0.333\textwidth]{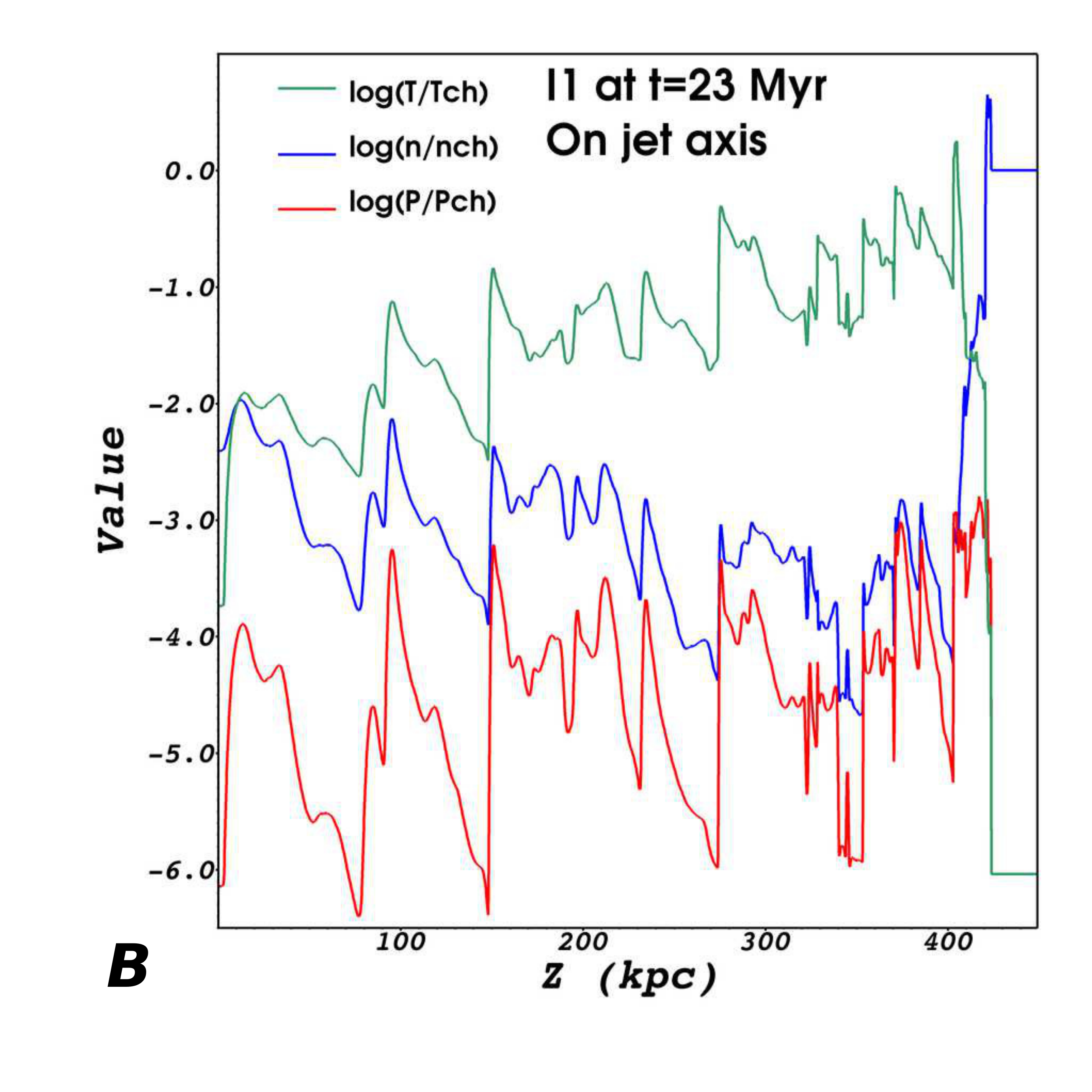} &
\includegraphics[clip=true,trim=2.5cm 1.5cm 1cm 1cm,width=0.333\textwidth]{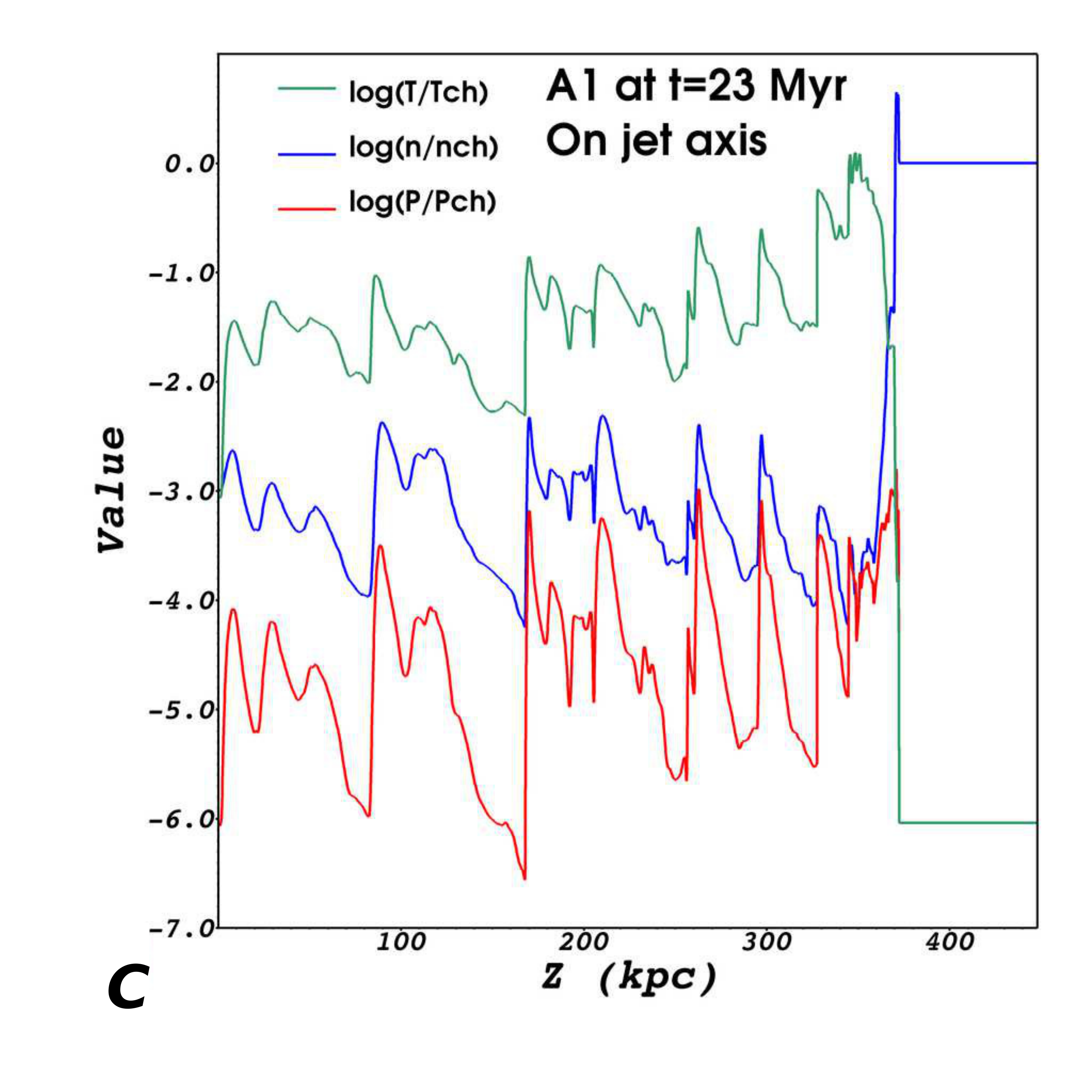} \\
\includegraphics[clip=true,trim=2.5cm 1.5cm 1cm 1cm,width=0.333\textwidth]{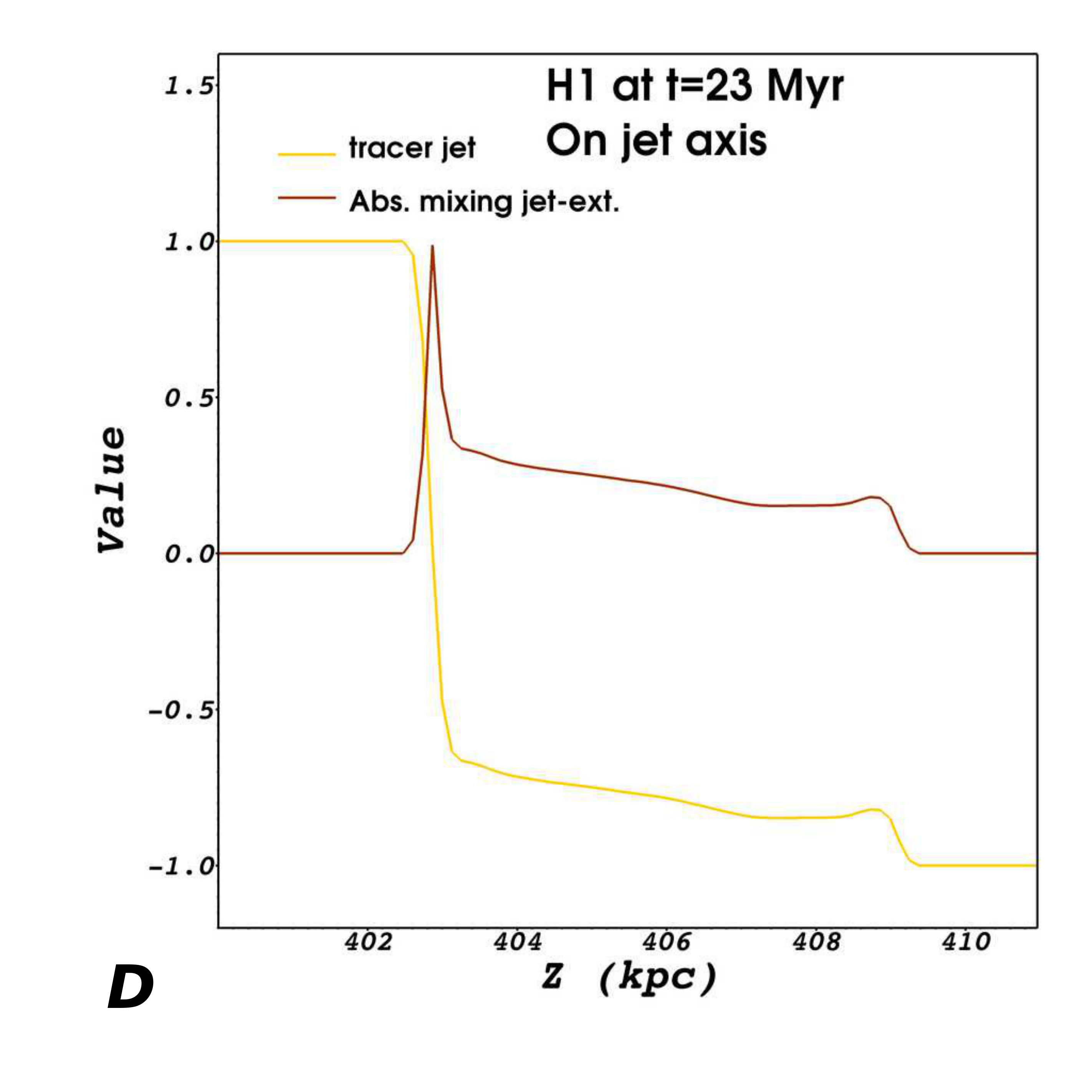} &
\includegraphics[clip=true,trim=2.5cm 1.5cm 1cm 1cm,width=0.333\textwidth]{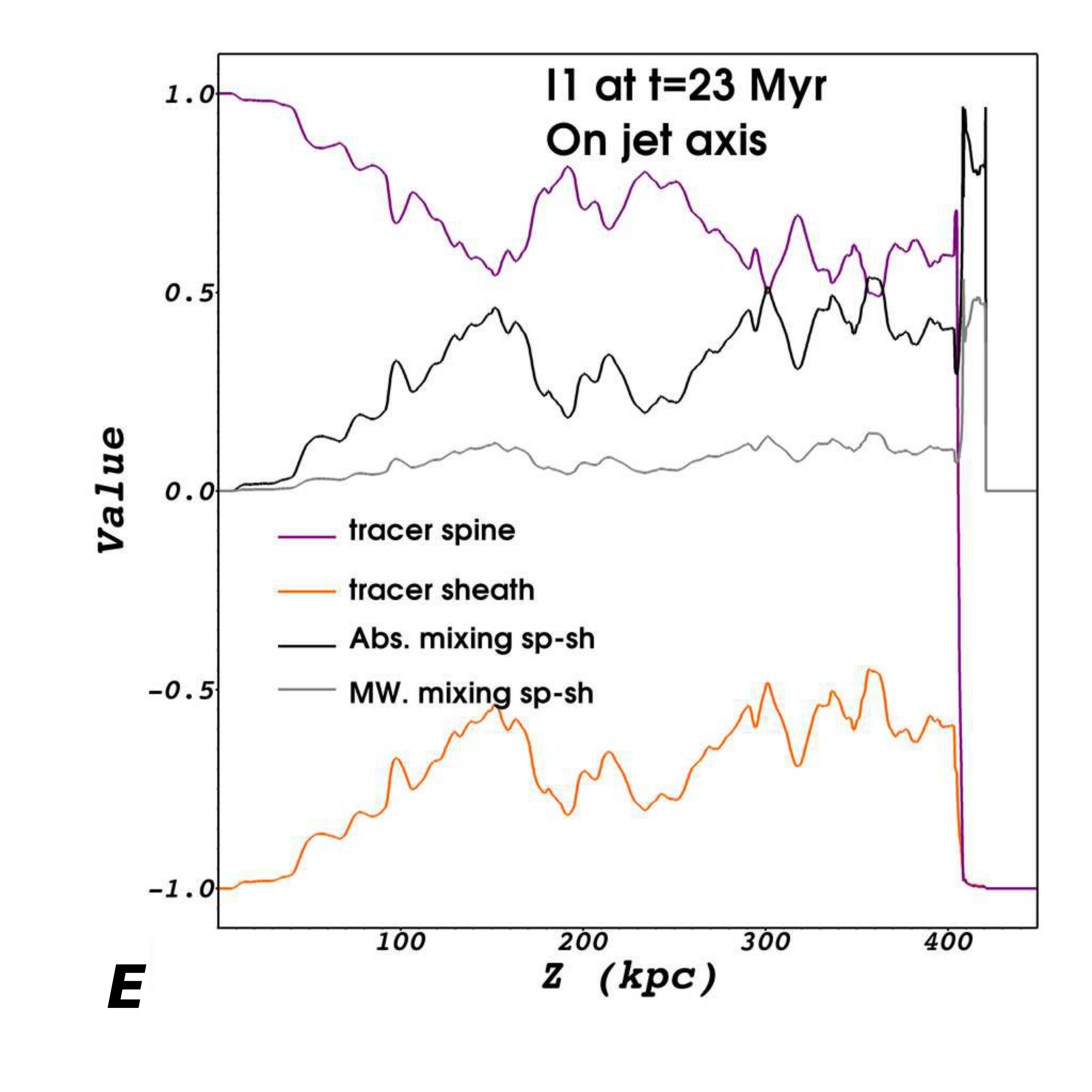} &
\includegraphics[clip=true,trim=2.5cm 1.5cm 1cm 1cm,width=0.333\textwidth]{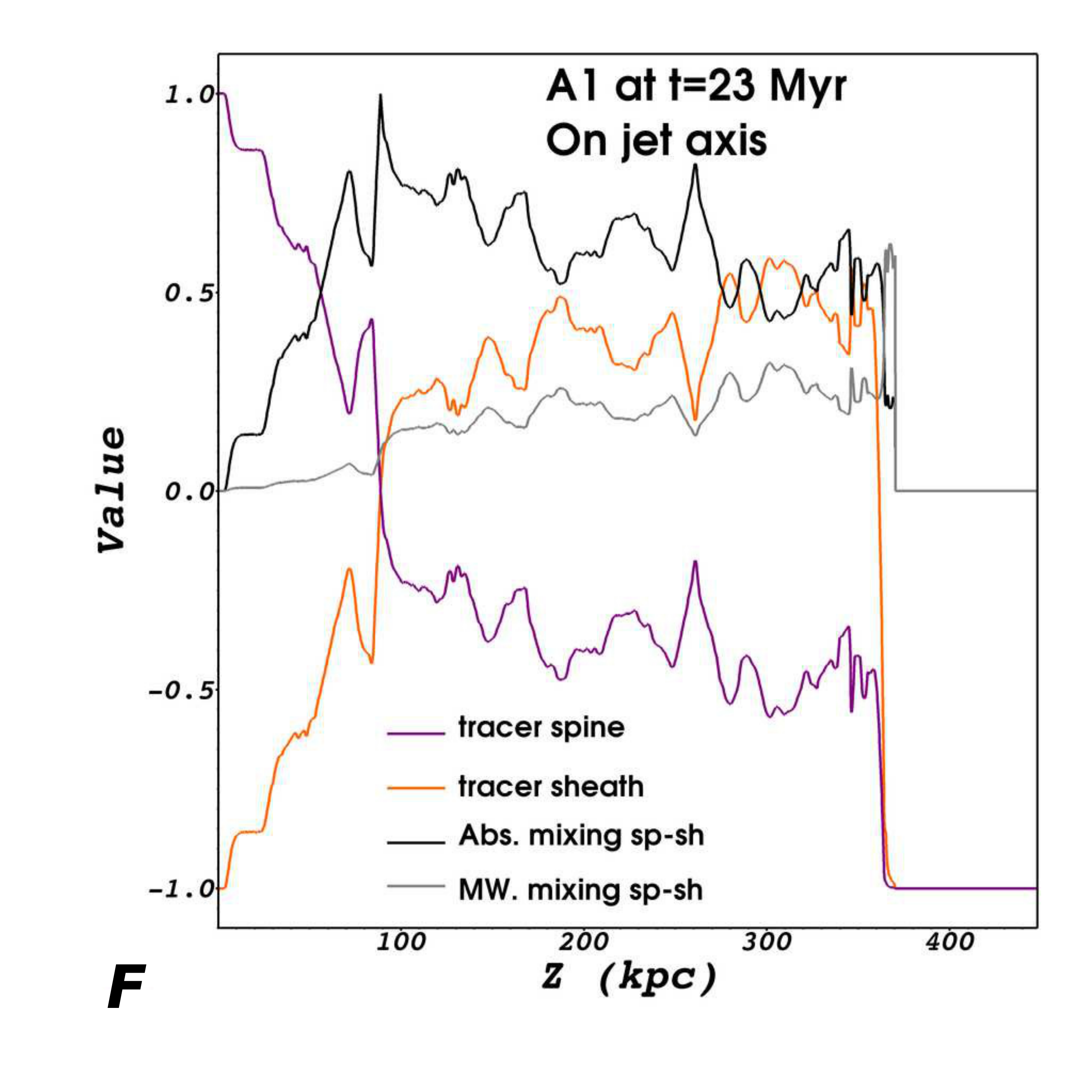}
\end{array}
$
\caption{Cuts along the jet axis (\mbox{$R=0$ kpc}) for the homogeneous jet $H1$ (left panels), the isothermal
jet $I1$ (centre panels) and the isochoric jet $A1$ (right panels) at $t=22.8$ Myr.
Top panels {\bf (plots A, B, C)} show the ($\log_{10}$ of) the number density in units of the
characteristic number density \mbox{$n_{\rm ch}$} (blue); the gas pressure in units of the characteristic
pressure $P_{\rm ch} = 1.50 \times 10^{-6}\;{\rm erg \; cm^{-3}}$ (red); and the thermal temperature $T$ in
units of the characteristic temperature $T_{\rm ch} = 1.09 \times 10^{13}$ K (green). The lower panel
{\bf (plots D, E, F)} shows tracer values and absolute and mass-weighted mixing. For the homogeneous jet
({\bf plot D}), only the jet-head is shown, between \mbox{$400 \le Z \le 410$ kpc}. This is the
only part along the jet axis where these variables show some variation. For the isothermal and isochoric jets
{\bf (plots E} and {\bf F}), the tracer values of jet spine material $\theta^{\rm sp}$ (purple), the tracer of
jet sheath material $\theta^{\rm sh}$ (orange), as well as the amount of internal absolute mixing (black) and
the amount of internal mass-weighted mixing (gray) between jet spine and jet sheath material are shown.
}
  \label{fig:CrossCuts}

\end{figure*}

In figures \ref{fig:CrossCuts} and \ref{fig:ContourPlots}, the simulations for the three steady jet
models $H1$, $I1$ and $A1$ are shown at the end of the simulation, corresponding to a time
\mbox{$t_{\rm max} = 22.8$ Myr}. Moreover, most variables that we show in our line
plots are dimensionless, given in units of the characteristic variables. Therefore, we indicated their values on
the vertical axis as ``{\em Value}".

The top panel of figure \mbox{\ref{fig:CrossCuts} (line plots {\bf A, B} and {\bf C}}) shows a cut along the jet
axis \mbox{$R = 0$ kpc} of three quantities, namely: ($\log_{10}$ of) the gas pressure $P$;
the number density $n$; and the thermal temperature $T$, in units of their characteristic quantities.

The lower panel of figure \mbox{\ref{fig:CrossCuts} (line plots {\bf D, E} and {\bf F}}) show a cut along the jet
axis of the tracer values and the absolute and mass-weighted mixing factors $\Delta$ and $\Lambda$. 

Several internal shocks appear along the jet axes, as can be seen in line plot \ref{fig:geffTemp} and
line plots \ref{fig:CrossCuts} ({\bf A, B} and {\bf C}). These shocks can be recognized by their strong jumps in
density, temperature and pressure. Vortices that emerge at the jet-head break off at a certain point, after
which they travel down the cocoon, along the jet axis. As we will discuss in section
\ref{subsec:CocoonJetCoupling}, these vortices are responsible for the internal shocks occurring along the jet
axis. Roughly nine shocks are found in all three models $H1$, $I1$ and $A1$ along the jet axes at the final time
of simulation.

The first internal shock after the jet inlet occurs fairly quickly for the $H1$, $I1$, and $A1$ jets. This might
be somewhat unexpected since the jets are set up in pressure equilibrium with their surroundings. However, as
soon as the jets start to plow through the ambient medium, a (forward) bow shock and a reverse shock (the Mach
disk) form. Shocked ambient medium that crosses the bow shock and jet material that crosses the Mach disk are
shock-heated at the jet-head. The shocked gasses flow downstream, away from the jet-head. This causes the pressure
to rise to $P_{\rm co} \sim 10^2 \; P_{\rm am}$ inside the cocoon. Since the pressure is kept constant at
the jet inlet, the jet quickly becomes under-pressured compared to the cocoon. As a result, the jet is
compressed and a structure containing a compression fan and a shock is formed. This shock tries to re-establish
pressure equilibrium between the jet and the surrounding cocoon. These adjustment shocks are an artifact of the
boundary conditions applied in the simulations and will not occur in actual jets. However, a new pressure
equilibrium is formed fairly quickly after the jet inlet. Therefore, we expect the influence of the adjustment
shock on the further evolution of the systems to be small.

The global behavior of the internal shocks for the three models are similar: After the adjustment shock, a
second strong shock is located at \mbox{$Z \sim 75-90$ kpc} and a third strong shock occurs at
\mbox{$Z \sim 140-170$ kpc}. After these  three strong internal shocks more shocks follow. These shocks are
less strong than the previous shocks, the separation between these shocks is smaller and they show more chaotic
behavior, i.e. fluctuations in pressure, density and temperature along the jet axis. The final and strongest
shock occurs at the jet-head. In this Mach disk, the jet material is shock-heated to relativistic temperatures.

The main differences between the individual models are as follows: For the $H1$ jet, the adjustment shock occurs
at a slightly larger distance from the jet inlet \mbox{($Z \sim 30$ kpc)} than for the $I1$ and $A1$ jets
(both \mbox{$Z \sim 5$ kpc}). A second distinction that can be made is the form of the shocks. The $H1$ jet
shows clear distinct shocks, where each shock is represented by a single peak in density, pressure and
temperature. The $I1$ jet on the other hand has more variable behavior. Each shock is followed by a few
small variations in the density, pressure and temperature. Finally, the $A1$ jet shows the most variable
behavior: each individual strong shock is followed by a number of (typically 2) weaker shocks. Therefore,
based on the internal structure of the jets along the jet axis, the $H1$ jet is the most regular jet, the
$I1$ jet comes next and finally the $A1$ jet is the least regular jet.

  \subsection{Temperature along the jet axis}\label{subsec:TemperatureAxis}

The temperature along the jet axis for the three different jet models shows some general behavior: it is
strongly coupled to the occurrence of internal shocks. Each shock heats up the jet material. The rise in
temperature leads to a strong increase in pressure, which in turn causes the jet to expand sideways. This
expansion leads to a decrease in temperature, but along the jet axis, the temperature shows an overall increase
all the way up to the jet-head.

The behavior of the temperature for the individual models is as follows: The $H1$ jet at jet inlet
\mbox{($Z = 0$ kpc)} has a temperature of \mbox{$T \sim 2.1 \times 10^9$ K}. After the first (adjustment) shock
at \mbox{$Z \sim 30$ kpc}, the temperature rises to \mbox{$T \sim 5.5 \times 10^{10}$ K}. Then finally after the
Mach disk, the temperature rises up to \mbox{$T \sim 9.7 \times 10^{12}$ K}.

The temperature of the $I1$ jet also has a value of \mbox{$T \sim 2.1 \times 10^{9}$ K} at the jet inlet (as it
is set up). However, already after the first shock the temperature increases to
\mbox{$T \sim 1.4 \times 10^{11}$ K}. At the Mach disk the temperature rises up to
\mbox{$T \sim 1.4 \times 10^{13}$ K}, putting it well into the relativistic regime and making it the hottest jet
of the three models (as seen on the jet axis).

Finally, the temperature of the $A1$ jet at jet inlet is \mbox{$T \sim 9.1 \times 10^{9}$ K} (so higher than the
other jets by a factor \mbox{of $\sim 4.5$}). At the first shock, the temperature rises to
\mbox{$T \sim 3.9 \times 10^{11}$ K}. At the jet's head, the temperature reaches its highest value,
\mbox{$T \sim 1.1 \times 10^{13}$ K}, again putting it into the relativistic regime.

  \subsection{Mixing effects along the jet axis}\label{subsec:MixingAxis}

In this subsection we will discuss the amount of mixing {\em along the jet axis} (line plots
\mbox{\ref{fig:CrossCuts} {\bf D, E} and {\bf F}}). None of the three jets mix with shocked ambient medium from
the cocoon at a notable level at any point along the jet axis. The simulations show that only a very small
fraction of shocked ambient medium is entrained by the jet. In line plot \mbox{\ref{fig:CrossCuts} {\bf D}} for
the $H1$ jet, this can directly been seen from the fact that the absolute mixing factor \mbox{$\Delta \sim 0$},
except at the jet-head. In case of the spine-sheath jets (line plots
\mbox{\ref{fig:CrossCuts} {\bf E}} and {\bf F}), this is seen from the fact that the sum of the jet spine and
jet sheath tracers always add up to $\sim$ zero, so that each grid cell inside the jet, along the jet axis,
contains little or no entrained material from the ambient medium. Therefore, the jets maintain their stability
along the jet axis.

In case of the spine-sheath jets, however, the structural integrity of the jets is not necessarily maintained
along the jet axis. Even though no mixing with the ambient medium occurs along the jet axis, jet spine and jet
sheath material within the jets are capable of mixing due to effects such as the Kelvin-Helmholtz instability,
strong internal shocks or the formation of vortices. The next two subsections will treat these mixing effects
along the jet axis for the $I1$ jet and the $A1$ jet separately.

    \subsubsection{The isothermal jet $I1$}

Line plot \ref{fig:CrossCuts} {\bf E} shows the mixing behavior between jet spine and jet sheath for the
isothermal jet. At the jet inlet \mbox{($Z = 0$)}, the jet consists of pure jet spine material. This
corresponds with the initialization of the jet (\mbox{$\theta^{\rm sp} = + 1$ and $\theta^{\rm sh} = - 1$}).
Therefore, the absolute mixing factor, as well as the mass-weighted mixing factors are both zero
\mbox{($\Delta_{\rm sp-sh} = \Lambda_{\rm sp-sh} = 0$)}.

As one moves towards the jet-head, the amount of mixing increases. However, the internal shocks that occur along
the jet axis do not significantly correlate with the internal mixing between jet spine and jet sheath material.
There is gradual increase in the amount of mixing up to \mbox{$Z \sim 150$ kpc} from the jet inlet, where the
absolute mixing approaches a peak of \mbox{$\Delta_{\rm sp-sh} \sim 0.5$}. This peak is the only point that
coincides with a strong shock at \mbox{$Z \sim 150$ kpc}. Moving further towards the jet-head, the absolute
mixing varies slightly, but never exceeds \mbox{$\Delta_{\rm sp-sh} = 0.5$}. Then finally at the Mach disk, the
absolute mixing becomes maximal and attains values of \mbox{$\Delta_{\rm sp-sh} \longrightarrow 1$}.

The mass-weighted mixing is small and never exceeds \mbox{$\Lambda_{\rm sp-sh} \sim 0.1$}. Therefore,
despite the fact that there is some (internal) mixing between jet spine and jet sheath, the level of mixing is
only $\sim 10 \%$ compared to a fully homogeneous mixture.

The tracers don't intersect anywhere along the jet axis. The physical meaning of this is that the material on
the jet axis consists almost entirely of jet spine material, all the way up to the jet-head. We therefore expect
that, based on the amount of jet spine and jet sheath constituents, the spine-sheath jet structure can be
recognized all the way to the jet-head.

    \subsubsection{The isochoric jet $A1$}

As in the case of the isothermal jet, we see that the isochoric jet at inlet consists of pure jet spine material
(\mbox{line plot \ref{fig:CrossCuts} {\bf F}}). There are two features which clearly differ from the isothermal
case: First of all, we see a strong correlation between the location of the internal shocks and the increase in
the level of (absolute-, as well as mass-weighted) mixing. Secondly, the tracer values $\theta^{\rm sp}$ and
$\theta^{\rm sh}$ both quickly approach zero at the second strong shock \mbox{($Z \sim 82$ kpc)}. Therefore,
at that location, the mass fractions of jet spine and jet sheath are equal in those grid cells. After this strong
internal shock, the tracers $\theta^{\rm sp}$ and $\theta^{\rm sh}$ switch signs. This means that from that point,
moving towards the jet-head, the jet axis is dominated by jet sheath material. Therefore, the jet spine and jet
sheath have undergone strong internal mixing, where the absolute mixing factor attains values of
\mbox{$\Delta_{\rm sp-sh} \sim 0.5 - 1$}.

The mass-weighted mixing, on the other hand, remains fairly low \mbox{($\Lambda_{\rm sp-sh} \le 0.3$)},
reflecting the fact that even though there has been considerable mixing between jet spine and jet sheath
material along the jet axis, the mixture is far from homogeneous. This suggests that some radial structure
should still be recognizable, however, a distinction between a jet spine and a jet sheath will probably no
longer be visible. Section \ref{subsec:JetStructureandIntegrity} discusses the explanation for this difference
in the amount of mixing in more detail.

  \subsection{Radial cuts across the jets}\label{subsec:RadialCuts}

\begin{figure*}
$
\begin{array}{c c c}
\includegraphics[clip=true,trim=1cm 1.25cm 1.25cm 1.25cm,width=0.333\textwidth]{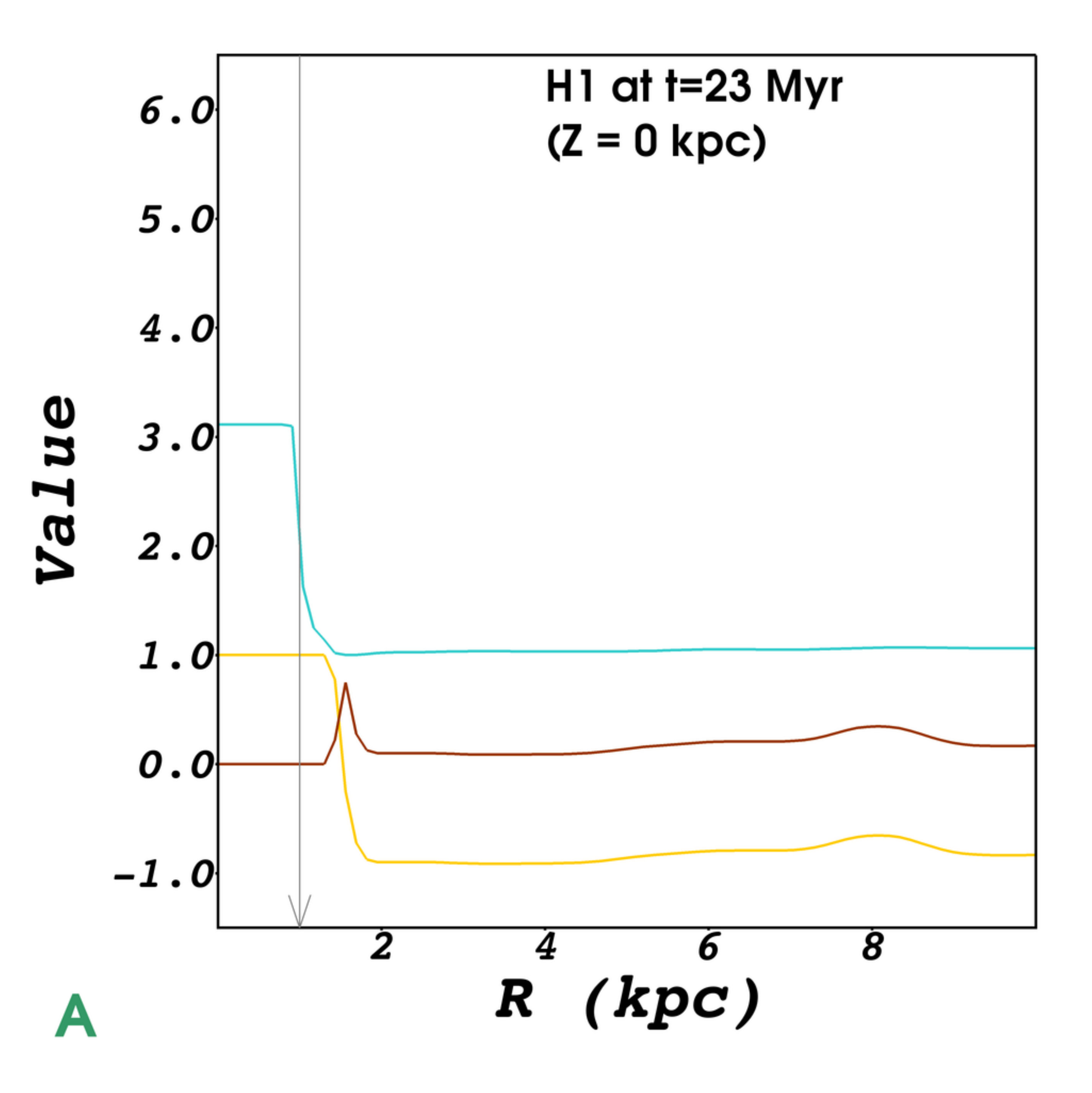}&
\includegraphics[clip=true,trim=1cm 1.25cm 1.25cm 1.25cm,width=0.333\textwidth]{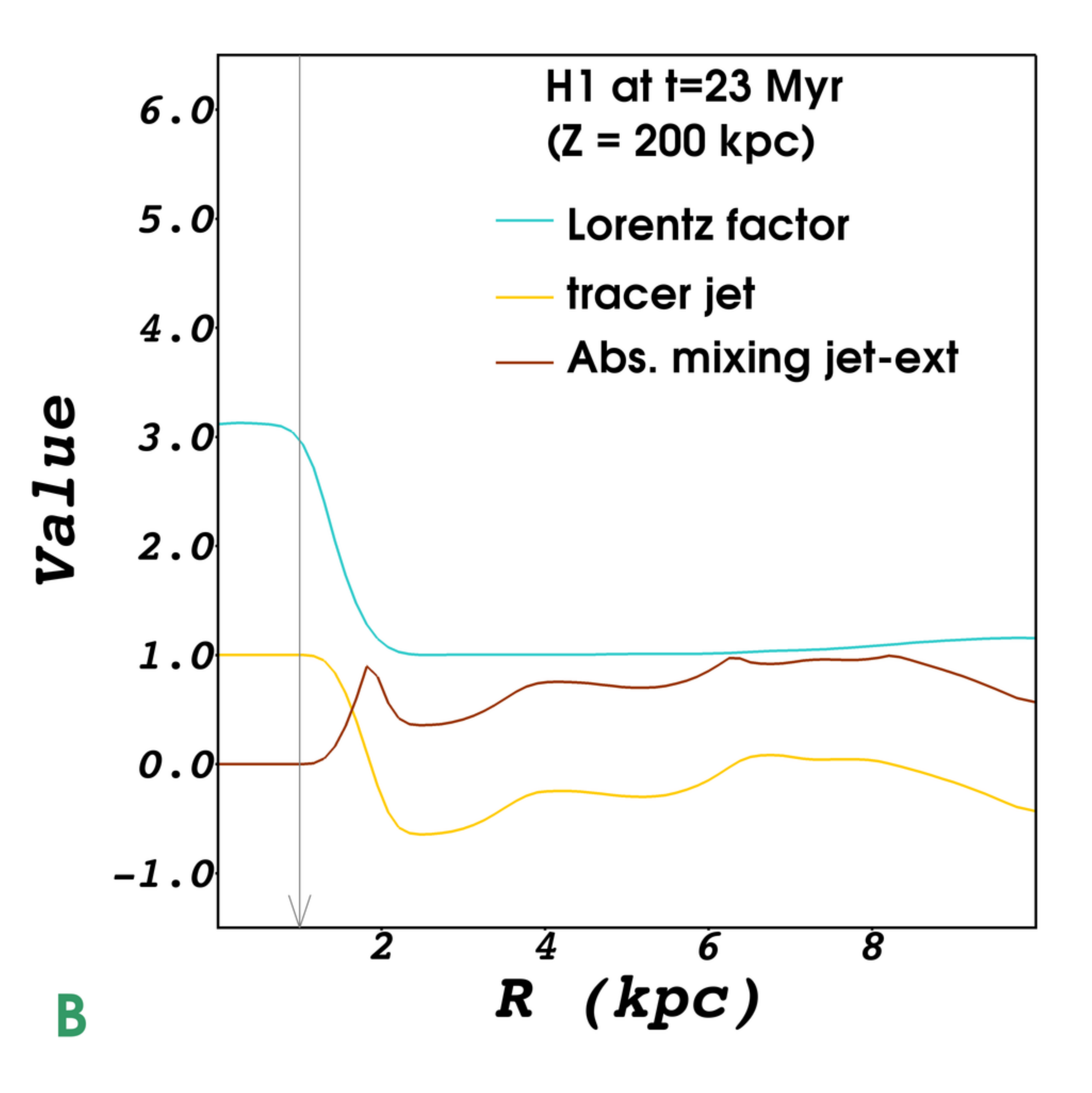}&
\includegraphics[clip=true,trim=1cm 1.25cm 1.25cm 1.25cm,width=0.333\textwidth]{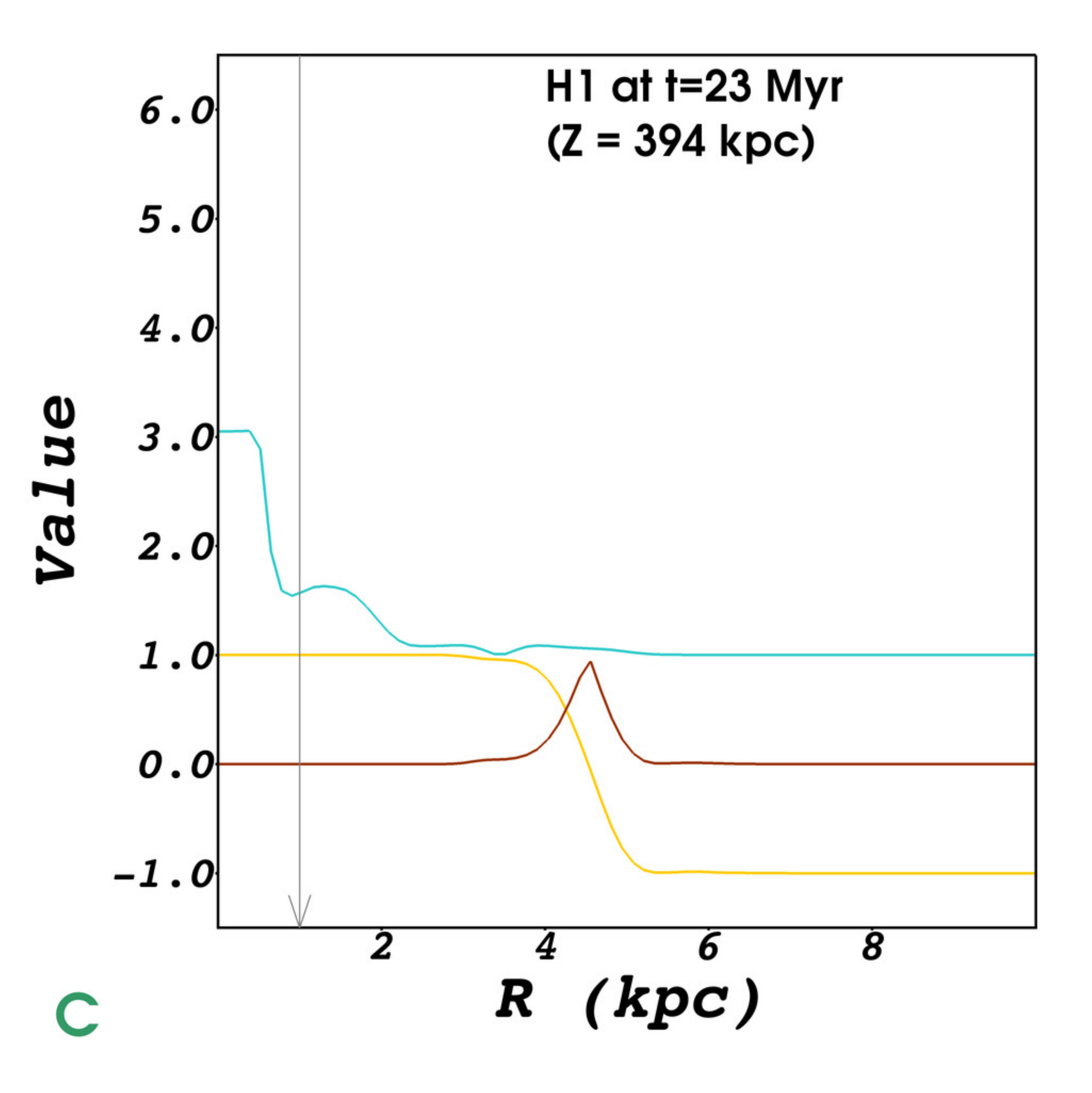}\\
\includegraphics[clip=true,trim=1cm 1.25cm 1.25cm 1.25cm,width=0.333\textwidth]{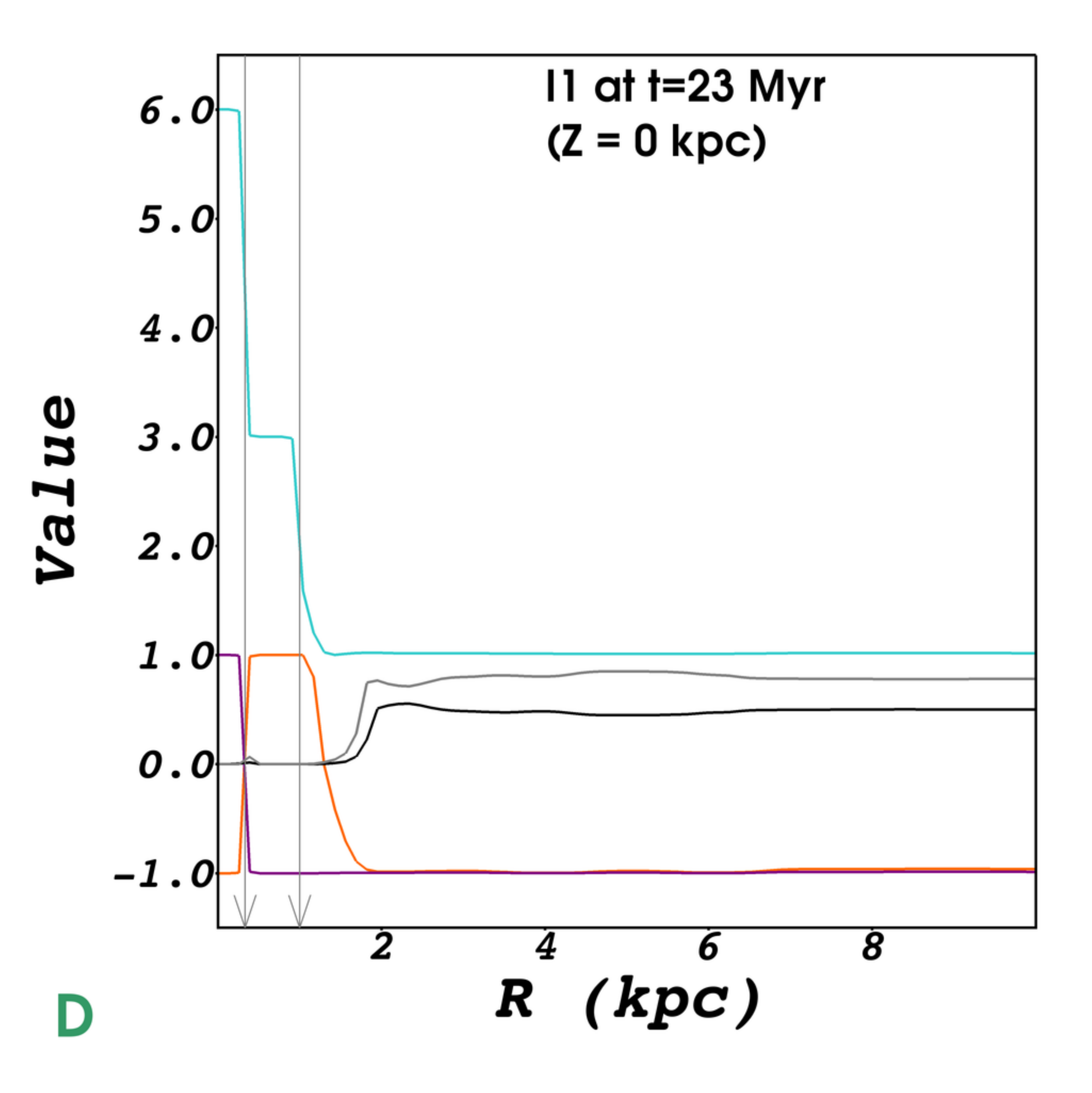}&
\includegraphics[clip=true,trim=1cm 1.25cm 1.25cm 1.25cm,width=0.333\textwidth]{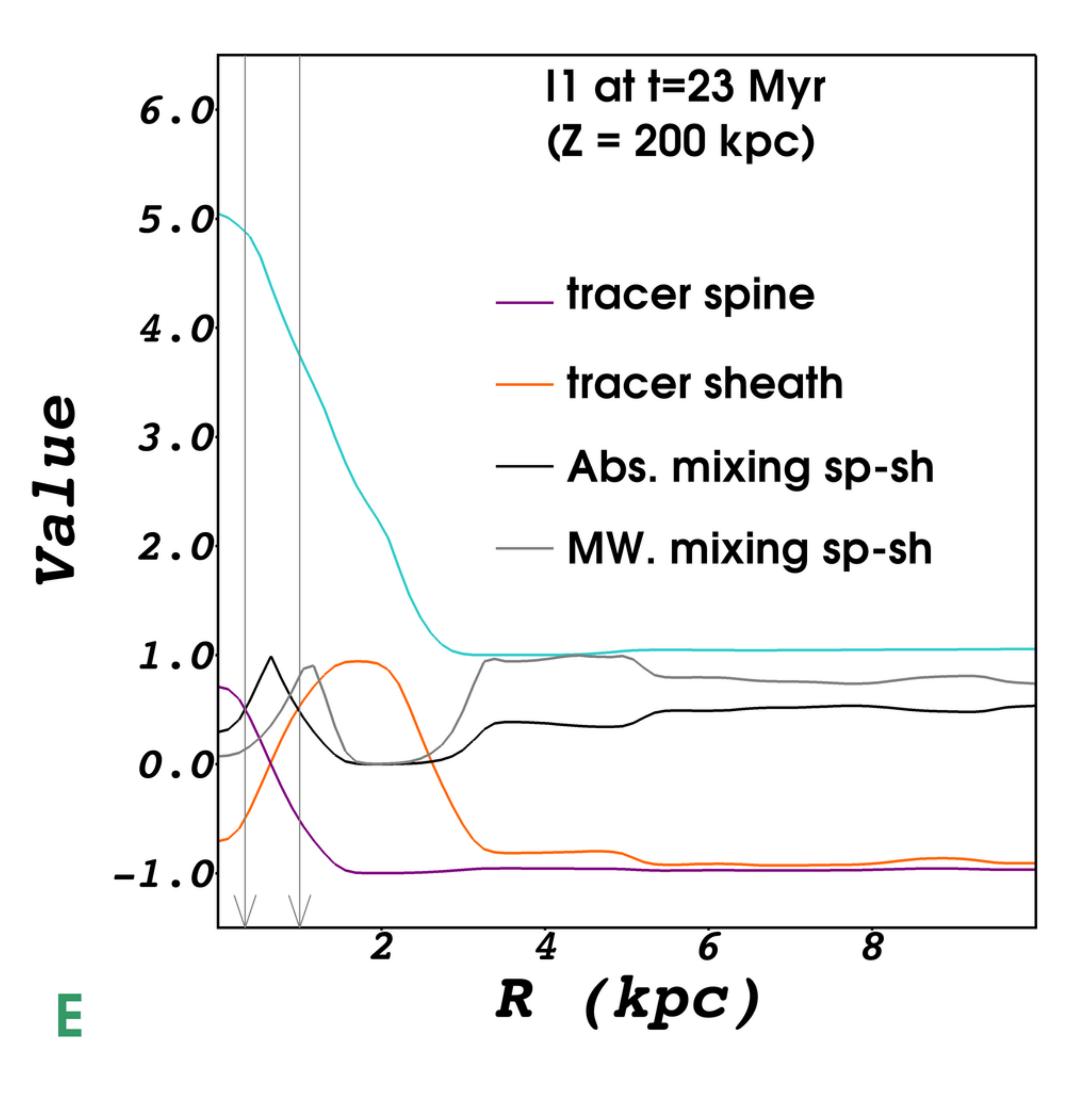}&
\includegraphics[clip=true,trim=1cm 1.25cm 1.25cm 1.25cm,width=0.333\textwidth]{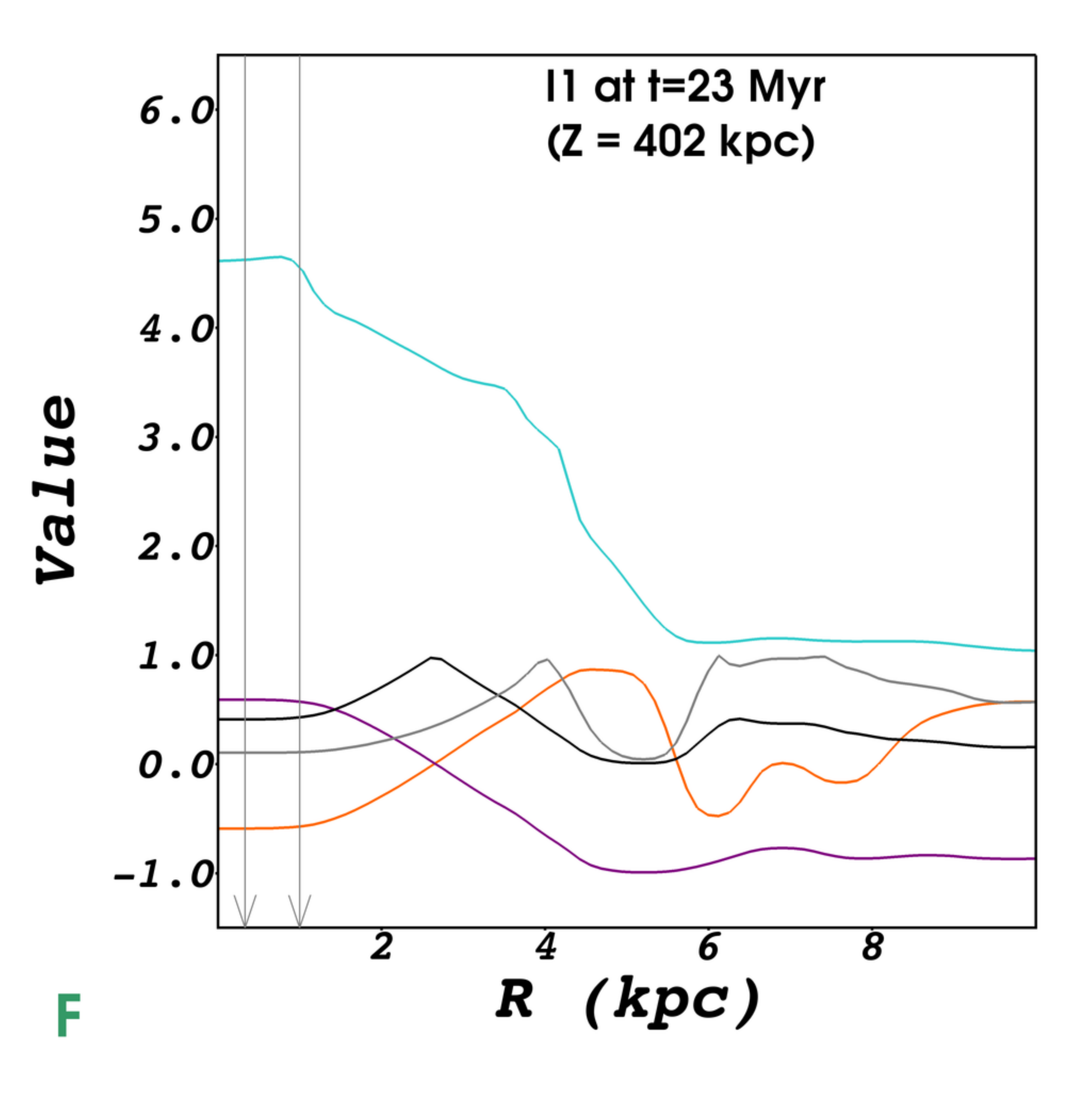}\\
\includegraphics[clip=true,trim=1cm 1.25cm 1.25cm 1.25cm,width=0.333\textwidth]{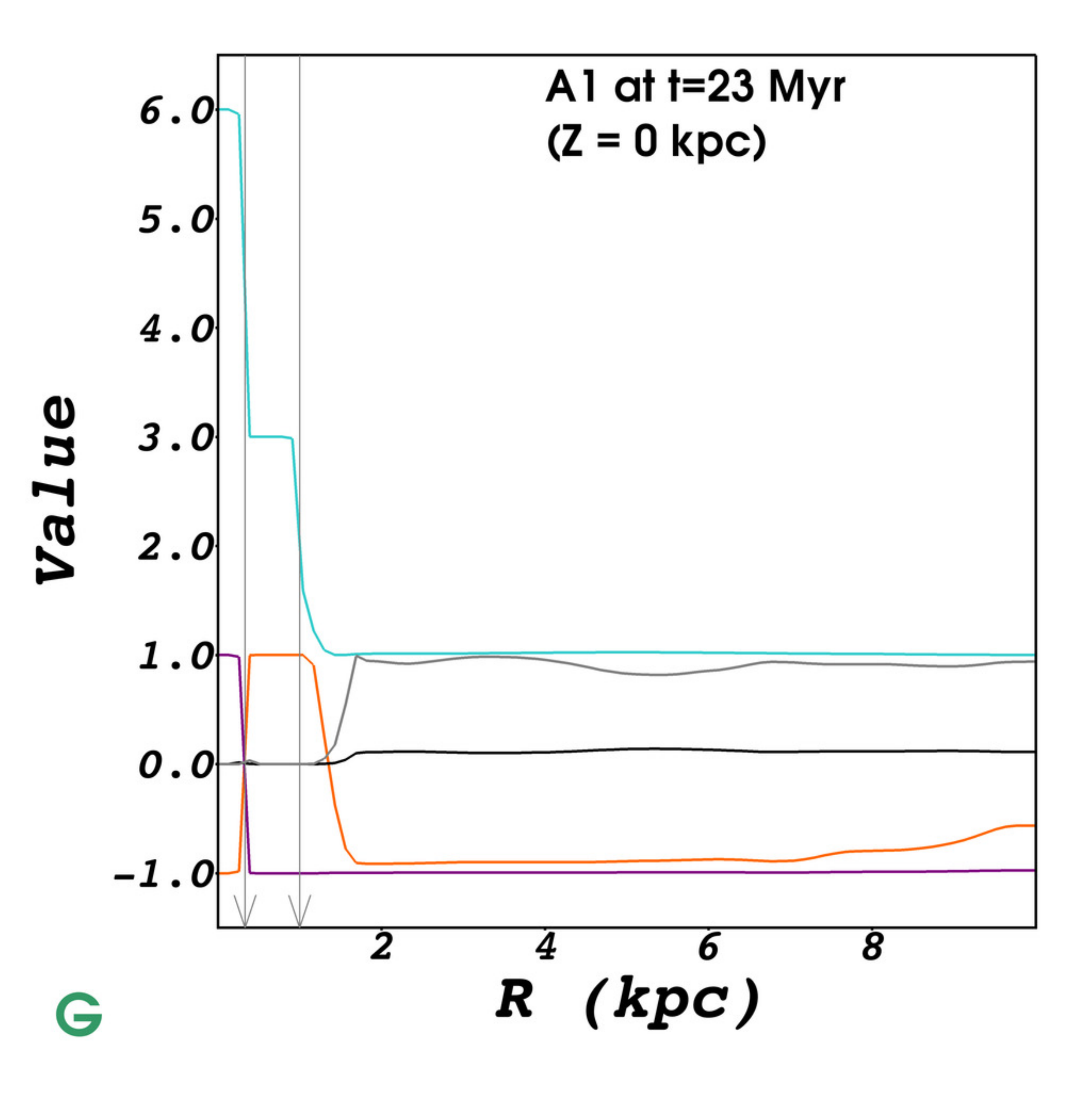}&
\includegraphics[clip=true,trim=1cm 1.25cm 1.25cm 1.25cm,width=0.333\textwidth]{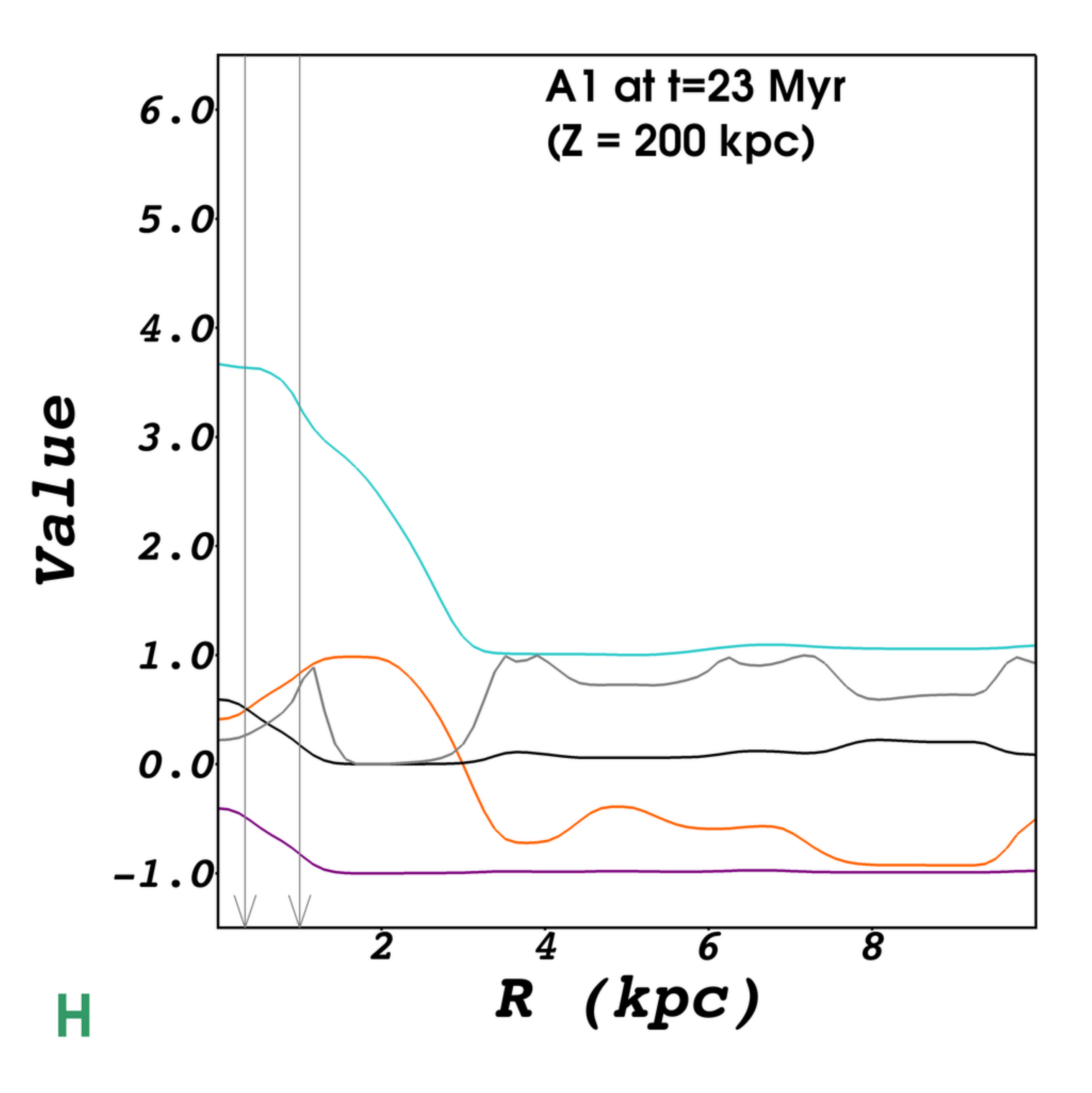}&
\includegraphics[clip=true,trim=1cm 1.25cm 1.25cm 1.25cm,width=0.333\textwidth]{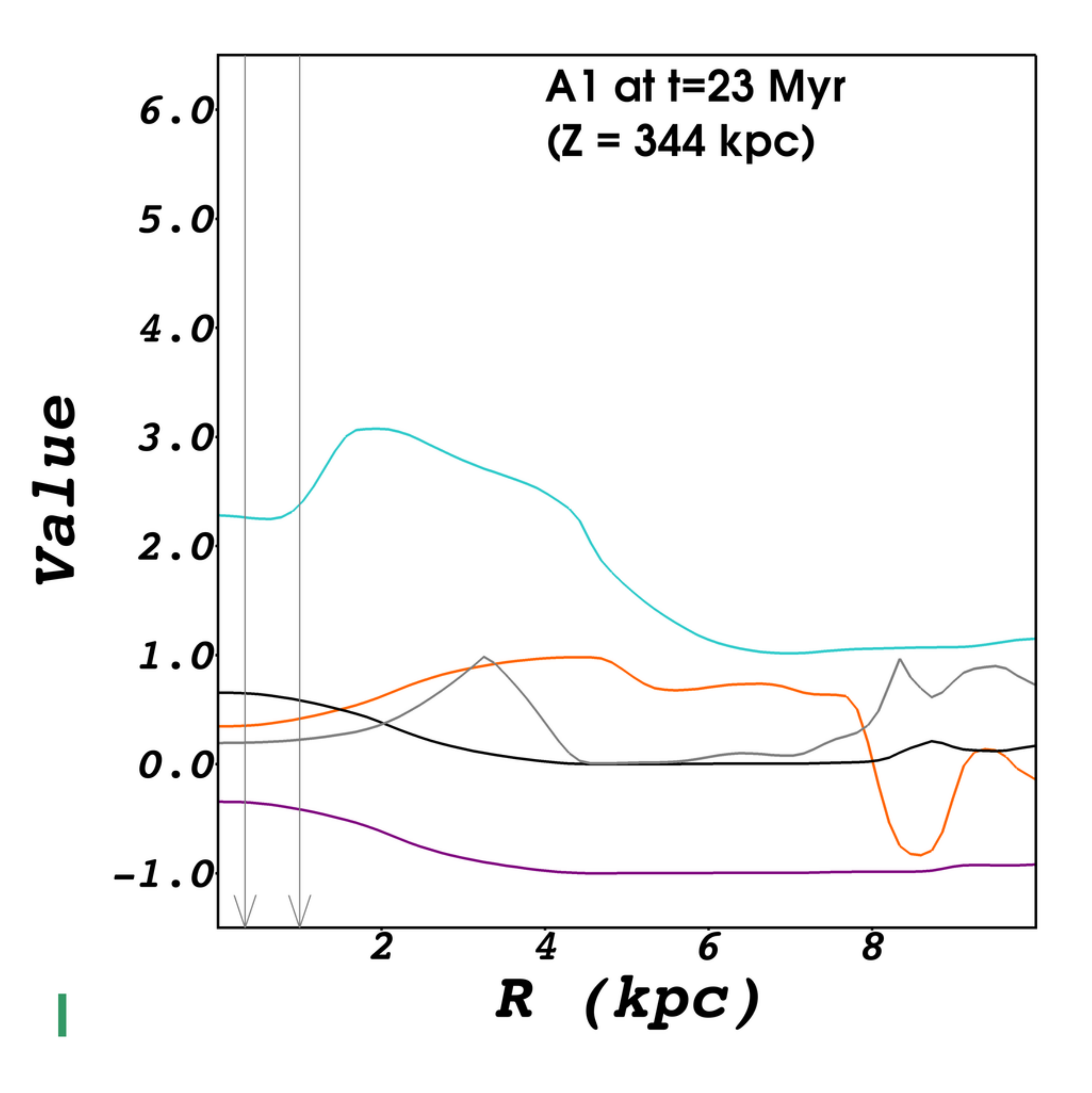}
\end{array}
$
\caption{Radial cuts along the cylindrical radial direction at three different heights, showing Lorentz factor;
tracer values; and the absolute- and mass-weighted mixing factor. The top row ({\bf plots A}, {\bf B} and
{\bf C}) shows radial cuts for model $H1$. The legend shown in figure {\bf B} applies to the top row, except
for the Lorentz factor which applies to all plots. The middle row ({\bf plots D}, {\bf E} and {\bf F}) shows
model $I1$. The legend shown in figure {\bf E} applies to this middle row and the bottom row. The bottom row
({\bf plots G}, {\bf H} and {\bf I}) shows model $A1$. In the first column, a radial cut is made at the jet inlet
at \mbox{$Z = 0$ kpc}. In the second column a radial cut is shown at a distance \mbox{$Z = 200$ kpc} from jet
inlet where all jets ($H1$, $I1$ and $A1$) have crossed three strong internal shocks. In the third column, a
radial cut is shown just below the hot spot (where the effective polytropic index drops below
\mbox{$\Gamma_{\rm eff} \le 1.417$}). The exact location of the hot spot for these three jet models differs. The
vertical arrows in all line plots ({\bf A} through {\bf I}) at \mbox{$R = 1$ kpc} mark the initial jet radius.
The vertical arrows at \mbox{$R \sim 0.3$ kpc} mark the initial jet spine - jet sheath interface.}
  \label{fig:RadialCuts}

\end{figure*}

In this section we will investigate the radial structure of the different jets by looking at {\em radial cuts} at
various heights. \mbox{Figure \ref{fig:RadialCuts}} shows radial cuts of the three individual jet models $H1$
(top row), $I1$ (middle row) and $A1$ (bottom row) at three different heights, all at the final time of
simulation \mbox{$t = 22.8$ Myr}.

The radial cuts cover the region \mbox{$R \le 10$ kpc}. Since the jets have an initial radius of
\mbox{$R_{\rm jt} = 1$ kpc}, these plots show the jet, as well as part of the surrounding cocoon.

The jet axis in figure \ref{fig:RadialCuts} is located at \mbox{$R = 0$ kpc} (left side of the line plots). From
the jet axis moving outwards, the jets are represented by jet material having a Lorentz factor
\mbox{$\gamma > 1$}. The jet boundary occurs at the point where the Lorentz factor becomes \mbox{$\gamma = 1$}.
Beyond the jet boundary lies part of the cocoon, containing a mixture of shocked jet (spine and sheath) and
shocked ambient medium. The transition between pure jet and cocoon is formed by a layer where the jet mixes with
the cocoon. With increasing mixing, we see the bulk Lorentz factor decreasing.

We have chosen to study the radial structure of the individual jets at three different heights, where we
expect different characteristics in the radial direction.

    \subsubsection{The jets at jet inlet}

The left column (figures \ref{fig:RadialCuts} {\bf A}, {\bf D} and {\bf G}) shows a cut at the jet inlet at
\mbox{$Z = 0$ kpc}, where the jets have just been injected into the domain. Here, the jets have only just had
the chance to interact with the cocoon and have not passed any strong shocks. Therefore, the jets are still
completely regular.

Figure \ref{fig:RadialCuts} {\bf A} shows the homogeneous jet. Up to its initial jet radius
\mbox{$R_{\rm jt} = 1$ kpc}, it has Lorentz factor $\gamma_{\rm jt} = 3.11$ and consists of pure jet material
$\theta = + 1$. There is a sharp transition between jet and cocoon at $R_{\rm jt}$, where the absolute
mixing with the cocoon quickly rises and falls off again.

Figures \ref{fig:RadialCuts} {\bf D} and {\bf G} show the isothermal jet and the isochoric jet respectively.
These jets have both been initiated with the same tracer values (pure jet spine material from jet
axis up to \mbox{$R_{\rm sp} = R_{\rm jt}/3$} and pure jet sheath material from \mbox{$R_{\rm sp}$} up to
\mbox{$R_{\rm jt}$}) and Lorentz factors (\mbox{$\gamma_{\rm sp} = 6$} for the jet spine and
\mbox{$\gamma_{\rm sh} = 3$} for the jet sheath) in radial direction. This is clearly seen in the plots.
Moreover, within the jets, there is very little internal mixing between jet spine and jet sheath material.

    \subsubsection{The jets after three strong shocks}

The centre column \mbox{(\ref{fig:RadialCuts} {\bf B}, {\bf E} and {\bf H})} shows a radial cut at
\mbox{$Z = 200$ kpc}. We consider this height because we know (from the previous subsection, where we discussed
the mixing behavior along the jet axis) that all three jets have passed three strong internal shocks at that
height, and that the spine-sheath jets have had the chance to undergo strong internal mixing. This height is
approximately half way from jet-inlet to jet-head.

Figure \ref{fig:RadialCuts} {\bf B} shows the homogeneous jet. We see that the jet has not significantly
changed its structure compared to that at jet inlet. The jet is still completely intact up to its initial jet
radius $R_{\rm jt}$. The main difference is that the transition layer is now wider, so that the jet has
effectively broadened to a radius of $\sim 2 R_{\rm jt}$.

In the isothermal jet (\ref{fig:RadialCuts} {\bf E}) and the isochoric jet ({\ref{fig:RadialCuts} {\bf H}}),
considerable internal mixing between jet spine and jet sheath has occurred. The radial cut of the $I1$ jet
shows that the inner part of the jet is still dominated by jet spine material. However, (as we also saw in the
previous subsection) the inner part of the $A1$ jet is already dominated by jet sheath material. Therefore, the
radial structure of the isothermal jet is less easily disrupted by internal shocks than that of the isochoric jet.

Looking at the radial cuts, moving from jet axis outwards to jet boundary, two forms of mixing occur. Already
at the jet axis, jet spine and jet sheath mix internally. Both jets show a peak where the mass-weighted mixing
becomes large (\mbox{$\Lambda_{\rm sp-sh} \sim 1$}). This mixing results in a smooth radial profile of the
Lorentz factor: The Lorentz factor at the jet axis has decreased compared to its initial value at jet inlet,
resulting in \mbox{$\gamma_{\rm sh} < \gamma < \gamma_{\rm sp}$}. From the axis moving outwards, the Lorentz
factor decreases with a decreasing amount of jet spine material, up to the point where only pure jet sheath
material is present with \mbox{$\gamma \sim \gamma_{\rm sh}$}. Moving out even further, the jet and the cocoon
start to mix. This mixing further decreases the Lorentz factor, up to the jet boundary where $\gamma = 1$.
Therefore, based on Lorentz factor alone, a spine-sheath jet structure will be hard to detect. However, looking
at the abundance of jet spine and jet sheath material, a distinction can still be made. This distinction will be
most prominent for the isothermal jet, where the jet core is still dominated by jet spine material, while the
surrounding layer is dominated by jet sheath material.

At this height, the $I1$ jet has broadened to approximately \mbox{$\sim 2.5 R_{\rm jt}$}. The $A1$ jet, on the
other hand has broadened to approximately \mbox{$\sim 3 R_{\rm jt}$}.

    \subsubsection{The jets just before the Mach disk}

The right column (figures \ref{fig:RadialCuts} {\bf C}, {\bf F} and {\bf I}) shows cuts just before the jets
cross the Mach disk, and therefore just before the hot spot. The height of this point varies between the
individual models. We have chosen to look at the radial structure of the jet at this height, because we are
interested in its behavior just before the jet is terminated at the final shock and because we want to know
if a spine-sheath jet structure can survive all the way up to the jet-head. 

Figure \ref{fig:RadialCuts} {\bf C} shows the radial cut of the homogeneous jet at a height of
\mbox{$Z = 394 $ kpc}. It is striking to see that even though the jet has passed \mbox{$\sim 9$} shocks, the
integrity of the core of the jet is still intact, with pure jet material and a Lorentz factor equal to its
initial value. Moving outward in radial direction, the Lorentz factor decreases (less smoothly than at
\mbox{$Z = 200$ kpc}), up to a radius of approximately $\sim 5 R_{\rm jt}$ which marks the jet boundary. The
strongest mixing between jet and cocoon occurs in a transition layer between 4 and 5 kpc from the jet axis.

Figure \ref{fig:RadialCuts} {\bf F} shows a radial cut of the isothermal jet at a height of \mbox{$Z = 402$ kpc}.
The radius of the jet has broadened to approximately \mbox{$\sim 6 R_{\rm jt}$}. It can be seen that the core of
the jet is still dominated by jet spine material and the outer part still by jet sheath material. Within the
jet, the jet spine and jet sheath have mixed slightly more than at \mbox{$Z = 200$ kpc} and the transition
layers (the regions of strong internal jet mixing, as well as mixing between jet and ambient medium) have
broadened. The internal mixing between jet spine and jet sheath, as well as the mixing between the jet and the
cocoon are reflected in the radial profile of the Lorentz factor: The Lorentz factor decreases with increasing
distance from the jet axis, but as with the homogeneous jet, the decrease occurs less smoothly than at
\mbox{$Z = 200$ kpc}. The distinction between jet spine and jet sheath for the isothermal jet, based on Lorentz
factor was already lost at \mbox{$Z = 200$ kpc}. This has not changed at the current height. However, based on
the abundance of jet spine and jet sheath material, a clear distinction can still be made. In this regard we
conclude that the isothermal spine-sheath jet structure survives all the way up to the jet-head.

Finally, figure \ref{fig:RadialCuts} {\bf I} shows the radial cut of the isochoric jet at \mbox{$Z = 344$ kpc}.
The radius of the jet has broadened to approximately \mbox{$\sim 7 R_{\rm jt}$}, making it the widest jet of the
three models, just before the hot spot. The jet spine and jet sheath material at the jet's centre have internally
mixed slightly more than at \mbox{$Z = 200$ kpc}. The amount of jet spine material very gradually decreases with
increasing distance from the jet axis up to \mbox{$R \sim 4.5$ kpc}, from which point on there is only pure
jet sheath material present that immediately starts to mix with the surrounding cocoon. The Lorentz factor at
this point even behaves rather counter-intuitively. Where one would expect a {\em higher} Lorentz factor at the
jet's centre, the central part now has a {\em lower} Lorentz factor than the surrounding jet sheath. The reason
for this becomes clear when we consider the jet flow at a distance slightly further from the jet-head. The
simulation shows that there is a strong internal shock just before the hot spot at \mbox{$Z \sim 340$ kpc}. As
mentioned before, a shock decelerates the jet flow, and only sufficiently far from the shock is the jet material
able to re-accelerate due to pressure gradients when the jet re-establishes pressure equilibrium. In this case,
the internal shock is so close to the hot spot, that the central part of the jet has not been able to fully
re-accelerate. The outer part of the jet, however, is less strongly shocked by this internal shock. It is
slightly deflected, and the Lorentz factor of the outer part of the jet is not significantly influenced,
resulting in a higher value than its central part.

The spine-sheath jet structure becomes unrecognizable well before one reaches the hot spot: The jet spine and
jet sheath material have almost completely mixed internally with each other. We conclude that the isochoric jet
can not maintain its spine-sheath jet structure up to large distances from the central engine.

  \subsection{Flow properties in the jet-heads}
\label{subsec:FlowProperties}

\begin{figure*}
$
\begin{array}{c c c}
\includegraphics[clip=true,trim=2cm 2cm 1cm 2.5cm,width=0.333\textwidth]{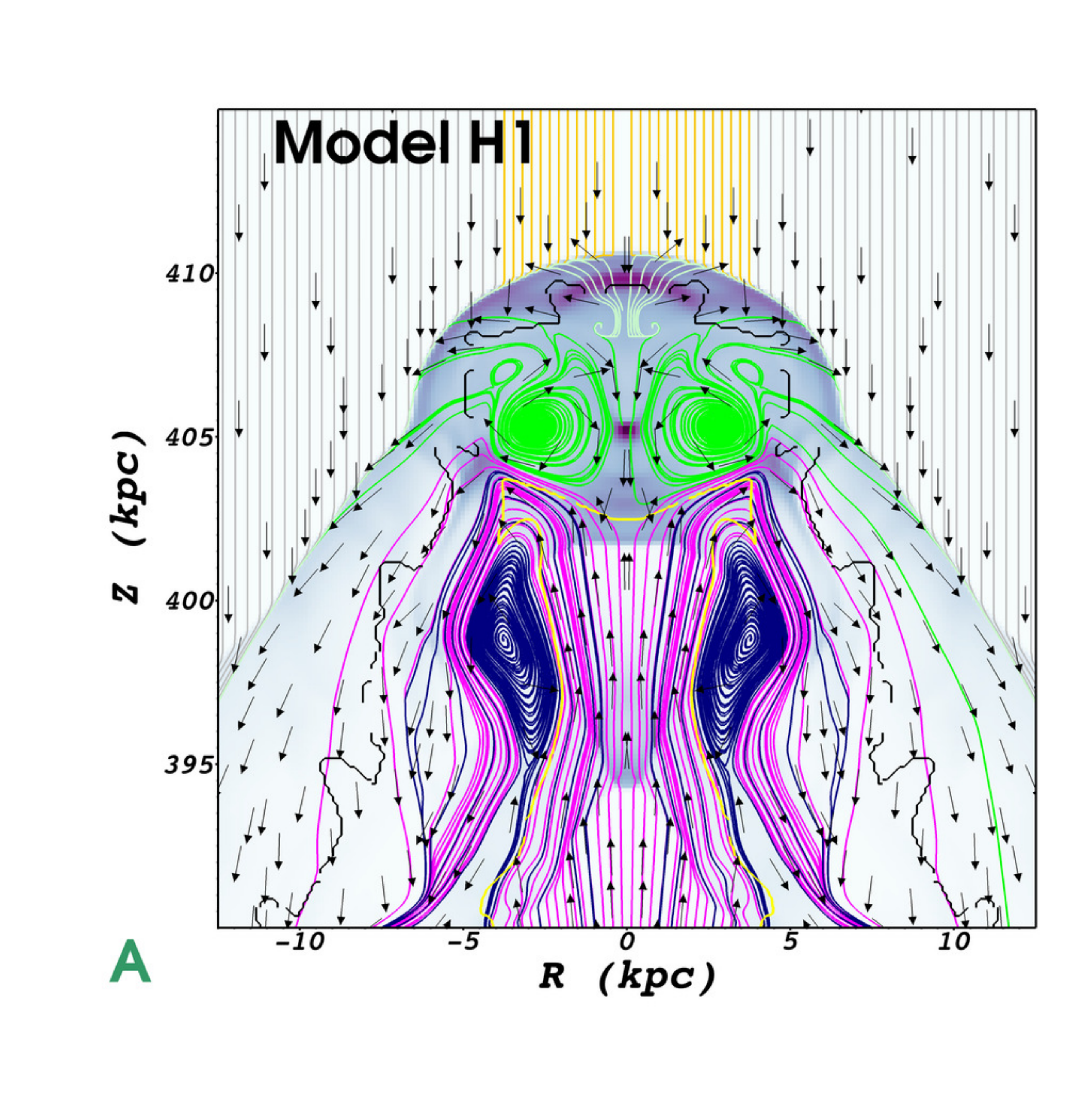} &
\includegraphics[clip=true,trim=2cm 2cm 1cm 2.5cm,width=0.333\textwidth]{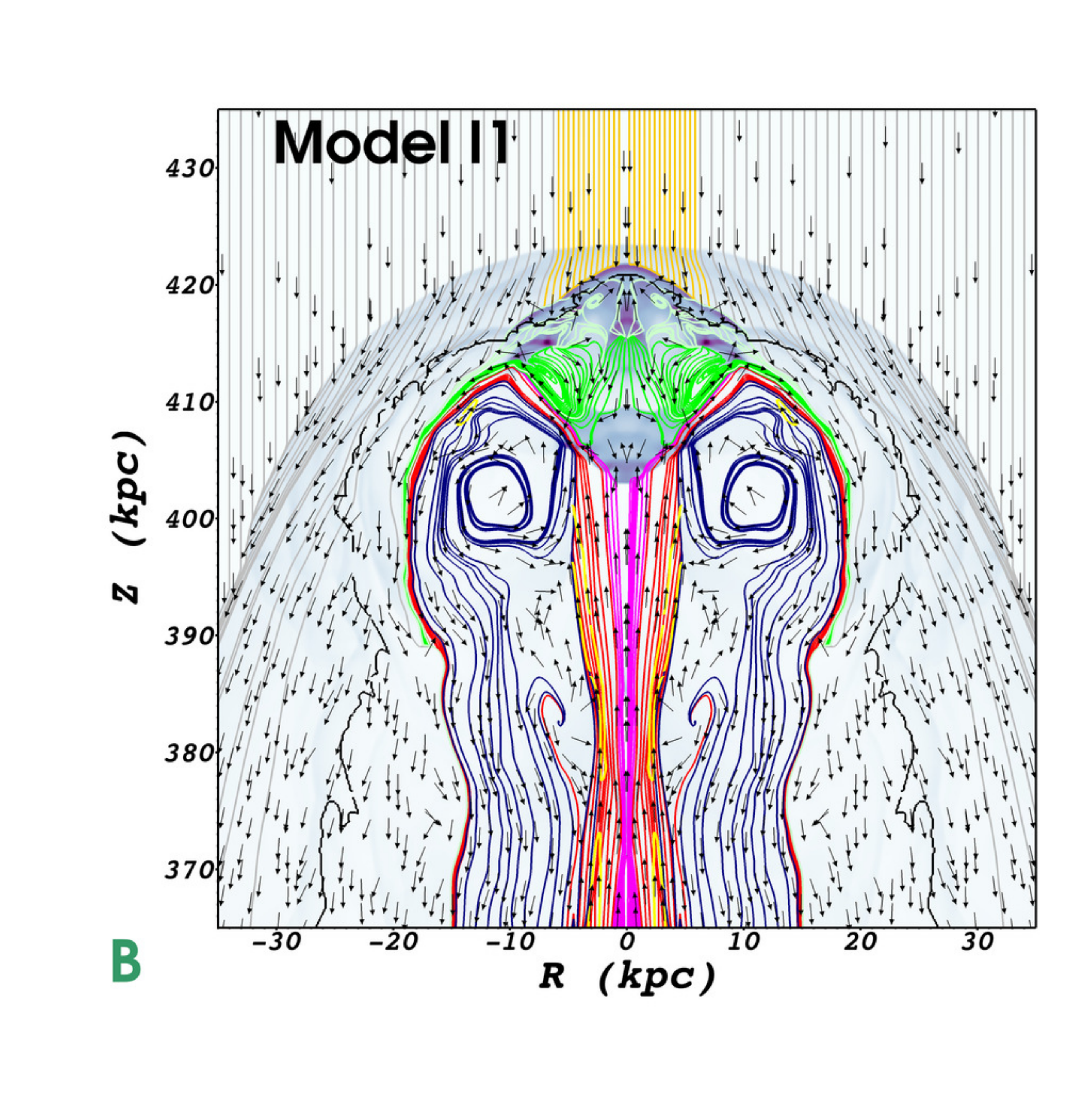} &
\includegraphics[clip=true,trim=2cm 2cm 1cm 2.5cm,width=0.333\textwidth]{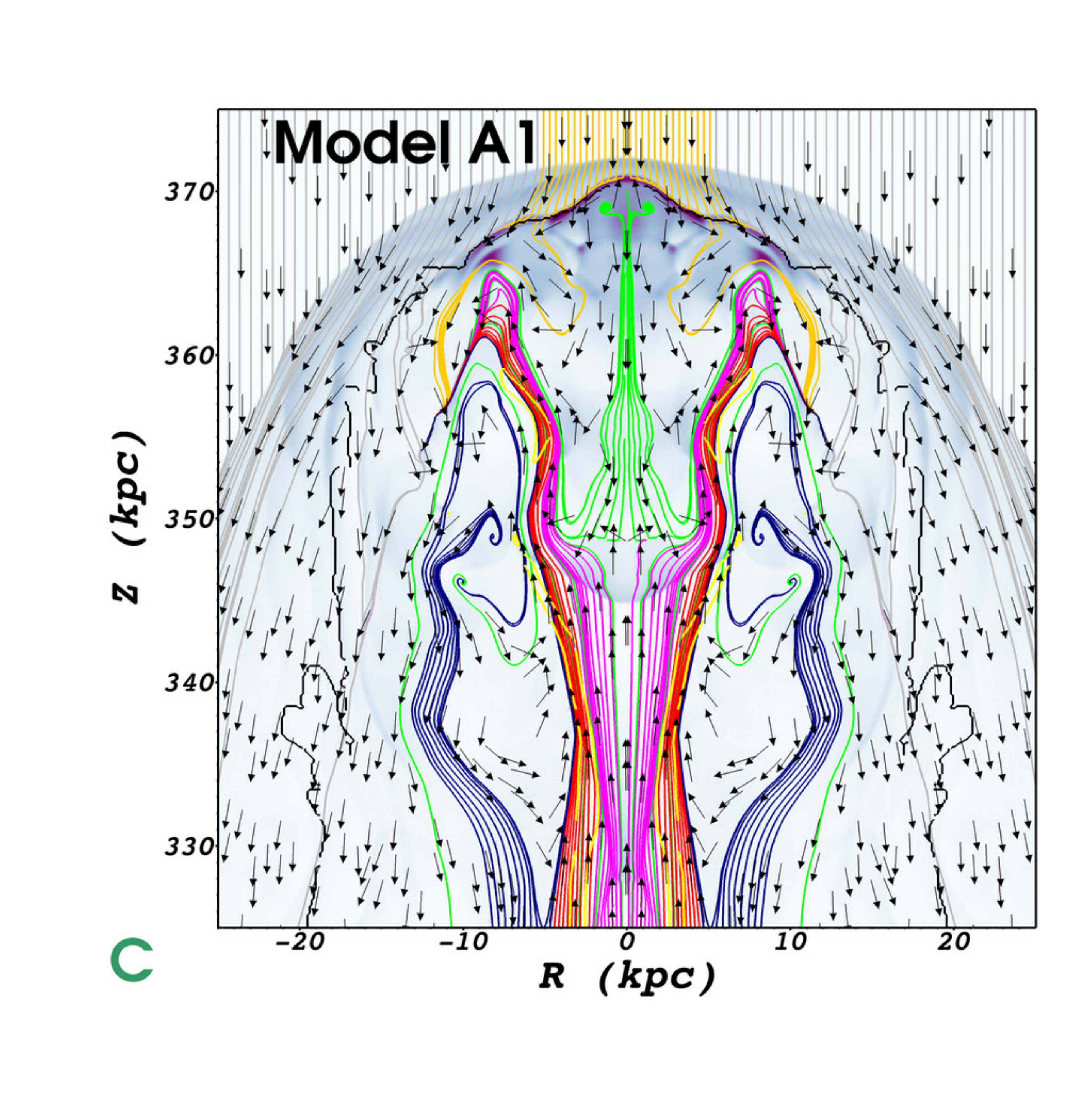}
\end{array}
$
\caption{Close-up of the jet-head structure for the models $H1$, $I1$ and $A1$ at the final time of
simulation, $t = 22.8$ Myr. Plot {\bf A} shows the homogeneous jet, {\bf B} the isothermal jet and {\bf C}
the isochoric jet. Various regions of the flow (in the frame of the advancing Mach disk) have been
depicted by different flow line colors (see section \ref{subsec:FlowProperties} for a full explanation). The main
features in the plots are: the jet flow; the effective impact area of the ambient medium that impacts the jet;
and regions of vortex formation and turbulent mixing.}
  \label{fig:jetheads}

\end{figure*}

In figures \ref{fig:jetheads} ({\bf A, B} and {\bf C}), close-ups of the jet-heads are shown at the final
time of simulation, $t = 22.8$ Myr. The blue color scale on the background represents the pressure, which
clearly shows the bow shock of the jet-head and the Mach disk where the jet flow is terminated.

Velocity unit vectors are also drawn in the rest-frame of the advancing Mach disk. They show the direction of the
undisturbed ambient medium flowing into the cocoon from the top of the plots, the direction of the inflowing jet
material from the bottom of the plot and the direction of the flows within the cocoon itself. As expected, the
velocity goes to zero at the Mach disk.

Flow lines (also as measured in the rest frame of the Mach disk) of various different regions marked
with different colors are also shown.
\begin{description}

\item {\em Gray} and {\em dark yellow} lines that start at the top mark the inflowing undisturbed ambient medium.
Here, the dark yellow lines mark the effective impact area, the values of which are calculated in
\mbox{section \ref{subsec:EffImpactArea})}.

\item {\em Green} flow lines (just below the bow shock) represent the shocked jet material and shocked
ambient medium material. These flow lines show the exact behavior of the mixing of jet material and
shocked ambient medium material at the top of the jet-head.

\item {\em Pink} and {\em red} flow lines (starting at the bottom centre) represent the jet flow. For the
homogeneous jet $H1$ in figure \mbox{\ref{fig:jetheads} {\bf A}}, the entire jet flow is represented by pink lines.
For the structured jets $I1$ and $A1$ in figures \mbox{\ref{fig:jetheads} {\bf B}} and
\mbox{\ref{fig:jetheads} {\bf C}}, the red lines mark the part of the jet that contains the maximum mass fraction
of jet sheath material across the cross section of the jet \mbox{(which is on the order of $\sim 90\% - 100\%$)}.
The pink lines mark the inner part on the jet flow where jet spine material is present. This region consists of a
mixture of jet spine and jet sheath material.

\item {Dark blue} lines mark the region of back-flowing jet material that lies between the jet and the bulk
of the back-flowing jet material. The morphology of these flow lines for the three models varies significantly
from one model to the next, but in all three cases vortices are clearly seen.
\end{description}

Finally, two line contours are shown. The first contour is the black line within the bow shock. This contour
encloses the region where jet material is found (which is derived from the presence of jet tracer material)
\footnote{The intermittency of the black contour is a result of the finite resolution of the simulations.
Regardless, it still marks the boundary of the jet material containing region to good approximation.}.
The second is a yellow contour that resides within the jet flow. It marks the boundary of the jet flow, which we
have taken to contain the largest mass fraction of jet material within the grid cells (which is
\mbox{$\sim 90\% - 100\%$}).

\vspace{0.5 cm}
We see that for the homogeneous jet $H1$ the pressure gradient at the top of the jet-head in plot
\ref{fig:jetheads} {\bf A} is so large that the flow lines are strongly deflected outward along the bow shock.
However, for the two jet-heads of the spine-sheath jets $I1$ and $A1$ in plots {\bf B} and {\bf C}, the
pressure gradients at the bow shock are less strong, allowing the flow lines to penetrate the cocoon before
they are deflected outwards. This can be explained by considering the Mach disk for the jet-heads of $H1$,
$I1$ and $A1$. The Mach disk of the homogeneous jet $H1$ has a much larger surface area than the Mach disk
of the structured jets $I1$ and $A1$. For the homogeneous jet, the entire jet flow is shocked by the Mach disk,
but in case of the structured jets only the inner part of the jet flow is shocked by the Mach disk. This results
in the fact that the pressure downstream of the Mach disk for the homogeneous jet becomes significantly
higher than in case of the structured jets. The jet sheath material (the red flow lines) is able to propagate
further into the jet-head structure than the jet spine material. When this jet sheath material then eventually
encounter regions with high enough pressure, it is deflected and flows away from the jet-head, back into the
cocoon.

By considering the bulk of back-flowing shocked jet material, we see that the black contour (enclosing the jet
material) does not coincide very well with these (outermost) flow lines, but mostly lies outside this region.
In the ideal scenario where no instabilities would develop at the jet-head, there would be a clear sudden
transition between shocked jet material and shocked ambient medium material, separated by the contact
discontinuity. In that case, the black contour would exactly correspond to the contact discontinuity. However,
in the realistic case with instabilities, there will be a layer with a certain thickness instead of a contact
discontinuity. The surface enclosing the bulk of back-flowing jet material corresponds to what we call the
contact discontinuity. The layer surrounding that contact discontinuity can best be considered as shocked ambient
medium material 'contaminated' with shocked jet material.

  \subsection{Large-scale behavior of jets and cocoons}\label{subsec:LargeScale}


\begin{figure*}
$
\begin{array}{c c c}
\includegraphics[clip=true,trim=2.5cm 1.5cm 1cm 1cm,width=0.333\textwidth]{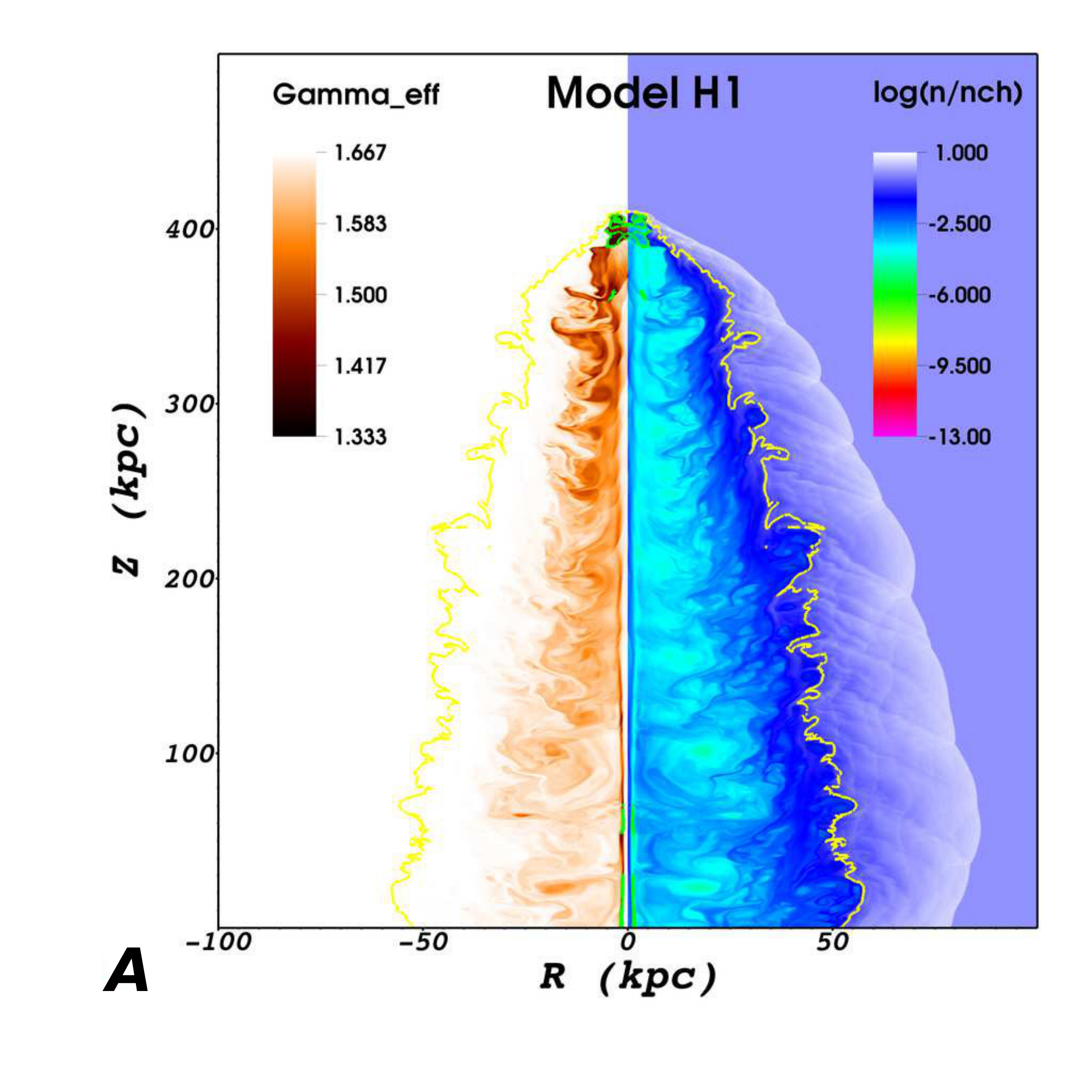} &
\includegraphics[clip=true,trim=2.5cm 1.5cm 1cm 1cm,width=0.333\textwidth]{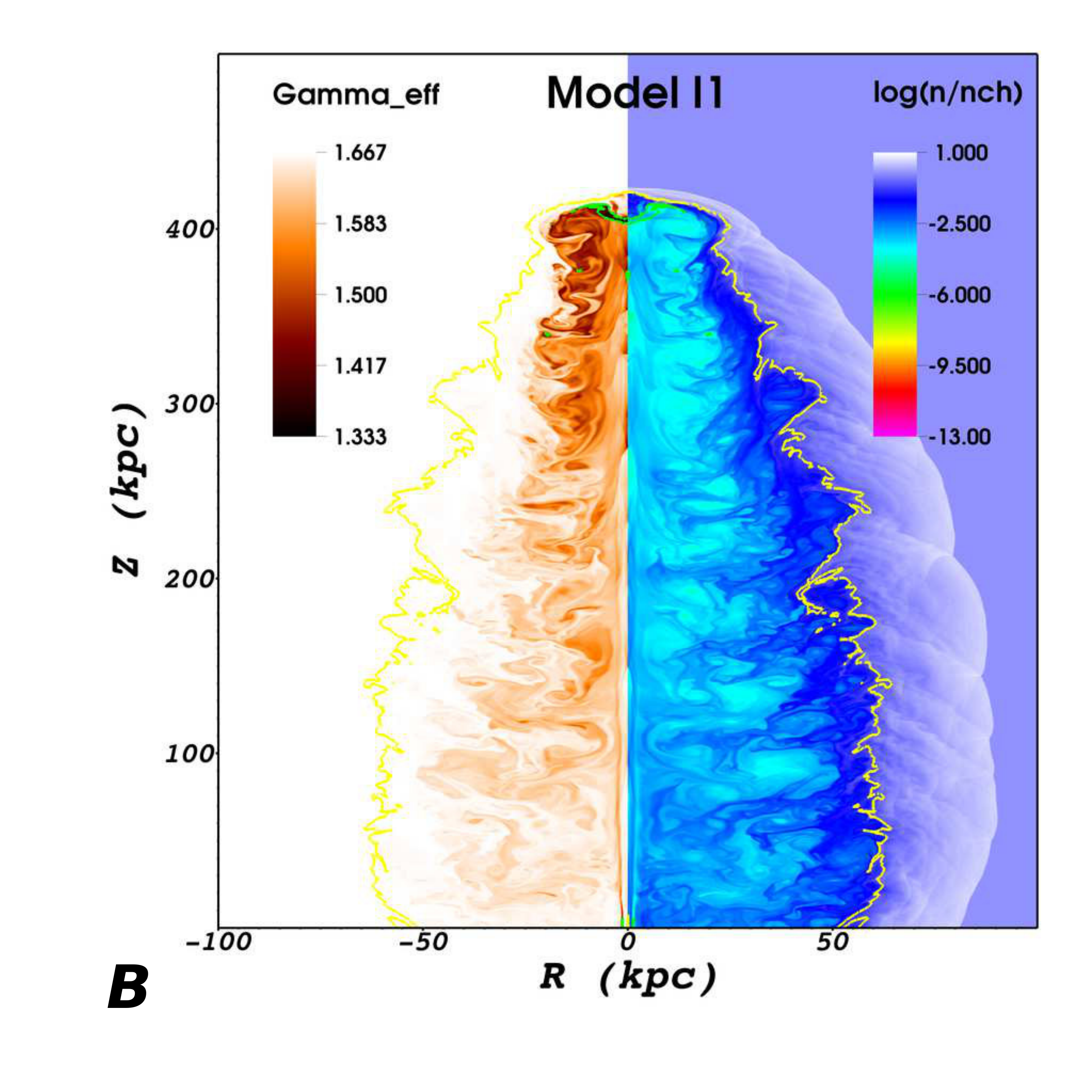} &
\includegraphics[clip=true,trim=2.5cm 1.5cm 1cm 1cm,width=0.333\textwidth]{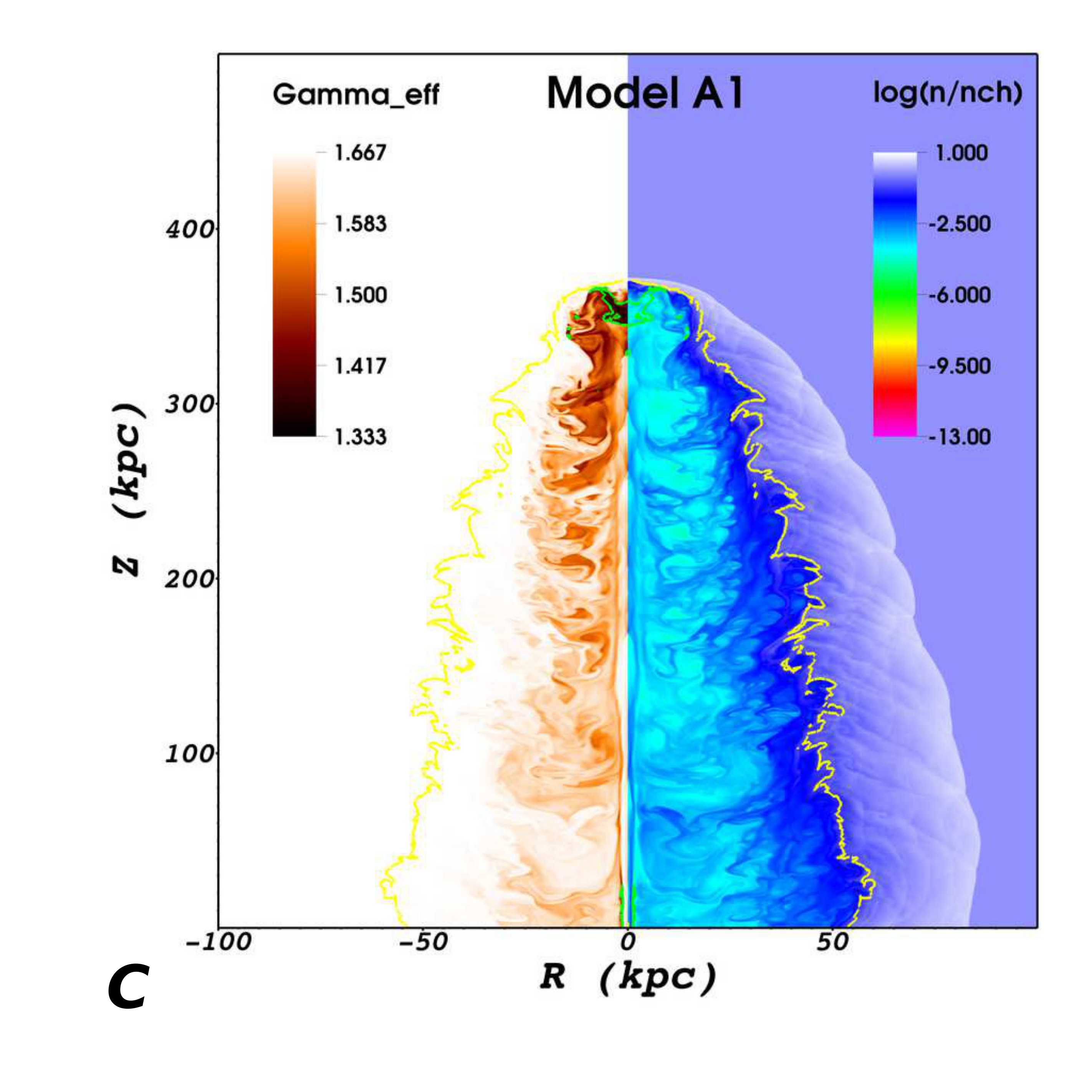} \\
\includegraphics[clip=true,trim=2.5cm 1.5cm 1cm 1cm,width=0.333\textwidth]{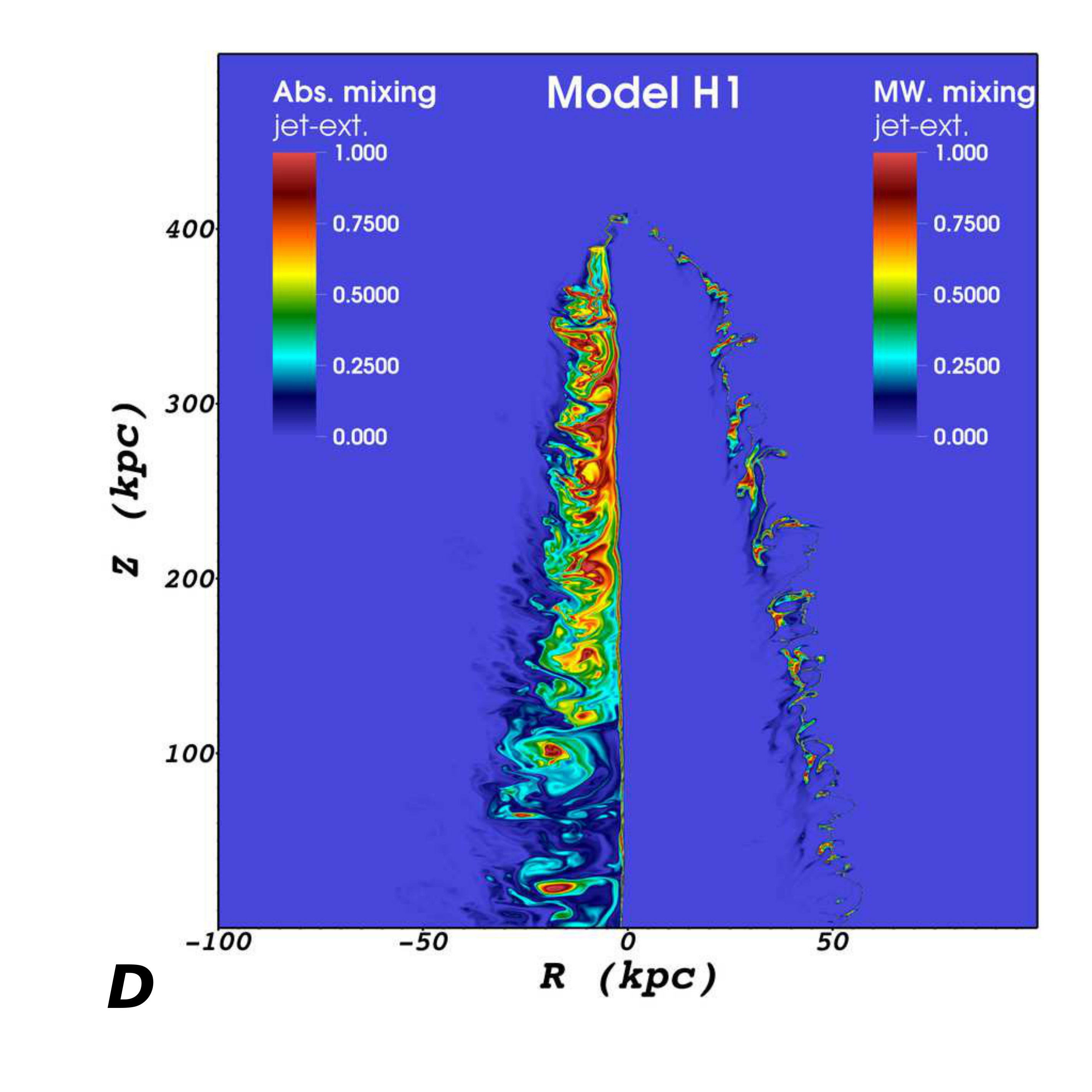} &
\includegraphics[clip=true,trim=2.5cm 1.5cm 1cm 1cm,width=0.333\textwidth]{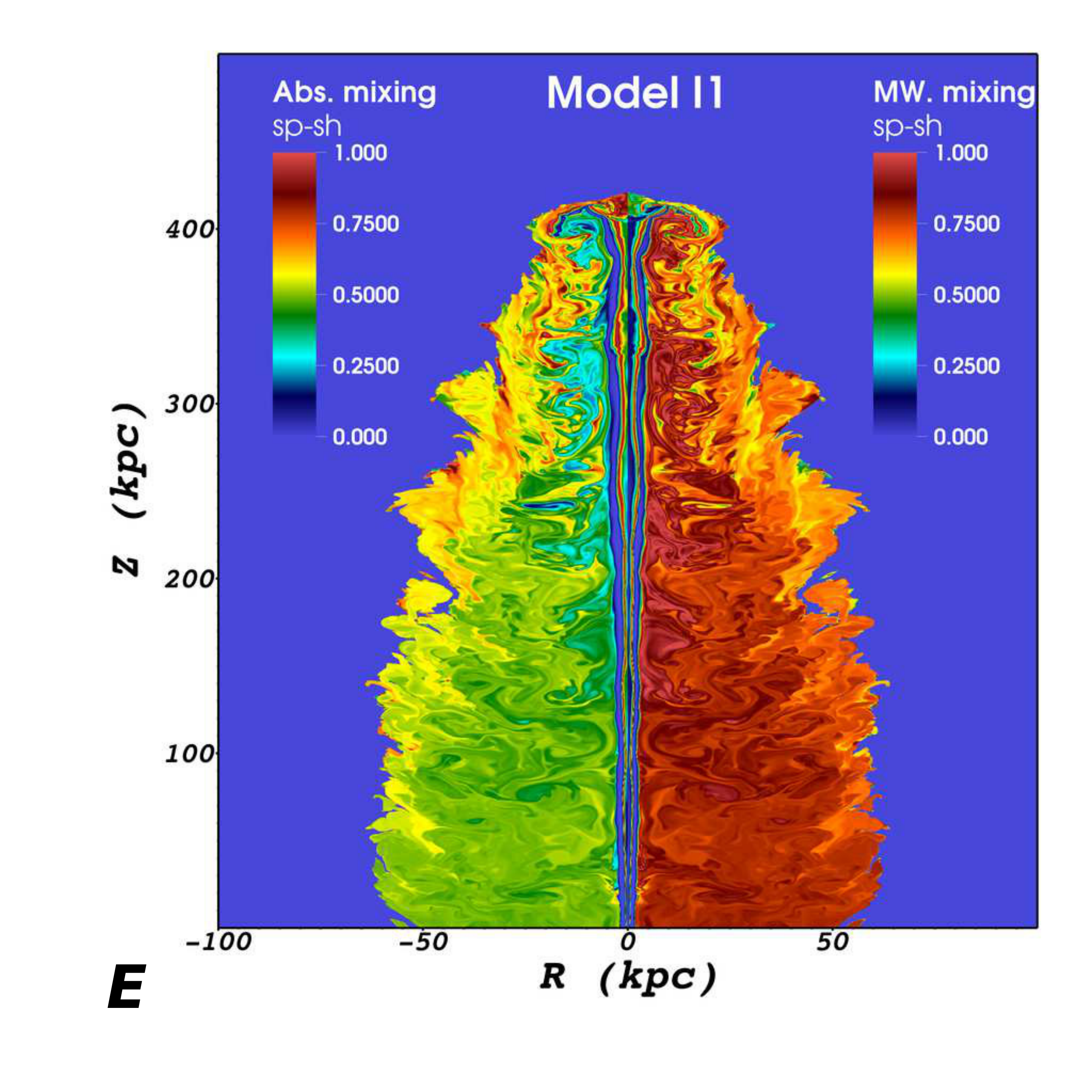} &
\includegraphics[clip=true,trim=2.5cm 1.5cm 1cm 1cm,width=0.333\textwidth]{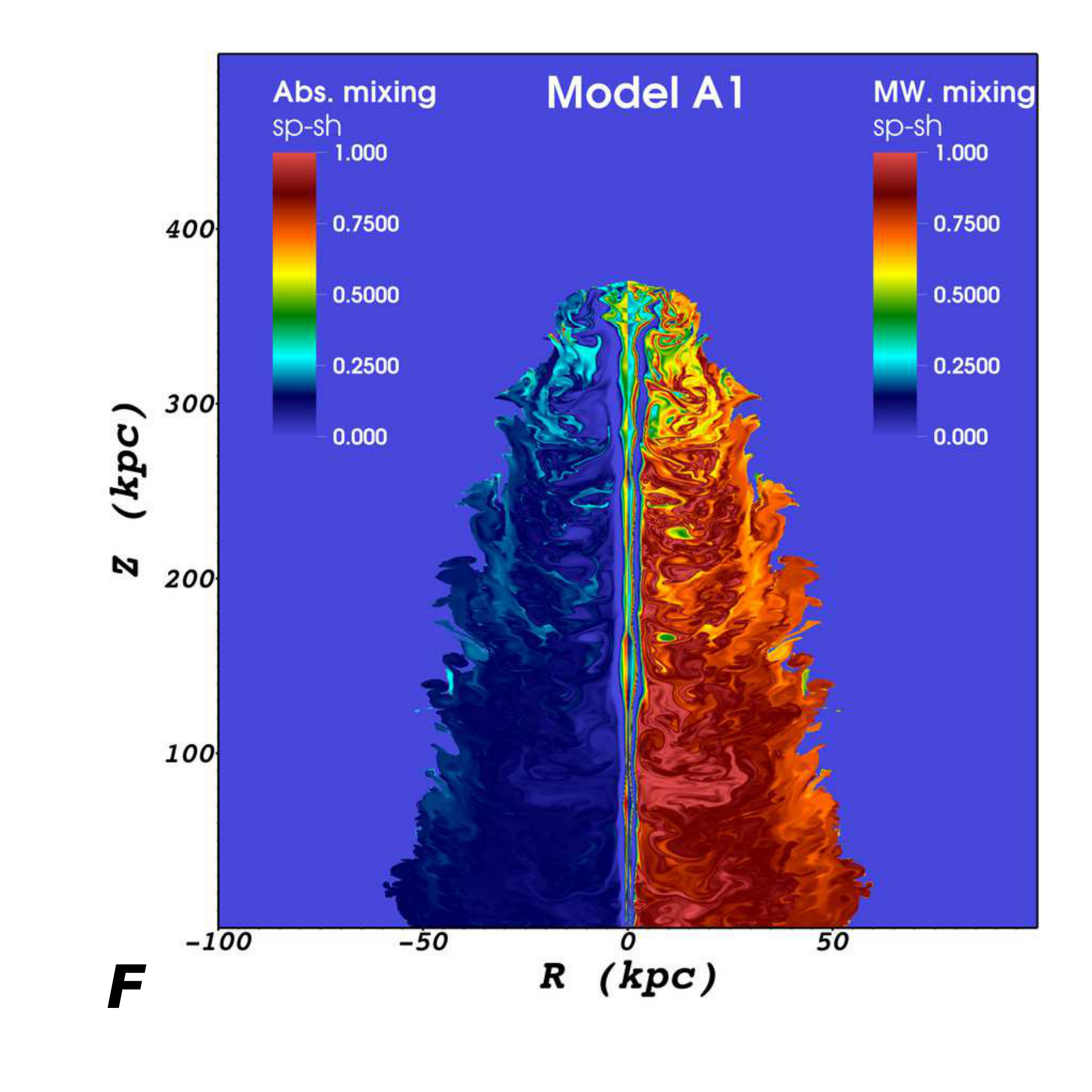}
\end{array}
$
\caption{Contour plots of the homogeneous jet ($H1$, panels on the left hand side), the isothermal jet
($I1$, centre panels) and the isochoric jet ($A1$, panels on the right hand side) after 22.8 Myr.
Note: the $R-$axis has been stretched by a factor of 2.5 in order to enhance the visibility of the
actual jet behavior.
\newline
{\bf Contour plots (A, B and C)} show the effective polytropic index $\Gamma_{\rm eff}$ (left hand side of the
plots) and the $\log_{10}$ of the number density in terms of the characteristic number density
\mbox{$n_{\rm ch} = 10^{-3} {\rm cm^{-3}}$} (right hand side of the plots). The yellow contour marks the
boundary where jet material (shocked, as well as unshocked) is found. The green contour (mainly found at
the jet-head) encloses relativistic gas with \mbox{$\Gamma_{\rm eff} \le 1.417$}.
\newline
{\bf Contour plots (D, E and F)} show the absolute mixing ($\Delta$) on the left side of the plots and
the mass-weighted mixing ($\Lambda$) on the right side of plots. For the homogeneous jet ($H1$), the
mixing between (shocked) jet material and shocked ambient medium (denoted as "ext") is shown, as described by
equations \equref{eq:Delta} and \equref{eq:Lambda}. For the isothermal jet ($I1$) and the isochoric jet ($A1$),
absolute mixing $\Delta_{\rm sp-sh}$ and mass-weighted  mixing $\Lambda_{\rm sp-sh}$ between (shocked,
as well as unshocked) spine material and sheath material is shown, as described by equations
\equref{eq:DeltaAB} and \equref{eq:LambdaAB}.
}

  \label{fig:ContourPlots}
\end{figure*}

In the next sections, we will focus on the behavior of material inside the cocoons. We will discuss properties
such  as cocoon shape, the distribution of densities and relativistic gas, mixing effects of shocked jet (spine
and sheath) material and shocked ambient medium. Finally, we will link these large-scale cocoon properties to
the properties that occur within the jets, the so-called effect of cocoon-jet coupling.

Figure \ref{fig:ContourPlots} shows the contour plots of the jets and cocoons as they have developed after the
full time of simulation $t=22.8$ Myr. The top panel shows the distribution of number density ($n/n_{\rm ch}$) and
relativistic gas ($\Gamma_{\rm eff}$), while the lower panel shows various forms of mixing. In all these contour
plots, the $R-$axis has been stretched by a factor of 2.5 in order to enhance the visibility of the jets. The
yellow contour marks the interface between the region consisting of purely shocked ambient medium and the
region that contains a mixture of shocked jet material and shocked ambient medium. The green contour (mainly
found at the jet-head) encloses relativistically hot regions \mbox{($\Gamma_{\rm eff} \le 1.417$)}. A summary
with the most notable characteristics of each individual model can be found in
\mbox{table \ref{tab:H1I1A1Characterisics}}.

  \subsection{Global morphology of the cocoons}\label{subsec:GlobalMorphology}

After 22.8 Myr, the average distance traveled by the jets is $Z = 400$ kpc. Individual models differ from this
mean value by a few tens of kpc. The isochoric jet $A1$ has traveled the shortest distance (\mbox{370 kpc});
the homogeneous jet $H1$ comes next (\mbox{409 kpc}); and the isothermal jet $I1$ has traveled the largest distance
(\mbox{421 kpc}). The averaged maximum width of the cocoons is \mbox{$R \sim 87$ kpc} at an average distance of
\mbox{$\sim 343$ kpc} from the jet-head.

All cocoons have a quasi-parabolic shape, from the jet-head down to the point where the cocoon reaches a maximum
width. At larger distances from the jet-head (closer to the jet inlet), the width of the cocoon slightly
decreases. The simulations show that the pressure in the cocoon near the jet-head is somewhat higher than the
pressure at larger distances from the jet-head. The associated pressure gradient causes the gas within the cocoon
to accelerate in the direction away from the jet-head. However, at large distances from the jet head, this effect
is no longer seen. If we had chosen reflective boundary conditions, instead of free outflow boundary conditions,
then this would have lead to a pile up of cocoon material near the lower boundary, leading to a broader cocoon
base. Therefore, we note that the decrease in cocoon width near the jet inlet might just be a boundary effect.

Overall, the isothermal jet has the broadest cocoon. The heads of the jets show clear differences between
the individual models: The jet-heads of the isothermal and isochoric jets have a wide and round shape (see
contour plots \ref{fig:ContourPlots} ({\bf B} and {\bf C}) and for a blow-up of the jet-head
\ref{fig:jetheads} ({\bf B} and {\bf C}). The homogeneous jet, on the other hand,
shows a sharply peaked jet-head \mbox{(see \ref{fig:ContourPlots} {\bf A} and \ref{fig:jetheads} {\bf A} ).}

  \subsection{Density distribution within the cocoons}\label{subsec:DensityDistribution}

The cocoons can be divided into three regions: The first region is the outermost part of the cocoon, which we
denote by $R_1$. This region consists of pure shocked ambient medium and is found between the outer edge
(bow-shock) of the cocoon and the yellow contour that encloses jet material. The other two regions $R_2$ and
$R_3$ both contain shocked jet material. We denote the inner part of the cocoon by $R_3$, the part where most of
the shocked jet material is flowing down the cocoon. Finally, there is a transition layer between the
regions $R_1$ and $R_3$, which we denote by $R_2$.

The number density in the region $R_{1}$ is typically on the order of \mbox{$n \sim 4 - 5 \; n_{\rm ch}$}
(where we initiated \mbox{$n_{\rm am} = n_{\rm ch} = 10^{-3} \; \rmn{cm^{-3}}$}). The compression ratio,
which can be calculated from shock conditions in the case of a strongly shocked non-relativistic gas
\mbox{($\Gamma = 5/3$)} takes the value $r=\frac{\Gamma+1}{\Gamma-1} = 4$. In case of a relativistically hot gas
\mbox{($\Gamma = 4/3$)}, the compression ratio will be $r = 7$
\footnote{The simulations show that the cocoons expand with sub-relativistic velocities, which allow us to
consider these expressions for the compression ratio. In case of shocks propagating at relativistic speeds,
the compression ratio will also depend on the Lorentz factor (see for instance \citealt{Eerten2010}).}.
Since we are dealing with a polytropic index, based on a Synge-like equation of state which interpolates between
these two values, and the fact that this shocked ambient medium has a polytropic index slightly less than 5/3,
a compression ratio slightly higher than 4 is expected.

In the region $R_{2}$, the shocked jet material has undergone strong mass-weighted mixing with the shocked
ambient medium. The typical number density in this region is on the order of \mbox{$n \sim 10^{-2}\; n_{\rm ch}$}.
As a practical definition we define region $R_2$ by the condition \mbox{$\Lambda \ge 0.01$}, where $\Lambda$ is
the mass-weighted mixing between (shocked) jet material and shocked ambient medium, defined in equation
\equref{eq:Lambda}.

The innermost region $R_{3}$ mostly consists of shocked jet material and contains a small amount of shocked
ambient medium material. This region $R_3$ contains material with the lowest density within the cocoon which is
typically \mbox{$\sim 10^{-3} \; n_{\rm ch}$}. We define this region by the condition \mbox{$\Lambda < 0.01$},
but the values tend to become very small (typically $\Lambda \le 10^{-4}$).

  \subsection{Distribution of relativistic gas}\label{subsec:DistributionRelGas}

In all three models, the effective polytropic index tends to obtain its lowest values (and therefore becomes more
relativistically hot) at the innermost part of the cocoon, region $R_3$, where the mass densities are low. Most
relativistic gas is found near the jet-head, where the jet material has just gone through the Mach disk. It
gradually becomes less relativistic with increasing distance from the jet-head. This is due to the expansion and
associated cooling of the material.

To distinguish those regions that are relativistically hot, we look for gas with $\Gamma_{\rm eff} \le 1.417$
(the green contour in the contour plots \ref{fig:ContourPlots} {\bf A, B}, and {\bf C}). As can be seen in these
contour plots, there are two distinct regions that are relativistically hot. The first is a very thin region
found at the interface between jet and cocoon, near the jet inlet and might be caused by spurious effects
in the lower boundary cells, next to the jet inlet. The second region containing relativistic gas is at the
jet-head where the jet flow is terminated and goes through the Mach disk. This is also seen in line plots
\ref{fig:geffTemp} and \ref{fig:CrossCuts} ({\bf A, B} and {\bf C}), where the gas becomes relativistically hot
when the temperature rises up to $T \sim T_{\rm ch}$. This relativistically hot region corresponds to the hot
spot of the jet. The hot spot of the $H1$ jet has an elongated shape, whereas the $I1$ and $A1$ jets show a
concave (bowl) shape that has a dip near the jet's centre. In \mbox{section \ref{subsec:JetHeadandMixing}}, we
will discuss the morphology of the $U$-shaped hot spot in more detail.

  \subsection{Mixing effects in the cocoons}\label{subsec:MixingCocoon}

Since we are dealing with a structureless homogeneous jet, as well as with jets with a spine-sheath jet structure,
we can consider different kinds of mixing effects. The $H1$ jet only consists of a single jet constituent, so
we shall consider the mixing between this (shocked, as well as unshocked) jet material and the shocked ambient
medium. The $I1$ and $A1$ jets have a spine-sheath jet structure. There, we will be concerned with the internal
mixing between jet spine and jet sheath material within the jet itself, as well as the mixing between shocked 
(spine and sheath) jet material in the surrounding cocoon. We shall consider each individual model separately.

    \subsubsection{Mixing in the cocoon for the homogeneous jet model $H1$}

Contour plot \ref{fig:ContourPlots} {\bf D} shows the absolute and mass-weighted mixing factors $\Delta$ and
$\Lambda$ between jet material and shocked ambient medium. A very thin layer of strongly mixed material resides
along the jet's boundary and persists all the way up to the jet's head.

Moreover, strong absolute mixing is found in the inner region of the cocoon where the densities are low
\mbox{($n \sim 10^{-3} \; n_{\rm ch}$)}. In these regions, the mass fractions of shocked ambient medium
material and jet material are nearly equal. However, outside the yellow contour marking the boundary between
shocked jet and shocked ambient, densities of the shocked ambient medium are on the order of
\mbox{$\sim 4 \; n_{\rm ch}$}. Therefore, regions that contain strongly mixed material only contain a tiny
fraction of the available amount of shocked ambient medium \mbox{($\sim 0.1 - 1\%$)}. This is also reflected in
the behavior of the mass-weighted mixing. It shows that virtually no mass-weighted mixing takes place in the
inner region of the cocoon, \mbox{$\Lambda \le 10^{-4}$}. Only very thin filaments of (mass-weighted) well-mixed
material are found adjacent to the yellow contour, where the two constituents are in close contact.

    \subsubsection{Mixing in the cocoons for the spine-sheath jet models}

Contour plots \ref{fig:ContourPlots} ({\bf E} and {\bf F}) show the mixing factors $\Delta_{\rm sp-sh}$ and
$\Lambda_{\rm sp-sh}$ between spine and sheath material as it occurs within the jet, as well as in the
surrounding cocoon. All mixing takes place within the yellow contour shown in contour plot in figures
\ref{fig:ContourPlots} {\bf B} and \ref{fig:ContourPlots} {\bf C} respectively. As soon as the jet material
has crossed the Mach disk, the shocked spine and sheath material have already mixed by a considerable amount.
This shocked spine and shocked sheath material continues to mix further as one moves back away from the jet-head
due to the effect of vortices and turbulent flows.

    \subsubsection{Shocked spine and sheath mixing: the isothermal jet $I1$}

For the $I1$ jet, the absolute mixing factor $\Delta_{\rm sp-sh}$ in the cocoon is strongest near the jet-head,
where absolute mixing $\Delta_{\rm sp-sh} \sim 1$. Here, the largest variations in mixing occur, and with
increasing distance from the jet-head, the absolute mixing saturates towards a fairly homogeneous state of
$\Delta_{\rm sp-sh} \sim 0.5$ at the lower regions of the cocoon.

The mass-weighted mixing $\Lambda_{\rm sp-sh}$ for the $I1$ jet, on the other hand, shows that near the jet's
head, the shocked spine and shocked sheath material in the cocoon have mixed to some extent, but the amount
of mass-weighted mixing varies largely near the vortices. The mixing increases with increasing distance from the
jet-head. Strong vortices are found close to the jet axis, where the mass-weighted mixing approaches
\mbox{$\Lambda_{\rm sp-sh} \longrightarrow 1$}. Most vortices have dissolved at distances larger than 291 kpc
from the jet-head. There, the mixture has become fairy homogeneous with mass-weighted mixing approaching
\mbox{$\Lambda_{\rm sp-sh} \sim 0.8 - 1$}.

    \subsubsection{Shocked spine and sheath mixing: the isochoric jet $A1$}

In contrast to the isothermal jet, the absolute mixing in the cocoon of the isochoric $A1$ jet is fairly weak
(see contour plot \ref{fig:ContourPlots} {\bf F}). Even at the jet-head, the absolute mixing hardly exceeds
\mbox{$\Delta_{\rm sp-sh} = 0.5$}. As soon as shocked jet material starts to flow away from the jet-head,
the absolute mixing quickly drops down to $\Delta_{\rm sp-sh} \sim 0.1$ and becomes a fairly homogeneous
mixture at distances larger than 100 kpc from the jet-head. The main reason that the absolute mixing for the
$A1$ jet is weaker than the absolute mixing for the $I1$ jet is the fact that the density contrast between
jet spine and jet sheath for the $A1$ jet is larger than that of the $I1$ jet.

The fact that absolute mixing does not always give the correct intuitive sense for the amount of homogeneity is
well reflected in the right panel of contour plot \ref{fig:ContourPlots} {\bf F}, where the mass-weighted
mixing shows a completely different picture of the amount of mixing. Material that crosses the Mach disk is
weakly mixed, but as the jet material flows back into the cocoon, filaments of strongly mixed material
($\Lambda_{\rm sp-sh} \sim 1$) are carried along by vortices and turbulent flows. These regions of strong
mass-weighted mixing increase in size and moving down the cocoon, the jet material saturates to a fairly
homogeneous mixture, attaining values of \mbox{$\Lambda_{\rm sp-sh} \sim$ 0.9 - 1}.

  \subsection{Cocoon-jet coupling}
\label{subsec:CocoonJetCoupling}

\begin{figure}
\includegraphics[clip=true,trim=2cm 2cm 1cm 2.5cm,width=0.5\textwidth]{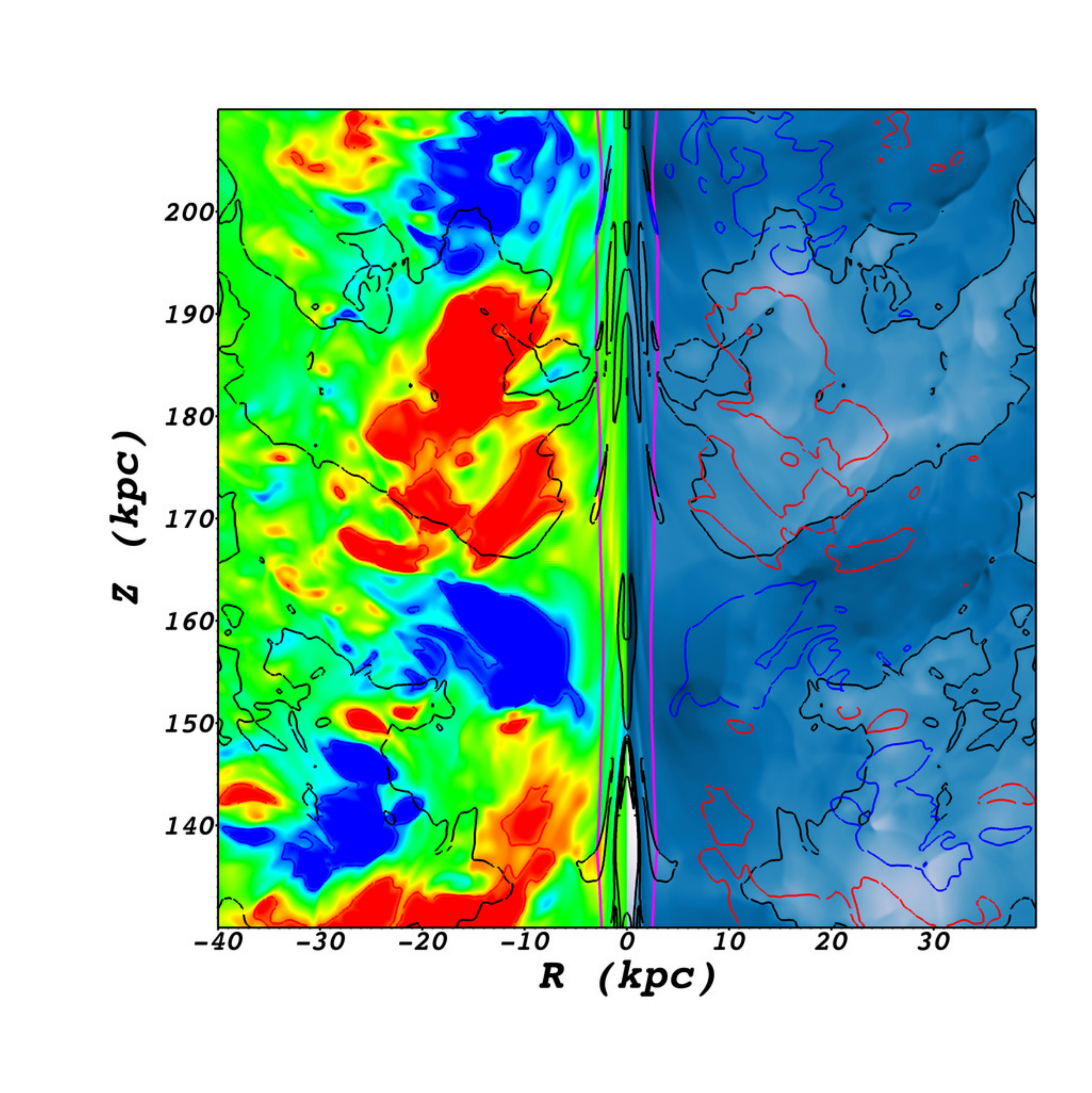}

\caption{A close up of the isothermal ($I1$) jet in direct contact with the surrounding cocoon. It shows the
connection between internal shocks within the jet (shown by the black contours along the jet centre) and vortices
in the cocoon. The pink contour from the centre bottom to the centre top of the plot marks the boundary of the
jet. {\bf Left panel}: The red-green-blue color scale denotes the
velocity of gas in the {\em radial direction}. The {\em red} regions mark gas that is moving away from the jet
and {\em blue} regions mark gas that is moving towards the jet. The red and blue contours (left and right panel)
denote material moving with $V_{\rm R} = \pm 0.08 c$ respectively.
{\bf Right panel:} The gray-blue color scale denotes the ($\log_{10}$ of the) gas pressure of the cocoon and jet
material. Darker regions correspond to higher pressure. High pressure regions compress the jet and cause
internal shocks to emerge. These high pressure regions correlate with gas moving towards the jet, whereas low
pressure regions correlate with gas moving away from the jet.}
  \label{fig:Cocoon-jet-coupling}
\end{figure}

At the jet-head, where the jet impacts the ambient medium, a complex structure of flow lines forms (see
figures \mbox{\ref{fig:jetheads}} {\bf A, B} and {\bf C} ). When the jet flow is terminated at the Mach disk,
and material is deflected away from the jet-head into the cocoon, vortices will form. These vortices are able
to reside at the jet-head for a certain amount of time, but eventually they will break off the jet-head
structure and move down the cocoon, entrained by mainly shocked back-flowing jet material. After a vortex is shed
by the jet-head, a new vortex starts to form and the cycle repeats itself.

The shed vortices are advected down the cocoon and create pressure waves within the cocoon.
\mbox{Figure \ref{fig:Cocoon-jet-coupling}} shows a close up of part of the $I1$ jet, denoted by the pink
contour, and its surrounding cocoon. The left panel of the plot shows the radial velocity component $V_{\rm R}$.
The red regions are moving away from the jet \mbox{($V_{\rm R}>0$)} and the blue regions are moving towards the
jet \mbox{($V_{\rm R}<0$)}. The alternating pattern in radial velocity that emerges is caused by the train of
vortices that have broken off the jet-head in a quasi-periodic and regular way. The right panel shows the
($\log_{10}$) of the gas pressure. Dark colors correspond to high pressure, and light colors to low pressure. The
red and blue contours indicate radial velocity \mbox{($V_{\rm R} = \pm 0.08 c$ respectively)}. Finally, the black
contours indicate pressure with values \mbox{$\log_{10}(P/P_{\rm ch}) = \{-6, \; 5.3, \; 4.7, \; -4\}$}.

When we compare the right panel to the left panel in \mbox{figure \ref{fig:Cocoon-jet-coupling}}, we find that
the regions of high pressure correlate well with gas moving towards the jet and that the regions of low
pressure correlate with gas moving away from the jet. This means that the alternating pattern in radial
velocity directly corresponds with the pressure waves traveling down the cocoon. Moreover, it can be seen that
at the sites where the high pressure regions are in contact with the jet boundary, the jet is compressed, which
consequently leads to the formation of a strong internal shock.

All three models show this correspondence between the formation of strong internal shocks and the pressure
waves caused by the back-flowing vortices. Therefore, we can directly translate the number of internal shocks
along the jet axis to the number of back-flowing vortices in the cocoon. The jets $H1$, $I1$ and $A1$ all have
approximately 9 internal shocks at the final time of simulation, $t = 22.8$ Myr. We find a fairly
regular pattern in the pressure waves traveling down the cocoon, which points towards an approximately constant
vortex cycle time, $t_{\rm vortex}$. We find \mbox{$t_{\rm vortex} \sim 2.5$ Myr}. With the assumption
of constant vortex cycle time, we can approximate the number of shocks $N_{\rm s}$ that have formed along the
jet axis, at a time $t$ after the jet has been injected by:

\be
N_{\rm s} \approx \frac{t}{t_{\rm vortex}} \; = \; 0.40 \; t \; ,
\ee
with $t$ in Myr. At time $t$, a cocoon with length $L_{\rm co} = \beta_{\rm hd} \; t$ will have formed, where
$\beta_{\rm hd}$ is the jet-head advance speed. Therefore, the relation between cocoon length and number of
shocks is also given by:

\be
N_{\rm s} \approx \frac{L_{\rm co}}{d_{\rm vortex}} \; ,
\ee
where $d_{\rm vortex} = \beta_{\rm hd} \; t_{\rm vortex}$ is the average distance between two vortices, for which
we find an average \mbox{$d_{\rm vortex} \sim 44$ kpc}.

One should note however that in a more realistic case where a jet is simulated in full 3$D$, using
magnetohydrodynamics, other types of instabilities might alter the behavior of the vortices at the jet-head.
In that case, the pressure waves traveling down the cocoon might be more irregular than in case of the 
$2.5D$ simulations that are described in this paper so that a typical vortex cycle time might not apply.
However, regardless of the type of instabilities, pressure waves are inevitably created that cause pressure
fluctuations in the cocoon. An increase of cocoon pressure will lead to jet constriction and the formation of
internal shocks.

\section{Discussion} \label{sec:Discussion} 

  \subsection{Will the jets appear as FRI or FRII jets?}
\label{subsec:FRIFRII}

Since all three jet models $H1$, $I1$ and $A1$ have been given a typical FRII jet power
(\mbox{$L_{\rm jt} = 4-5 \times 10^{46}$ erg s$^{-1}$}), one might expect that the jets will
further evolve as FRII jets as well. Indeed, we find that global features such as the length and the width of
the cocoon, the number of internal shocks along the jet axis and the stability of the jets do not show large
variations between the individual models. All three jets maintain their stability all the way up to the jet
head. Moreover, regions with relativistic gas are found downstream of the Mach disk which can be identified
with the hot spots of the jets. Collimated and undisrupted jets with hot spots at their jet-heads are two
typical signatures for FRII jets. Therefore, based on these signatures we conclude that the $H1$, $I1$ and
$A1$ jets (initiated as FRII jets) will all continue to further develop as FRII jets. 

  \subsection{Jet-head structure and the mixing of components}
\label{subsec:JetHeadandMixing}

For the homogeneous jet, the structure of the jet-head and the general behavior of the flow lines can be
fairly easily understood. As soon as it impacts the ambient medium, a bow shock and Mach disk are formed.
All jet material is shocked equally strong throughout the cross section of the jet. The shocked jet material
and the shocked ambient medium both have a high pressure and the associated pressure gradients cause the
shocked material to flow away from the jet-head. Vortices are created in this high pressure region, which
cause shocked jet material and shocked ambient medium material to mix strongly. This behavior can be seen in
the flow patterns of figure \mbox{\ref{fig:jetheads} {\bf A}}.

For jets with structure consisting of a fast moving jet spine and slower moving jet sheath, the formation,
structure, and evolution of the jet-head is more complex. In this case, the jet spine (with a higher bulk
velocity) initially impacts the shocked ambient medium before the jet sheath does. Material from the jet spine
will be shocked by the Mach disk and, together with the shocked ambient medium, form a preceding substructure
in the jet-head. Material in this substructure flows away from the bow shock due to pressure gradients, causing
turbulence and possibly vortices where shocked spine material and shocked ambient medium mix strongly.

The jet sheath, on the other hand, doesn't impact the shocked ambient medium directly, but it impacts this
preceding jet-head substructure. The pressure in the preceding jet-head will be high because the Mach disk and
the bow shock are both strong shocks. The shocked jet spine expands sideways after the Mach disk due to this
pressure jump. Because the jet sheath has slightly higher density, but lower inertia, it can be more easily
displaced than the jet spine. It is therefore pushed outwards, while still propagating towards the top of the
jet-head. Further out it is deflected due to the high pressure in the preceding jet-head, causing the shocked
sheath material to flow back into the cocoon. This flow pattern causes the hot spot to obtain a concave shape,
which is seen in both the models $I1$ and $A1$.

The back-flowing shocked sheath creates more vortices. At these vortices, shocked spine, shocked sheath and
shocked ambient medium material all mix strongly. As the vortices evolve and shocked material flows away from
the jet-head, the different constituents eventually evolve into an approximate homogeneous mixture at large
distances from the jet-head. Figure \ref{fig:Cocoon-jet-coupling} shows vortices in the cocoon of the $I1$ jet
as they are moving downstream from the jet-head. Figures \mbox{\ref{fig:ContourPlots} ({\bf E}} and {\bf F})
show that at large distances from the jet-head, where shocked gas from the cocoon has gone through much
turbulence, the mixture of shocked spine-sheath material in the cocoon has become approximately homogeneous.

  \subsection{Effective impact surface area of the ambient medium}
\label{subsec:EffImpactArea}

Even though the assumption of using an area-weighted average for the analytical prediction of the jet-head
advance speed is a rather simple one, the values obtained are remarkably close to the actual values at the
start-up phase of the simulations. However, not long after this start-up phase ($\gtrsim$ 2.5 Myr), we see the
advance speed of the jet-heads of all three models decline to a new, fairly constant value (although the advance
speed for model $I1$ continues to decline slightly during the entire run of the simulation). Taking an average
over the final 2.8 Myr shows that the jet-heads only propagate with \mbox{$\sim$ 15\% - 25\%} of their initial
velocity.

The total momentum discharge through the Mach disk (also see \citealt{Rosen1999}) is given by:

\be
  Q_{\rm jt} = \left[ \left( e + P \right)\gamma_{\rm MD}^2\beta_{\rm MD}^2 + P \right]_{\rm jt}\; A_{\rm jt}
\ee
where as before $e$ is the total internal energy density, $P$ is the pressure, $\gamma_{\rm MD}$ and
$\beta_{\rm MD}$ are the bulk jet Lorentz factor and velocity respectively as measured in the frame of the
Mach disk, and finally $A_{\rm jt}$ is the discharge surface area of the jet. One important note must be made.
The conditions in the jet, just before the Mach disk will significantly differ from those at the jet inlet. For
example, the surface of the cross section of the jet will be larger; the mass density, pressure and temperature
will differ; and the velocity profile will also have changed. However, the total momentum discharge must
approximately be constant at all $Z$, so we can safely use the parameters at jet inlet.
\lskip
\lskip
We explain the quickly decelerating jet-head advance speed by assuming that the impact surface area
\mbox{$A_{\rm jt}$} of the jet over which the momentum discharge takes place and the impact surface area of the
ambient medium $A_{\rm am}$ are not equal: \mbox{$A_{\rm jt} \ne A_{\rm am}$}
\footnote{There is also a small fraction of ambient medium material entrained by the jet. However, this
fraction is so small ($\sim 1\%$), we expect this effect not to influence the deceleration of the jet.}.
If we take these unequal surface areas into account in equation \equref{eq:JHAS}, the analytical prediction
for the jet-head advance speed becomes:

\be
\beta_{\rm hd} = \frac{\gamma_{\rm jt}\sqrt{\eta_{\rm R}}\; \beta_{\rm jt}}{\displaystyle
\Omega + \gamma_{\rm jt}\sqrt{\eta_{\rm R}}} \; ,
\label{eq:JHAS2}
\ee
where $\Omega = \sqrt{\frac{A_{\rm am}}{A_{\rm jt}}}$ is the square root of the ratio of impact surface areas.
In these $2.5D$ simulations, the impact surface area of the jet is a disk $A_{\rm jt} = \pi R_{\rm jt}^2$ with
radius $R_{\rm jt}$, and the impact surface area of the ambient medium is a disk
\mbox{$A_{\rm am} = \pi R_{\rm am}^2$} with radius $R_{\rm am}$. From this we find for the effective impact
radius:

\be
R_{\rm am} = \gamma_{\rm jt} \sqrt{\eta_{\rm R}} \left( \frac{\beta_{\rm jt}}{\beta_{\rm HD}} - 1
\right) R_{\rm jt} \; ,
\label{eq:Rext}
\ee
where $\beta_{\rm HD}$ is now the actual measured jet-head advance speed. We substituted the results for the
jet-head advance speed we found from the simulations and listed the effective impact radii in
\mbox{table \ref{tab:JHAS}}. As can be seen in figure \mbox{\ref{fig:jetheads} ({\bf A}, {\bf B} and {\bf C})},
the yellow flow lines represent the effective impact radius for the three jet models. The size of the impact
area corresponds very well to the (projected) size of the turbulent region of shocked gas in the jet-head
(the green flow lines), and is approximately similar to the cross section of the jet, just before the Mach disk.

The effective impact area is $\Omega^2$ times as large as in the case predicted by simple theory. This is 16 for
the homogeneous jet, 30 for the isochoric jet and even 40 for the isothermal jet, which shows that AGN jets are
capable of shock-heating fairly large regions of the surrounding intergalactic medium.

  \subsection{Jet structure and transverse jet integrity}
\label{subsec:JetStructureandIntegrity}

    \subsubsection{The effect of a radial variation in density $\rho(R)$ and relativistic inertia
                   $\gamma^2 \rho h(R)$}
\label{subsubsec:EffectOfrho}

In this section, we discuss the stability of the jet, as well as the integrity of the jet in the {\em radial}
direction. The choice for the radial initialization of a jet strongly determines its further evolution as
it propagates into the ambient medium. In particular, the difference between the three models of the radial
density variation $\rho(R)$ and the relativistic inertia (perpendicular to the jet flow) $\gamma^2 \rho h(R)$
play an important role. \mbox{Table \ref{tab:shockpropagation}} shows the values of the density and relativistic
inertia at the jet axis \mbox{$R=0$}; the jet spine-sheath interface \mbox{$R=R_{\rm sp}$} and the jet
boundary \mbox{$R=R_{\rm jt}$}.

As discussed in \mbox{section \ref{subsec:CocoonJetCoupling}} on cocoon-jet coupling, pressure
gradients in the cocoon surrounding the jet can cause the jet to be compressed. Such a compression causes a shock
to form in the centre part of the jet (therefore, in the case of structured spine-sheath jet, in the jet spine).
Such an internal shock is capable of propagating through the jet itself, in poloidal, as well as in the radial
direction. After a strong internal shock has occurred in the jet centre, the post-shock jet material gets heated,
causing the jet material to expand sideways. The influence of different density- and relativistic inertia
variations on the propagation of such a shock, and its relation to internal mixing between jet spine and jet
sheath material shall be discussed for each individual model.

\paragraph*{The homogeneous jet ($H1$)}
maintains its radial integrity almost entirely up to the jet-head. At the jet head, the jet material has crossed
9 strong internal shocks, but these have not disrupted the jet flow, or lead to a large amount of mixing between
jet and surrounding cocoon material. This can be explained as the result of the uniform density and Lorentz
factor over the entire cross section of the jet. This leads to a uniform shock strength, and the post-shock
sideways expansion of the jet material will be similar at all radii as the relativistic inertia,
\mbox{$\gamma^2 h \rho$} with \mbox{$h \sim 1$}, is constant across the entire cross section (also see
\mbox{tabel \ref{tab:shockpropagation}}).

The absence of a radial density- and relativistic inertia variation results in a stable jet with a stable radial
structural integrity (see for example line plot \mbox{\ref{fig:CrossCuts} {\bf D}} and the radial cuts in 
\mbox{figure \ref{fig:RadialCuts} ({\bf A}, {\bf B} and {\bf C})}.

\paragraph*{The isothermal jet ($I1$)}
maintains a considerable amount of its radial structural integrity up to large distances from the jet inlet. Its
density varies smoothly in the radial direction of the jet (see \mbox{figure \ref{fig:Initial} {\bf C}})
\footnote{Note that this is a consequence of choosing the isothermal jet to have a constant temperature across
the entire jet cross section.}.
When a shock emerges in the jet spine due to a pressure variation in the cocoon, these shocks propagate outwards.
When the shock reaches the jet spine-sheath interface, the shock will not be reflected, since the density
variation across this interface is zero. Moreover, the inertia of the jet sheath is a factor 4 less than that of
the jet spine (see \mbox{table \ref{tab:shockpropagation}}). Therefore, the jet sheath will easily be deflected
outwards by the shock-heated jet spine material.

The absence of shock reflections at the jet spine-sheath interface, together with a jet sheath that has lower
relativistic inertia than the jet spine yield inefficient internal mixing between the jet spine and jet sheath
material. Therefore, a considerable amount of the radial structural integrity will be maintained up to large
distances from the jet inlet. This behavior can be seen in line plot \mbox{\ref{fig:CrossCuts} {\bf E}} and from
the radial cuts in \mbox{figure \ref{fig:RadialCuts} ({\bf D}, {\bf E} and {\bf F})}.

\paragraph*{The isochoric jet ($A1$)}
loses its radial structure fairly quickly after the jet is injected into the system. As with the isothermal jet,
shock develop in the jet spine and propagate outwards. However, in this case, when such a shock reaches the jet
spine-sheath interface, the shocks are largely reflected because of a jump in density by a factor of 5 in jet
sheath. Moreover, the jet sheath has slightly higher relativistic inertia than the jet spine, which enhances the
effect of the shock reflections (see \mbox{table \ref{tab:shockpropagation}}). Each internal shock reflection
internally mixes the jet spine and jet sheath more, and therefore soon after inlet, the clear spine-sheath jet
structure will be lost. This behavior can be seen in line plot \mbox{\ref{fig:CrossCuts} {\bf F}}
and the radial cuts in \mbox{figure \ref{fig:RadialCuts} ({\bf G}, {\bf H} and {\bf I})}.

\nskip

The relation between shock reflections and internal jet spine-sheath mixing is explained as follows:
Flow lines that cross a shock are deflected away from the shock normal. Such a deflection
will create vorticity in the jet flow. A shock front that encounters a medium with a higher density (as is the
case for the jet spine-sheath interface of the isochoric jet) will partially be reflected and partially be
transmitted. A higher density contrast will result in a larger part of the shock being reflected. Near such a
shock reflection, the creation of vorticity will be strong due to the strong deflection of the flow
lines. At the interface between jet spine and jet sheath, vorticity is directly responsible for the mixing of the
jet constituents. Moreover, the emergence of vorticity at the jet spine-sheath interface is capable of triggering
the Kelvin-Helmholtz instability. This instability itself creates vorticity and will eventually lead to a
turbulent flow.

Therefore, it is expected that the internal mixing between the jet spine and the jet sheath in jets with a large
density jump across the jet spine-sheath interface will be stronger than in the case where this density jump is
small, or non-existent.

    \subsubsection{Critical azimuthal velocity}
\label{subsubsec:CriticalAzimuthalV}

{\renewcommand{\arraystretch}{1.3}
\begin{table}
\caption{Variation of density and relativistic inertia for the jets $H1$, $I1$ and $A1$, in the radial direction.}
  \rowcolors{1}{white}{lightgray}
    \begin{tabular}{l l c c c} \hline
     {\bf Quantity (\mbox{$10^{-3} m_{\rm p}n_{\rm ch}$})}
                                         & {\bf at radius}      & {\bf H1} & {\bf I1} & {\bf A1}
      \\ \hline
     {\em Density} $\rho$                & $R = 0$              & 4.55     & 3.31     & 1.00
     \\
     \phantom{}                          & $R = R_{\rm sp}^{-}$ & 4.55     & 4.89     & 1.00
     \\
     \phantom{}                          & $R = R_{\rm sp}^{+}$ & 4.55     & 4.89     & 5.00
     \\
     \phantom{}                          & $R = R_{\rm jt}$     & 4.55     & 5.00     & 5.00
     \\
     \phantom{} & \phantom{} & \phantom{} & \phantom{} & \phantom{}
     \\
     {\em Rel. inertia} $\gamma^2\rho h$ & $R = 0$              & 44.1     & 119.0    & 36.0
     \\
     \phantom{}                          & $R = R_{\rm sp}^{-}$ & 44.1     & 176.0    & 36.0
     \\
     \phantom{}                          & $R = R_{\rm sp}^{+}$ & 44.1     & 44.0     & 45.0
     \\
     \phantom{}                          & $R = R_{\rm jt}$     & 44.1     & 45.0     & 45.0
     \\ \hline
    \end{tabular}

\medskip
Table showing the density $\rho$ and the relativistic inertia $\gamma^2\rho h$ of the initial radial jet profile,
both in units of \mbox{$10^{-3} m_{\rm p}n_{\rm ch}$}, at different distances from the jet axis:
\mbox{$R = 0$} corresponds to the jet axis; \mbox{$R = R_{\rm sp}^{-}$} corresponds to the point just inside of
the jet spine-sheath interface; \mbox{$R = R_{\rm sp}^{+}$} corresponds to the point just outside of the
jet spine-sheath interface; and finally \mbox{$R = R_{\rm jt}$} corresponds to the jet boundary.
    \label{tab:shockpropagation}
\end{table}
}

\begin{figure*}
$
\begin{array}{c c}
\includegraphics[width=0.5\textwidth]{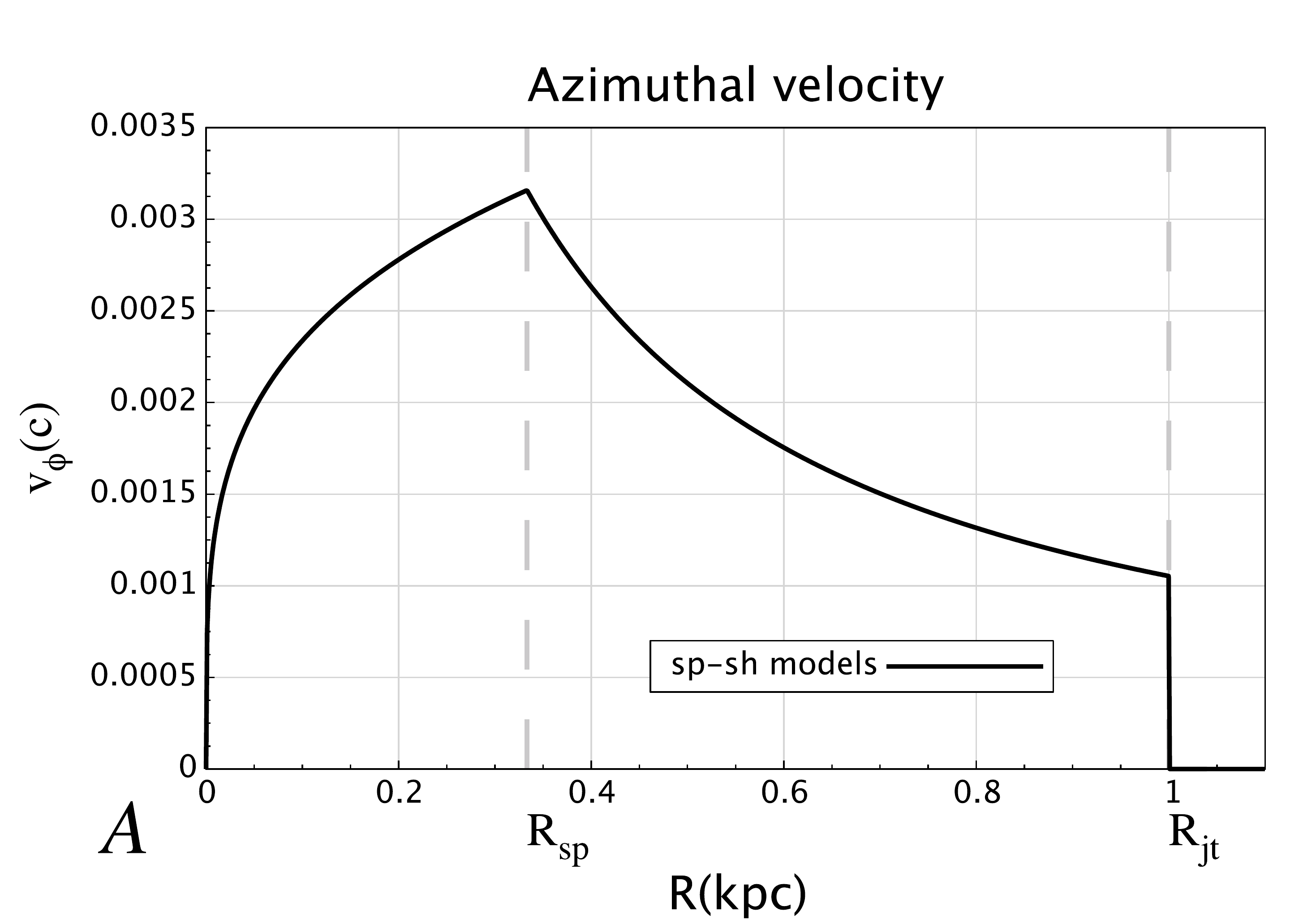} &
\includegraphics[width=0.5\textwidth]{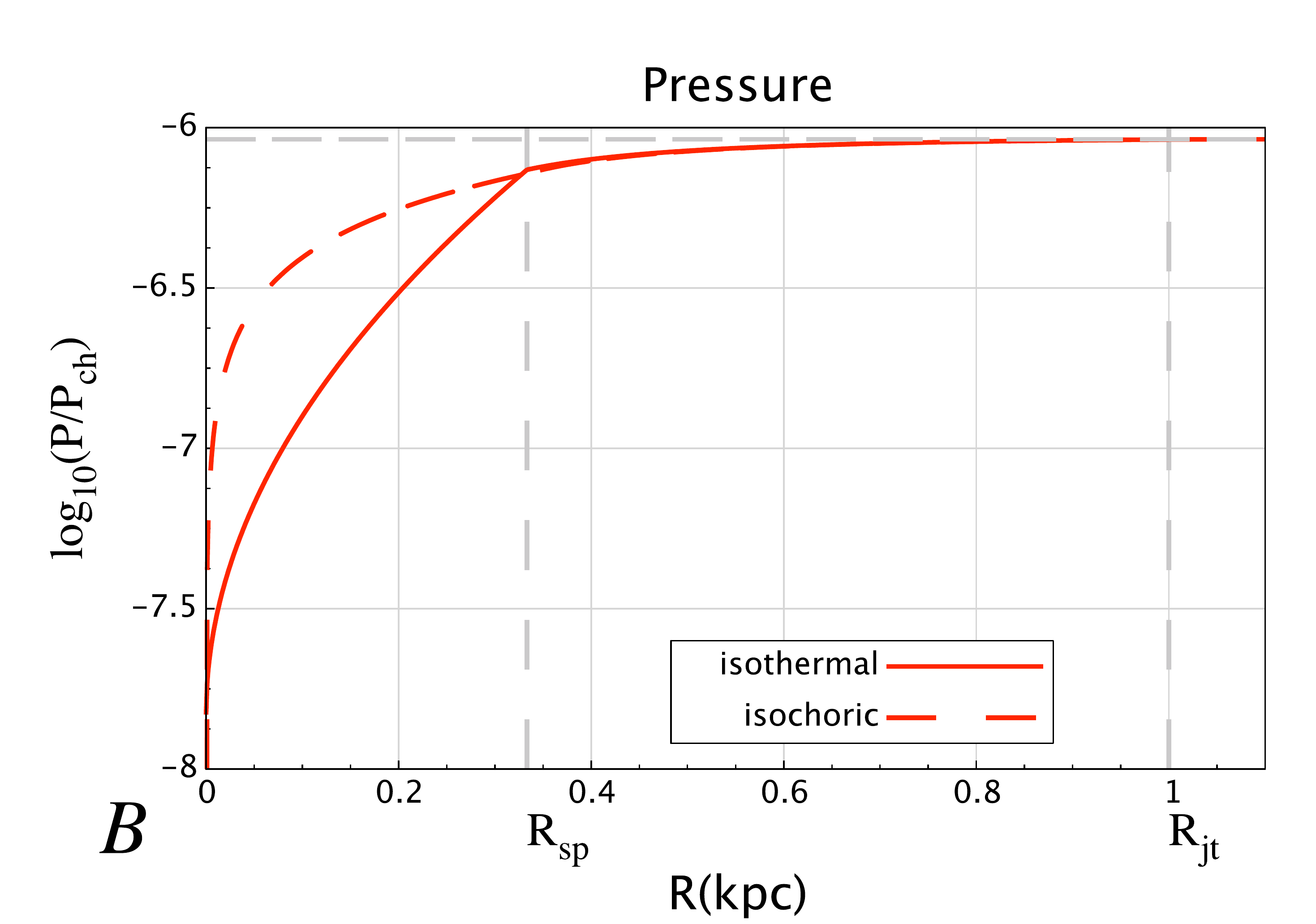} \\
\includegraphics[width=0.5\textwidth]{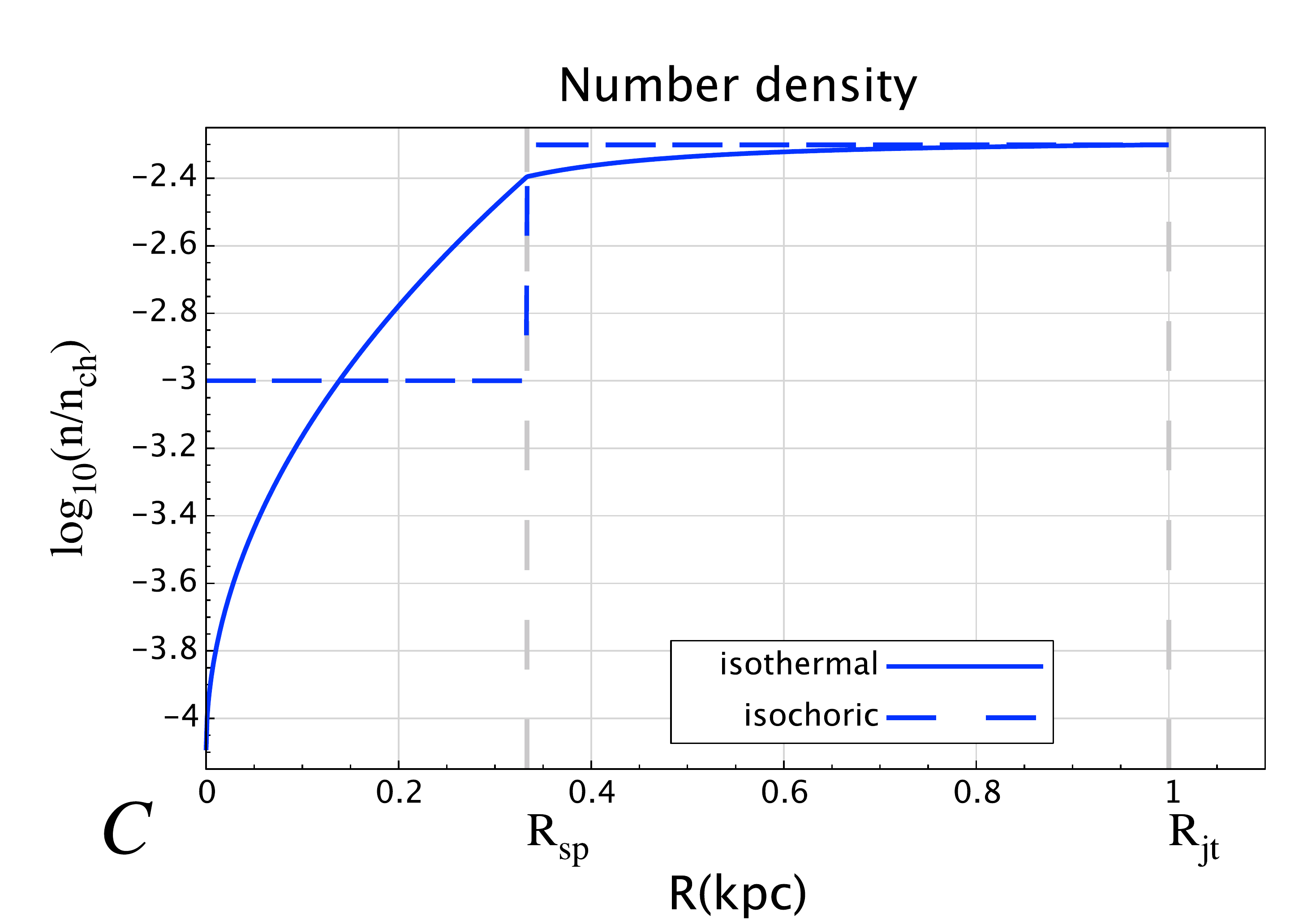} &
\includegraphics[width=0.5\textwidth]{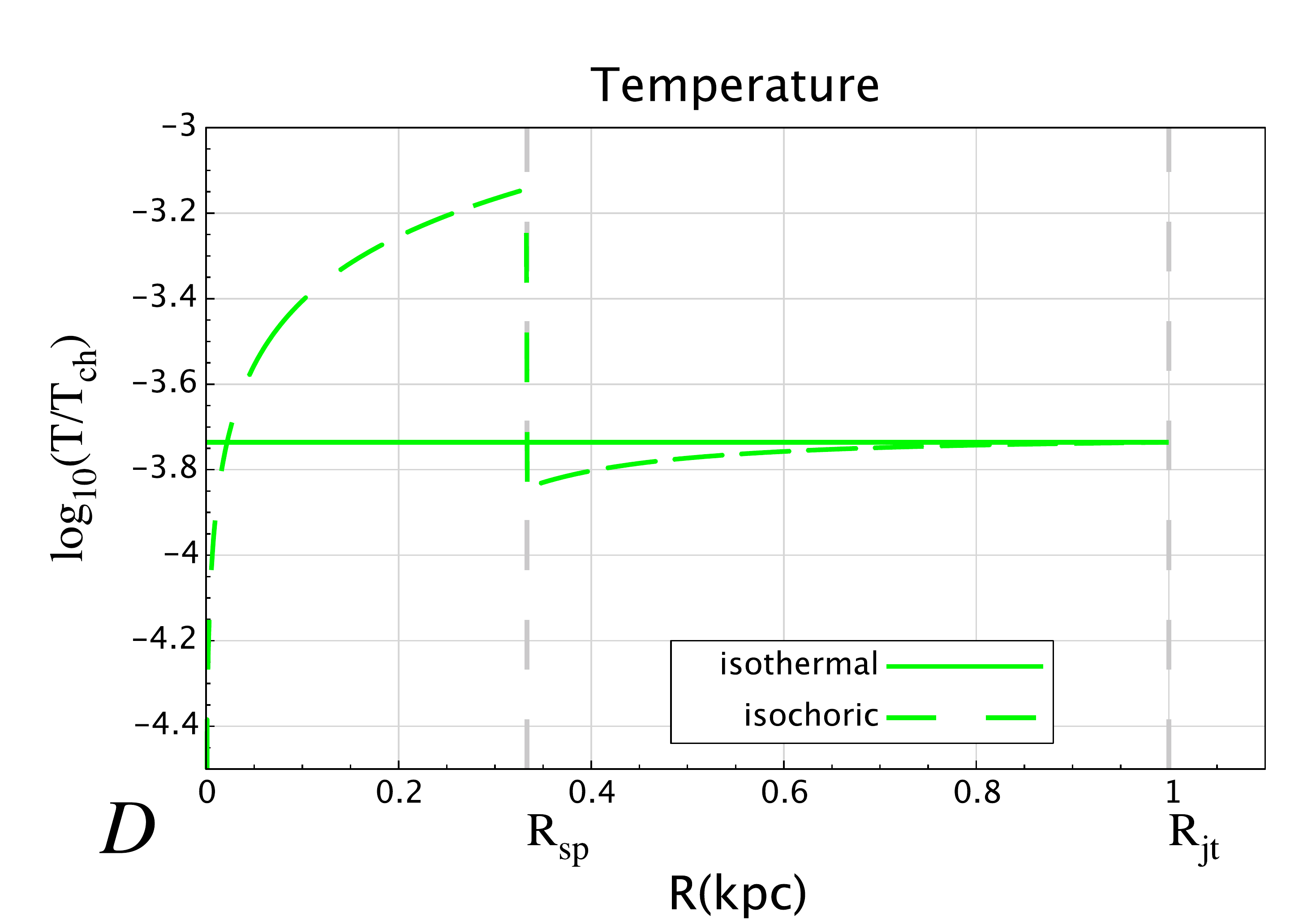}
\end{array}
$

\caption{Initial transverse jet profiles for the isothermal jet (solid lines) and the piecewise isochoric jet
(dashed lines), {\em but now in the case of} \mbox{$V_{\phi} = V_{\rm \phi, C}^{\rm A} = 3.159 \times 10^{-3} c$}.
The cross cuts show in {\em black} the profile of azimuthal rotation $v_{\phi}(R)$ \mbox{(figure {\bf A})}. This
rotation profile has been used for both the isothermal and the isochoric jet model; In {\em red}, the $\log_{10}$
of the pressure $P$ in units of the characteristic pressure
\mbox{$P_{\rm ch} = 1.50\times 10^{-6}$ erg cm$^{-3}$} \mbox{(figure {\bf B})}; In {\em blue} the $\log_{10}$ of
the number density $n$ in units of the characteristic number density \mbox{$n_{\rm ch} = 10^{-3}$ cm$^{-3}$}
\mbox{(figure {\bf C})}; and in {\em green} the $\log_{10}$ of the thermal temperature $T$ in units of the
characteristic temperature \mbox{$T_{\rm ch} = 1.09\times 10^{13}$ K} \mbox{(figure {\bf D})} of the jet. In
addition, the images show the jet radius at \mbox{$R_{\rm jt} = 1$ kpc} and the jet spine radius at
\mbox{$R_{\rm sp} = R_{\rm jt}/3$} as the two vertical dashed lines. The pressure of the ambient medium is
denoted by the dashed horizontal line in \mbox{figure {\bf B}}.}

  \label{fig:InitialCrit}
\end{figure*}

Regardless of the physical mechanism (for example, constant angular velocity $\Omega$, or constant specific
angular momentum $\lambda$) that eventually leads to a certain amount of rotation at a certain distance from
the central engine, there is another physical restriction that determines an upper-limit for the amount of
rotation, the so-called {\em critical azimuthal velocity}, $V_{\rm \phi, C}$. The critical azimuthal velocity
follows from demanding the gas pressure on the jet axis to remain positive, \mbox{$P(R=0) \ge 0$}.

In case of the isothermal jet, we derived in section \ref{subsubsec:Pressure_I} that this
translates into the condition \mbox{$V_{\phi} < \sqrt{1 - V_{z}^2}$}. Since the jet spine of the
structured jets has a Lorentz factor \mbox{$\gamma_{\rm sp} = 6$}, the critical azimuthal velocity of the
isothermal jet equals \mbox{$V_{\rm \phi, C}^{\rm I} = 0.167$}.

The condition for the pressure to remain positive on the jet axis in case of the isochoric jet (section
\ref{subsubsec:Pressure_A}), leads to an expression that cannot be solved analytically. We have made a
second-order Taylor expansion in \mbox{$(\Delta V^2) = V_{\phi}^2 - V_{\phi,0}^2$} for this expression to
obtain an accurate value for the critical azimuthal velocity. Here, the value of $V_{\phi,0}$ is assumed to
lie close to the true value of the critical azimuthal velocity. We determined the critical azimuthal velocity
for the isochoric jet to be: \mbox{$V_{\rm \phi, C}^{\rm A} = 3.159 \times 10^{-3}$}.

Figure \mbox{\ref{fig:InitialCrit} ({\bf A}, {\bf B}, {\bf C} and {\bf D})} shows the initial radial jet
profiles for both the isothermal jet, as well as the isochoric jet as they would appear for a critical
azimuthal velocity \mbox{$V_{\phi} = V_{\rm \phi, C}^{\rm A} = 3.159 \times 10^{-3}$}. Clear differences can be
seen between these radial jet profiles, and the ones shown in figure
\mbox{\ref{fig:Initial} ({\bf A}, {\bf B}, {\bf C} and {\bf D})}, especially in the jet spine region.

Our choices for the self-similarity constants $a_{\rm sp}$ and $a_{\rm sh}$ have been discussed in section
\ref{sec:Theory}. We chose the self-similarity constant in the jet sheath $a_{\rm sh}=-2$, motivated by constant
specific angular momentum. The self-similarity constant of the spine needs to be positive, however, its choice
\mbox{$a_{\rm sp} = 1/2$} was somewhat arbitrary. The criterion for internal stability
(\mbox{$\{a_{\rm sp},a_{\rm sh}\}>-2$}) does, however, allow for different values for $a_{\rm sp}$ and
$a_{\rm sh}$. For completeness' sake, the authors have studied the effect of choosing different sets
\mbox{$\{a_{\rm sp},a_{\rm sh}\}$} on the initial radial jet profiles. It turns out that with the present
choice of the maximum azimuthal velocity, $V_{\phi}$, different (but still realistic) values for
\mbox{$\{a_{\rm sp},a_{\rm sh}\}$} do not strongly influence the resulting radial jet profiles, for both the
isothermal jet, as well as for the isochoric jet. We therefore argue that with our present choice,
\mbox{$a_{\rm sp} = 1/2$}; \mbox{$a_{\rm sh} = -2$} and \mbox{$V_{\phi} = 1\times 10^{-3}$}, the resulting jet
profiles should be representative for the actual isothermal and isochoric jets, even in the case where the
actual self-similarity constants of the jets differ (modestly) from the values that were used in this paper.
These specific jet models are relatively insensitive to different choices for \mbox{$\{a_{\rm sp},a_{\rm sh}\}$}. 

However, strong responses to the choice of \mbox{$\{a_{\rm sp},a_{\rm sh}\}$} do occur as one lets the maximum
azimuthal velocity approach the critical azimuthal velocity. For a more complete understanding of both the
fundamental properties of, as well as the fundamental differences between the isothermal and isochoric jet models
(for example looking at mixing effects and jet integrity), the case where the maximum azimuthal velocity
constant approaches the critical azimuthal velocity, \mbox{$V_{\phi} \longrightarrow V_{\rm \phi,C}$}, should be
studied as well.

    \subsubsection{The effect of cylindrical symmetry on mixing effects}
    \label{subsubsec:EffectOfcylsymm}

As a last note, it needs to be mentioned that the amount of mixing presented in this paper should be taken
with some caution. Performing simulations in axisymmetric $2.5D$ models has the advantage of producing
"some aspects" of $3D$ behavior, while keeping the amount of computation time manageable. However, in full $3D$,
in addition to the type of instabilities that occur in this $2.5D$ study, also different kind of instabilities
will occur, which will automatically lead to more turbulence and possibly less stable jets. As we discussed,
turbulence is the primary mechanism behind the amount of mixing (along the jet axis, as well as in the
surrounding cocoon). It is therefore expected that $2.5D$ models likely give an underestimation as compared to
the amount of mixing that would occur in full $3D$ simulations. 

{\renewcommand{\arraystretch}{1.3}
\begin{table*}
  \caption{Summary of the most appreciable characteristics of the three jet models $H1$, $I1$ and $A1$.}
  \rowcolors{1}{white}{lightgray}
    \begin{tabular}{p{0.22\linewidth} | p{0.22\linewidth}  p{0.22\linewidth}  p{0.22\linewidth}} \hline
      {\bf Jets at} $\bm{t = 22.8}$ {\bf Myr} & {\bf Homogeneous} $\mathbf{H1}$ &
      {\bf Isothermal} $\mathbf{I1}$ & {\bf (Piecewise) isochoric} $\mathbf{A1}$
       \\ \hline
      Jet length & {\it 409 kpc} & {\it 421 kpc} & {\it 370 kpc}
      \\
      Maximum cocoon width & {\it 85 kpc} & {\it 90 kpc} & {\it 86 kpc}
      \\
      Appearance hot spot& {\it elongated} & {\it concave (bowl) shape} & {\it concave (bowl) shape}
      \\
      Internal shocks along jet axis& {\it 9} & {\it 9} & {\it 9}
      \\
      Appearance internal shocks & {\it single peaks in density and temperature} &
                                   {\it single peaks in density and temperature, followed by variable behavior} &
                                   {\it single peaks in density and temperature, followed by weaker shocks}
      \\
      Jet-head & {\it sharply peaked} & {\it wide round shape} & {\it wide round shape}
      \\
      Temperature of the Mach disk (on the jet axis) & {\it $9.7 \times 10^{12}$ K (relativistic)} &
                                                       {\it $1.4 \times 10^{13}$ K (relativistic)} &
                                                       {\it $1.1 \times 10^{13}$ K (relativistic)}
      \\
      Mixing within the cocoon & {\it  shocked jet- shocked ambient medium: large region of strong absolute
                                       mixing, but only little mass-weighted mixing. Far from a homogeneous
                                       mixture} &
                                 {\it  shocked spine - shocked sheath: considerable mixing within entire cocoon
                                       Approximately homogeneous mixture for distances larger than 291 kpc from
                                       jet-head with $\Lambda_{\rm sp-sh} \sim 0.8 - 1$} &
                                 {\it  shocked spine - shocked sheath: considerable mixing within entire cocoon
                                       Approximately homogeneous mixture for distances larger than 100 kpc from
                                       jet-head with $\Lambda_{rm sp-sh} \sim 0.9 - 1$}
      \\
      Mixing along jet axis & {\it jet-shocked ambient medium:  No notable mixing along entire axis, except at the
                                   jet-head} &
                              {\it jet spine - jet sheath: considerable absolute mixing
                                   (\mbox{$\Delta_{\rm sp-sh} < 0.5$}), but weak mass-weighted mixing
                                   (\mbox{$\Lambda_{\rm sp-sh} < 0.1$}). Predominantly
                                   jet spine material along jet axis} &
                              {\it jet spine - jet sheath: strong absolute mixing
                                   (\mbox{$\Delta_{\rm sp-sh} > 0.5$}), and notable mass-weighted mixing
                                   (\mbox{$\Lambda_{\rm sp-sh} < 0.3$}). Predominantly jet sheath material
                                   along jet axis for $Z > 82$ kpc} 
      \\
      Radial jet structure before hot spot & {\it jet integrity maintained all the way up to the hot spot} &
                                             {\it jet spine-sheath structure can still be recognized through
                                                  the abundance of the constituents in radial direction} &
                                             {\it jet spine-sheath structure can no longer be recognized}
      \\
      Jet-head advance speed after start-up phase & {\it 0.047} $c$ & {\it 0.032} $c$& {\it 0.035} $c$
      \\ \hline
    \end{tabular}

\medskip
The left column contains some of the characteristic features that have been discussed in the previous sections.
The second column shows these characteristics for model $H1$, the third for model $I1$ and the right column for
model $A1$.
    \label{tab:H1I1A1Characterisics}
\end{table*}
}

\section{Conclusions} \label{sec:Conclusions}

In this paper, we compared three different AGN jet models with a jet power of
\mbox{$L_{\rm jt} \sim 4-5 \times 10^{46}$ erg s$^{-1}$}, a typical energy for powerful FRII radio sources.
We simulated a homogeneous jet (denoted as $H1$) without radial structure or angular momentum, and two different
jet models with a radial spine-sheath jet structure carrying angular momentum. For both spine-sheath jets, the jet
spine has lower mass density and higher Lorentz factor than the jet sheath. The first spine-sheath jet (denoted
as $I1$) is a jet set up with an isothermal equation of state. The second spine-sheath jet (denoted as $A1$) is
a (piecewise) isochoric jet: set up with constant, but different density for jet spine and jet sheath, using a
polytropic index of $\Gamma = 5/3$.

We simulate these jets in the case of a steady scenario where the jets have been active for 22.8 Myr. The jets
are under-dense as compared to the ambient medium by a factor of \mbox{$\eta \sim (1 - 5) \times 10^{-3}$}. All
three jet models reach approximately the same distance of 400 kpc, where the individual models differ by a few tens
of kpc. At the final time of simulation, all three jets have developed approximately 9 strong internal shocks.
The emergence of these shocks can be directly linked to the shedding of vortices by the jet-head, causing
pressure waves to travel down the cocoon and compress the jet at the high pressure regions. At the Mach disk,
all three jets are shocked to relativistic temperatures. This relativistically hot gas is identified with
the hot spot of the jet. Based on the stability of the jets and the appearance of hot spots, we conclude that
all three jets will have developed as typical FRII jets.

We find that the homogeneous jet ($H1$) remains regular all the way up to the jet-head with the same Lorentz
factor as it was initiated at the jet inlet. It has an elongated hot spot and a fairly flat Mach disk.

The isothermal jet ($I1$) loses its structural integrity slowly with increasing distance from the jet inlet.
However, all the way up to the jet-head, the core of the jet is still predominantly made up of jet spine
material and the surrounding layer still consists predominantly of jet sheath material. The Lorentz factor of the
central part of the jet is still slightly higher than that of the surrounding jet sheath. Due to the fact that
the density contrast between jet spine and jet sheath is zero at the jet spine-sheath interface, strong internal
shocks are {\em not} reflected at this interface. Instead, shock-heated jet spine material merely displaces the
jet sheath outwards. This causes the jet spine and the jet sheath to internally mix relatively {\em inefficient}
within the isothermal jet.

The isochoric jet ($A1$) loses most of its structural integrity after the jet is injected into the system. After
crossing two strong internal shocks, the central part of the jet is dominated by jet sheath material and the
Lorentz factor is only slightly higher than that of the surrounding jet sheath. At the jet-head, the jet spine
material has been smeared out over a large part of the jet cross section and there is only a thin outer region
that is still made up predominantly of jet sheath material. This difference is caused by a strong jump in
density at the interface between the jet spine and the jet sheath. Shocks that occur within the jet spine are
reflected as soon as they encounter this jet spine-sheath interface. Every shock reflection (internally) mixes
the jet spine and jet sheath material further. This causes {\em efficient} internal mixing between jet spine and
jet sheath material in the isochoric jet.

Both the isothermal jet and the isochoric jet have a concave (bowl-shaped) hot spot. This is explained by
considering the complex flow behavior at the jet-head. For both structured jets we find that at each strong
internal shock, only the central part of the jet is shocked and that the jet sheath merely is radially deflected
by the shock-heated jet spine material. Therefore, at the jet-head, the jet sheath is not terminated at
the Mach disk, but continues to propagate further towards the top of the jet-head. Then, at some point the
pressure at the jet's head becomes large enough to deflect the jet sheath, from which point on it moves further
down the cocoon, away from the jet-head. This flow pattern of jet sheath material deforms the hot spot, giving it
the concave shape.

Finally, we find that the propagation speed of the jet-heads is less than predicted from simple theory. We
find that this is most probably caused by an enlarged, effective impact area of the ambient medium that
interacts with the jet. Taking this effect into account, we find that the effective impact area varies from
being 16 times as large (in case of the homogeneous jet), to 30 times as large (in case of the isochoric jet)
and can be up to 40 times as large (as in the case of the isothermal jet). The size of this effective impact
area corresponds well to the size of the hot spots and the vortices (projected on the plane perpendicular to
the jet axis) that make up part of the jet-head.

Since the homogeneous jet, the isothermal jet and the isochoric jet were all given the same jet power, and
in addition there were merely subtle differences in the radial pressure- and density profiles of the two
structured spine-sheath jets, we consequently found that a number of aspects of the jets and the cocoons (e.g.
cocoon length and width, number of internal shocks, temperature along the jet axis and the occurrence of an
enlarged effective impact area) are fairly similar. Regarding jet integrity, jet-head morphology and internal
mixing efficiency between the jet spine and the jet sheath, on the other hand, we found prominent differences
(see table \ref{tab:H1I1A1Characterisics} for the most notable features). It is therefore expected that with
increasing difference in parameters between the individual models (such as maximum azimuthal velocity, or
density contrasts), more prominent distinctive features will occur. The influence of radial jet stratification
on jet integrity, jet-head morphology and the development of internal shocks have become apparent. They should
therefore be taken into account when one tries to model jets in realistic scenarios.

  \subsection{Continuation of this work}
\label{subsec:Continuation}

In order to investigate the differences between the individual models further, different parts of parameter space
need to be considered. For example, close to the central engine, the density contrast between a jet spine and jet
sheath, (maximum) azimuthal velocity, or Lorentz factors might be much higher and there the influence of magnetic
fields should be involved as well. Another example of where other parts of parameter space apply is in the case
of jets coming from micro-quasars. Moreover, in the case where the central engine shows episodic behavior
(multiple subsequent outbursts), the properties of the ambient medium into which the jets are injected changes
dynamically. It is expected that this changing environment will have large consequences for the jet propagation,
jet stability and jet integrity. Comparing the isothermal jet model to the isochoric jet model for these other
cases might lead to new insights. An extension to the work done by \citet{Meliani2009}, where the evolution of
the cross section of an isochoric jet was described, might also be performed in case of the isothermal jet model.
A useful question would be if the relativistically enhanced, rotation-induced Rayleigh-Taylor instability leading
to internal mixing of jet spine and jet sheath material (in cases where cylindrical symmetry no longer
applies), will also occur in the isothermal jet model. Eventually simulating these models in 3D might yet reveal
more fundamental differences and characteristics (as for example in mixing effects) of the different jet models.

In the follow-up of this paper we will be concerned with the case of episodic activity. Again we will compare
the homogeneous jet, the isochoric jet and the isothermal jet. There we will also be concerned with jet
stability and jet integrity; jet-head advance speed; effective impact area; and mixing effects of (shocked, as
well as unshocked) spine and sheath material, between spine and sheath coming from the same outburst episode,
as well as mixing of components coming from different outbursts.

\section*{Acknowledgements}
This research is funded by the {\it Nederlandse Onderzoekschool Voor Astronomie} (NOVA).
S.M. also acknowledges support from The European Communities Seventh Framework Programme (FP7/2007-2013)
under grant agreement number ITN 215212 “Black Hole Universe”; Z.M. acknowledges financial support from
the PNHE; and R.K. acknowledges financial support from FWO-Vlaanderen, project G.02308.12.

\footnotesize{
\bibliographystyle{./mn2e} 
\bibliography{2012References}
}

\bsp

\appendix

\section[]{Passive scalar advection: The tracing of a constituent}\label{sec:Tracers}

A tracer $\theta(t,\bmath{r})$ that is transported along flow lines can only change under the influence
of diffusivity and external sources such as gravitation. In the case of ideal hydrodynamics and in the
case where there is no diffusivity (apart from numerical effects which are small), a tracer is
passively advected. In this case the tracer satisfies the equation:

\be
\frac{d\theta}{dt} = \partial_t\theta + \bm{v} \cdot \nabla \theta = 0 \; .\label{eq:TrAdv}
\ee
By assigning different tracer values to various constituents of the flow, one can distinguish these constituents
even in a complex flow geometry. The advection equation for the mass-density (the continuity equation) reads:
\be
\partial_t D + \nabla \cdot(\bm{v}\: D) = 0\; . \label{eq:continuity}
\ee

Since MPI-AMRVAC uses the conservative formulation of the fundamental equations \equref{eq:conservationlaw},
we use a conservative tracer equation. To that end we define:

\be
\tilde\theta(t,\bm{r}) = \theta(t,\bm{r}) \: D(t,\bm{r}) \; .
\ee
By writing out \equref{eq:TrAdv} in terms of the rescaled tracer $\tilde\theta(t,\bmath{r})$ and making use of the
continuity equation \equref{eq:continuity}, the tracer advection equation can be written as:

\be
\partial_t\tilde\theta + \nabla \cdot (\tilde\theta \: \bm{v})=0 \; ,
\ee
hence, the rescaled tracer equation is in conservative form. MPI-AMRVAC will advect this rescaled tracer as a
regular flux variable. To obtain the actual tracer we substitute back
\be
\theta(t,\bm{r}) = \frac{\tilde\theta(t,\bm{r})}{D(t,\bm{r})} \; .
\ee
We choose $\tilde\theta(t,\bmath{r})$ to lie in the range
\be
-D(t,\bm{r}) \le \tilde\theta \le +D(t,\bm{r}) \; .
\ee
The original tracer will then lie in the range

\be
-1 \le \theta(t,\bm{r}) \le +1 \; .
\ee

\label{lastpage}

\end{document}